%% file: main.tex
\documentclass[a4paper,11pt]{article}
\pdfoutput=1
\usepackage{jheppub}

\usepackage[T1]{fontenc} 

\usepackage{amssymb,amsmath,bm,bbm,natbib}
\usepackage{color}
\usepackage{slashed}
\usepackage{graphicx}
\usepackage[utf8]{inputenc}
\usepackage{url}
\usepackage{dsfont}
\usepackage{cancel}
\usepackage[dvipsnames]{xcolor}
\usepackage{colortbl}
\usepackage{rotating}
\usepackage[export]{adjustbox}
\usepackage{subcaption}
\usepackage{soul}

\definecolor{Gray}{gray}{0.7}
\setstcolor{ForestGreen}

\renewcommand{\arraystretch}{1.3}

\preprint{PSI-PR-22-24,  ZU-TH 33/22}
	\title{\boldmath Comprehensive Analysis of Charged Lepton Flavour Violation in the Symmetry Protected Type-I Seesaw}

\author[a,b]{Andreas Crivellin}

\author[a,b]{Fiona Kirk}
		
\author[a,b]{Claudio Andrea Manzari}

\affiliation[a]{Physik-Institut, Universit\"at Z\"urich, Winterthurerstrasse 190, CH--8057 Z\"urich, Switzerland}
\affiliation[b]{Paul Scherrer Institut, CH--5232 Villigen PSI, Switzerland}
	
\emailAdd{andreas.crivellin@cern.ch}
\emailAdd{fiona.kirk@psi.ch}
\emailAdd{claudioandrea.manzari@physik.uzh.ch}

\input{Conventions}

\abstract{
The type-I seesaw model is probably the most straightforward and best studied extension of the Standard Model that can account for the tiny active neutrino masses determined from neutrino oscillation data. In this article, we calculate the complete set of one-loop corrections to charged lepton flavour violating processes within this model. We give the results both using exact diagonalisation of the neutrino mass matrix, and at at leading order in the seesaw expansion (i.e.~$\mathcal{O}(v^2/M_R^2)$). Furthermore, we perform the matching onto the $SU(2)_L$ invariant Standard Model Effective Field Theory at the dimension~6 level. These results can be used as initial conditions for the renormalisation group evolution from the right-handed neutrino scale down to the scale of the physical processes, which resums large logarithms. In our numerical analysis, we study the inverse seesaw limit, i.e.~the symmetry protected type-I seesaw, where the Wilson coefficient of the Weinberg operator is zero such that sizeable neutrino Yukawas are permissible and relevant effects in charged lepton flavour violating observables are possible. We correlate the different charged lepton flavour violating processes, e.g.~$\ell\to\ell^\prime\gamma$, $\ell\to3\ell^\prime$, $\mu\to e$ conversion and $Z\to \ell\ell^\prime$, taking into account the constraints from electroweak precision observables and tests of lepton flavour universality.}
\newpage

\begin{document} 
\maketitle

\section{Introduction}
Since neutrinos are massless within the Standard Model (SM), any explanation of the non-vanishing neutrino masses determined from neutrino oscillation data must involve new particles. The most studied scenario in this context is the extension of the SM by right-handed neutrinos, which reproduces a situation similar to that in the quark and charged lepton sector, where each left-handed field has a right-handed counterpart. This allows for Yukawa interactions, which, after electroweak symmetry breaking, give rise to Dirac mass terms for neutrinos. This minimal extension of the SM, referred to as the $\nu$SM~\cite{Mohapatra:1998rq}, appeals due to its simplicity, however, it is often considered unnatural since extremely small Yukawa couplings would be necessary to reproduce the observed neutrino masses, which are at most at the eV scale. Furthermore, the small $\nu$SM Yukawa couplings do not lead to any observable new physics effects in precision observables. Indeed, within the $\nu$SM e.g.~charged lepton flavour violating processes suffer from a GIM-like suppression by the active neutrino masses, leading to branching ratios below $10^{-45}$, which are unobservable.
\smallskip

A more natural explanation of the smallness of the neutrino masses can be provided by seesaw mechanisms, such at the type I seesaw~\cite{Minkowski:1977sc,Gell-Mann:1979vob,Yanagida:1979as,Mohapatra:1979ia,Schechter:1980gr}, which assigns large Majorana masses, $\MR$,to the right-handed neutrinos. In this case, the light neutrinos masses turn out to be proportional to $\MD\MR^{-1}\MDtransp$, where $\MD$ denotes the Dirac mass term. Depending on the scale of $\MR$, the type-I seesaw model can be probed at colliders~\cite{delAguila:2007qnc,Kersten:2007vk,Deppisch:2015qwa,Atre:2009rg,Antusch:2014woa,Banerjee:2015gca,Antusch:2016vyf,Pascoli:2018heg,Chakraborty:2018khw,Das:2018usr,Mekala:2022cmm}, it can be used as a framework for leptogenesis~\cite{Fukugita:1986hr,Davidson:2008bu,Boyarsky:2009ix,Antusch:2017pkq} or the right-handed neutrinos can be viewed as dark matter candidates~\cite{Dodelson:1993je,Shi:1998km,Asaka:2005an,Shaposhnikov:2006xi,Boyarsky:2009ix}. The discovery potential of the generic type-I seesaw is, however, limited, since the smallness of the active neutrino masses, as inferred from neutrino oscillation data, excludes sizeable Yukawa couplings to TeV-scale right-handed neutrinos.
\smallskip

Sizeable Yukawa couplings are admissible if an (approximate) lepton number symmetry~\cite{Ilakovac:1994kj,Wyler:1982dd,Mohapatra:1986bd,Branco:1988ex,Gonzalez-Garcia:1988okv,Barbieri:2003qd,Raidal:2004vt,Kersten:2007vk,Shaposhnikov:2006nn,Abada:2007ux,Asaka:2008bj,Gavela:2009cd} is imposed, which suppresses the Wilson coefficient of the Weinberg operator~\cite{Weinberg:1979sa}, and therefore the observed neutrino masses. This strategy is adopted in the inverse seesaw model~\cite{Wyler:1982dd,Mohapatra:1986aw,Mohapatra:1986bd, Bernabeu:1987gr,Gonzalez-Garcia:1988okv}, therefore we refer to the limit with vanishing active neutrino masses as the inverse-seesaw limit. In this symmetry-protected type-I seesaw, admissibly sizeable Yukawa couplings can significantly modify the neutrino couplings to SM gauge bosons. At tree-level, this leads to effects in processes such as $\pi\to\ell\nu$, $\tau\to\mu\nu\nu$ $Z\to\nu\nu$ and beta decays~\cite{Lee:1977tib,Shrock:1980vy,Schechter:1980gr,Shrock:1980ct,Shrock:1981wq,Langacker:1988ur,Bilenky:1992wv,Nardi:1994iv,Tommasini:1995ii,Bergmann:1998rg,Loinaz:2002ep,Loinaz:2003gc,Loinaz:2004qc,Antusch:2006vwa,Antusch:2008tz,Biggio:2008in,Alonso:2012ji,Abada:2012mc,Akhmedov:2013hec,Basso:2013jka,Abada:2013aba,Antusch:2014woa,Antusch:2015mia,Abada:2015oba,Fernandez-Martinez:2016lgt,Chrzaszcz:2019inj,Abada:2015trh,Abada:2016awd,Bolton:2019pcu,Coutinho:2019aiy,Crivellin:2020ebi}, and at the loop level to effects in $\ell\to\ell^\prime\gamma$, $\ell\to 3\ell'$, and $Z\to \ell^+\ell^-$~\cite{Lee:1977tib,Shrock:1980vy,Schechter:1980gr,Shrock:1980ct,Shrock:1981wq,Langacker:1988ur,Bilenky:1992wv,Nardi:1994iv,Tommasini:1995ii,Bergmann:1998rg,Loinaz:2002ep,Loinaz:2003gc,Loinaz:2004qc,Antusch:2006vwa,Antusch:2008tz,Biggio:2008in,Alonso:2012ji,Abada:2012mc,Akhmedov:2013hec,Basso:2013jka,Abada:2013aba,Antusch:2014woa,Antusch:2015mia,Abada:2015oba,Abada:2015trh,Fernandez-Martinez:2016lgt,Abada:2016awd,Herrero:2018luu,Coy:2018bxr,Bolton:2019pcu,Chrzaszcz:2019inj,Coutinho:2019aiy,Crivellin:2020ebi,Hagedorn:2021ldq,Zhang:2021tsq,Abada:2021zcm,Abada:2022asx,Calderon:2022alb}, which have also been studied in the SM Effective Field Theory (SMEFT)~\cite{Weinberg:1979sa,Broncano:2002rw,Cirigliano:2005ck,Antusch:2005gp,Abada:2007ux,deGouvea:2007qla,Coy:2018bxr,Alcaide:2019pnf,DeVries:2020jbs,Aparici:2009fh,Zhang:2021tsq,Zhang:2021jdf,Li:2022ipc,Du:2022vso}.
\smallskip

In this article we perform a comprehensive analysis of charged lepton flavour violation in the symmetry protected type-I seesaw: We calculate these effects both using exact diagonalisation of the neutrino mass matrix and by expanding the amplitudes in $v^2 /\MR^2$, which corresponds to the seesaw limit. 
Furthermore, we match the type-I seesaw model onto the $SU(2)_L$ gauge invariant SMEFT, which allows for the use of renormalisation group improved perturbation theory that resums the large logarithms between the right-handed neutrino scale and the scale of the physical processes. We state our conventions for the type-I seesaw model in Sec.~\ref{sec:Model}. 
In Sec.~\ref{sec:SMEFT}, we give the 1-loop SMEFT matching conditions that are relevant for flavour observables, in Sec.~\ref{sec:Flavour} we list the fixed-order results for the flavour observables of interest, before performing our phenomenological analysis in  Sec.~\ref{sec:Pheno} and concluding Sec.~\ref{sec:Conclusions}. In the appendix we provide results using exact diagonalisation of the neutrino mass matrix and/or in a general $R_\xi$ gauge. 
\smallskip

\section{Setup}
\label{sec:Model}

In the most general type-I seesaw setup, the SM is supplemented by $n$ generations of right-handed neutrinos $N_R$, i.e.~by fermions that are singlets under the SM gauge group. These new fields can have Majorana mass terms, as well as Yukawa-like interactions with the lepton doublet $\LL=(\vL, \lL)$ and the Higgs doublet $\Phi$. In the interaction basis, the Lagrangian is given by
\begin{equation}
\LN=\vRbar i\slashed{\partial}\vR- \left(\LLbar \Yv \tilde{\Phi} \vR +\frac{1}{2}\vRcbar \MR \vR +\hc \right)\,,
	\label{eq:LN}
\end{equation}
where $c$ stands for charge conjugation and we have suppressed flavour indices for better readability, i.e.~$M_R$ is an $n\times n$ matrix, that, without loss of generality, can be chosen to be diagonal and real, $Y^\nu$ is a $3\times n$ matrix. After electroweak symmetry breaking, the Higgs doublet acquires a vacuum expectation value of $v/\sqrt{2}\approx 175$~GeV and takes the form
\begin{equation}
\Phi \equiv\begin{pmatrix}
\varphi^+\\
\frac{v+h+i\varphi^0}{\sqrt{2}}
\end{pmatrix},\qquad 
\tilde{\Phi}\equiv i\sigma_2 \Phi^*
= \begin{pmatrix}
\frac{v+h-i\varphi^0}{\sqrt{2}}\\
-\varphi^-
\end{pmatrix}\,,\label{eq:Higgs}
\end{equation}
where $\sigma_2$ is the second Pauli matrix. Electroweak symmetry breaking generates the $3\times n$ Dirac mass matrix
\begin{equation}
\MD = \frac{v}{\sqrt{2}}Y^\nu\,.\label{eq:MD}
\end{equation}
We can now write the mass terms resulting from \Eq{eq:LN} as
\begin{equation}
\mathcal{L}_N = -\frac{1}{2}\begin{pmatrix}
\vLbar & \vRcbar
\end{pmatrix}
\Mv
\begin{pmatrix}
\vL^c\\
\vR
\end{pmatrix}
+\hc \,,\label{eq:Mvint}
\end{equation}
with the mass matrix
\begin{equation}
\Mv=
\begin{pmatrix}
\niente{3} & \MD\\
\MDtransp & \MR
\end{pmatrix}\,.
\label{eq:Mv}
\end{equation}
\smallskip
Next we move to the mass eigenbasis in which $\Mv$ is diagonal,
\begin{align}
\mathcal{L}_N = -\frac{1}{2} \nLbar\Mdiag \nR +\hc\,,\label{eq:diagonalisation}
\end{align}
with
\begin{align}
\Mdiag=&\Ovdag
\Mv
\Ovconj
\equiv\begin{pmatrix}
\MA & 0\\
0 & \MS
\end{pmatrix}\,.\label{eq:Mdiag}
\end{align}
Here $V$ is a unitary $(3+n)\times(3+n)$ matrix, $\MA$ is a $3\times 3$ matrix containing the light neutrino masses, while $\MS$ is an $n\times n$ matrix with masses of $\mathcal{O}(\MR)$. Since the light neutrinos are mostly composed of the ones within the lepton doublet $L_i$, they are commonly referred to as \emph{active} neutrinos, whereas the heavy neutrinos, which are mostly aligned with the gauge singlets $N_R$, are known as \emph{sterile} neutrinos. The neutrino mass eigenstates (3+$n$ vectors) are defined as
\begin{align}
\nR
=
\Ovtransp
\begin{pmatrix}
\vL^c\\
\vR
\end{pmatrix}
\equiv
\begin{pmatrix}
\nAR\\
\nSR
\end{pmatrix}
,\qquad
\nLbar
={\begin{pmatrix}
\vLbar \\ \vRcbar 
\end{pmatrix}\!}^T
\Ov
\equiv
\begin{pmatrix}
\nALbar \\ \nSLbar 
\end{pmatrix}^T\,.
\label{eq:MassEigenstates}
\end{align}

\noindent 
In the following, we will consider the seesaw limit $v\ll M_R$ and expand our results in $v/\MR$. The full results, obtained by exact diagonalisation of the neutrino mass matrix, are given in Appendix~\ref{sec:FullResults}.

In a first step, we block-diagonalise $M_\nu$, such that $\tilde{M}_\nu = \mathrm{diag}\left(\MAblock, \MSblock\right)$, where $\MAblock$ is a $3\times 3$ matrix and $\MSblock$ is an $n\times n$ matrix in flavour space. 
At leading order in $v/M_R$, we find
\begin{align}
\Ov&=\begin{pmatrix}
\id{3} -\frac{1}{2}\MD \MR^{-2} \MDdag & \MD \MR^{-1}\\
- \MR^{-1} \MDdag & \id{n} +\mathcal{O}\left(\frac{v^2}{\MR^2}\right) 
\end{pmatrix}+\mathcal{O}\left(\frac{v^3}{\MR^3}\right) \,,\label{eq:Vexp}\\
\MAblock &=  - M_D M_R^{-1} M_D^T\,,\label{eq:mlexp}\\
\MSblock &= M_R +\mathcal{O}\left(\frac{v^2}{\MR^2}\right)\,.\label{eq:mhexp}
\end{align}
\smallskip

Note that the off-diagonal blocks induce \emph{active-sterile} mixing, while the correction to the upper-left block induces (apparent) PMNS unitarity violation. Since our focus will be on charged lepton flavour violation, to which the active neutrino masses do not contribute in any observable way, we assume 
\begin{align}
    \MAblock \approx -\MD \MR^{-1} \MDtransp \equiv 0\,,
    \label{eq:InverseSeesawCond}
\end{align}
which we will refer to as the \emph{inverse seesaw limit}~\cite{Wyler:1982dd,Mohapatra:1986bd,Gonzalez-Garcia:1988okv,Branco:1988ex,Kersten:2007vk,Shaposhnikov:2006nn,Coy:2018bxr} of the type-I seesaw. In this scenario, which is also known as the symmetry protected seesaw, the Wilson coefficient of the Weinberg operator is zero, implying that the neutrino mass matrix is automatically diagonal after block diagonalisation, and given by
\begin{align}
\Mdiag\approx \mathrm{diag}\left(0,0,0,M_{R,1},M_{R,2},M_{R,3}\right) + \mathcal{O}\left(\frac{v^2}{\MR^2}\right)\,,
\label{eq:Mvexp}
\end{align}
while $\mathcal{O}(1)$ Yukawa couplings remain possible.
\smallskip

In presence of a single sterile neutrino, Eq.~\eqref{eq:InverseSeesawCond} only has a trivial solution (i.e.~$Y^\nu=0$), while for two sterile neutrinos, the solutions to Eq.~\eqref{eq:InverseSeesawCond} are given by~\cite{Coy:2018bxr} 
\begin{align}
Y^\nu=
\begin{pmatrix}
\lambda_e & \pm i \lambda_e \sqrt{\frac{\MRi{2}}{\MRi{1}}}\\
\lambda_\mu & \pm i \lambda_\mu \sqrt{\frac{\MRi{2}}{\MRi{1}}}\\
\lambda_\tau & \pm i \lambda_\tau \sqrt{\frac{\MRi{2}}{\MRi{1}}}\\
\end{pmatrix}\,.
\label{eq:WeinberglessLimit2}
\end{align}
If three sterile neutrinos are added to the SM,
\begin{align}
Y^\nu=&\begin{pmatrix}
\lambda_e & z \lambda_e \sqrt{\frac{\MRi{2}}{\MRi{1}}} & \pm i\sqrt{1+z^2} \lambda_e\sqrt{\frac{\MRi{3}}{\MRi{1}}}\\
\lambda_\mu & z  \lambda_\mu \sqrt{\frac{\MRi{2}}{\MRi{1}}} & \pm i\sqrt{1+z^2} \lambda_\mu\sqrt{\frac{\MRi{3}}{\MRi{1}}}\\
\lambda_\tau & z  \lambda_\tau \sqrt{\frac{\MRi{2}}{\MRi{1}}} & \pm i\sqrt{1+z^2} \lambda_\tau\sqrt{\frac{\MRi{3}}{\MRi{1}}}\\
\end{pmatrix}\,.\label{eq:WeinberglessLimit}
\end{align}
Here $z$ is an arbitrary complex number, and $\lambda_e,\,\lambda_\mu$ and $\lambda_\tau$ can be chosen to be real and positive.
\smallskip
In the seesaw-expanded results, we will encounter the combination
\begin{align}
	\Sij{ij}=(\MD \MR^{-2}\MDdag)_{ij} 
	\label{eq:Sij}
\end{align}
or equivalently,
\begin{align}
	\Tij{ij}=(\Yv \MR^{-2}\Yvdag)_{ij}=\frac{2}{v^2}S_{ij}\,.
	\label{eq:Tij}
\end{align}
For two sterile neutrinos with degenerate masses $\MRi{1}=\MRi{2}=\MR$, $T_{ij}$ is given by
 \begin{align}
     \Tij{ij} =2  \frac{\lambda_i \lambda_j}{\MR^2}\,.
     \label{eq:TijExplicit2}
 \end{align}
We will also encounter the matrix products 
\begin{equation}
\left(\Yv \Yvdag \Yv\Yvdag\right)_{ij}\,\qquad \left(\Yv \Yvdag\right)_{ij}\left(\Yv \Yvdag\right)_{kl}\,. 
\end{equation}
If we apply the parametrisation in Eq.~\eqref{eq:WeinberglessLimit2} {and set $\MRi{1}=\MRi{2}=\MR$, $T_{ij}$}, we find
\begin{align}
    \left(\Yv \Yvdag \Yv \Yvdag\right)_{ij} &= 4 \lambda_i\lambda_j\sum_{k\in \{e,\mu,\tau\}} \!\!\!\!\lambda_k^2
    \label{eq:Y4Explicit2}\\
   \left(\Yv \Yvdag\right)_{ij} \left(\Yv \Yvdag\right)_{kl} &=
   4 \lambda_i \lambda _j \lambda_k \lambda_l\,.
   \label{eq:Y2Y2Explicit2}
\end{align}
If we take three sterile neutrinos with degenerate masses $\MRi{1}=\MRi{2}=\MR$,
and the Yukawa matrix of Eq.~\eqref{eq:WeinberglessLimit}, these matrix products take the form
 \begin{align}
     \Tij{ij} =& \left(1 + |z|^2 + |1 + z^2|\right)\frac{\lambda_i \lambda_j}{\MR^2}\,.
     \label{eq:TijExplicit}\\
    \left(\Yv \Yvdag \Yv \Yvdag\right)_{ij} =& \left(1+|z|^2+|1+z^2|\right)^2 \lambda_i\lambda_j\sum_{k\in \{e,\mu,\tau\}} \!\!\!\!\lambda_k^2
    \label{eq:Y4Explicit}\\
   \left(\Yv \Yvdag\right)_{ij} \left(\Yv \Yvdag\right)_{kl} =& \left(1+|z|^2+|1+z^2|\right)^2 \lambda_i \lambda_j \lambda_k \lambda_l\,.
   \label{eq:Y2Y2Explicit}
\end{align}
\smallskip
The presence of active-sterile mixing leads to tree-level modifications of the gauge boson couplings to the SM neutrinos. Defining the covariant derivative as \begin{equation}
D_{\mu}=\partial_{\mu}+ig_2W_{\mu}^I \tau^I+ig_1B_{\mu}Y\,,\label{eq:CoviD}
\end{equation}
where $\tau^I\equiv\sigma^I/2$ and $\sigma^I$ denote the Pauli matrices,
and introducing the Lagrangian of the neutral and charged current interactions after electroweak symmetry breaking,
\begin{equation}
\mathcal{L}_{W,Z}^{\ell,\nu}=\left({{\bar \ell }_i}g _{ij}^{\ell\nu}{\gamma ^\mu }{P_L}{\nu _j}\,{W_\mu } + \hc \right)
+ \left[ {{{\bar \ell }_i}{\gamma ^\mu }\left( {g _{ij}^{\ell L}{P_L} + g _{ij}^{\ell R}{P_R}} \right){\ell _j}
+ {{\bar \nu }_i}g _{ij}^\nu {\gamma ^\mu }{P_L}{\nu _j}} \right]{Z_\mu }\,,
\label{eq:WZLag}
\end{equation}
where $i$ and $j$ are flavour indices,
we identify the couplings
\begin{equation}
\begin{aligned}
g _{ij}^{\ell L} =& \frac{e}{2\sw \cw}\left( 1 - 2\sw^2\right)\delta _{ij}\,,\qquad &
g _{ij}^{\ell R} =& -\frac{e\sw}{\cw}\delta _{ij}\,,\\
g _{ij}^\nu  =&  - \frac{e}{2\sw\cw}\left( \delta _{ij} -\Sij{ij}\right)\,,\qquad &
g _{ij}^{\ell\nu} =&  - \frac{e}{\sqrt 2  \sw}\left( \delta _{ij} -\frac{1}{2}\Sij{ij}\right)\,.
\end{aligned}
\label{eq:WZcouplings}
\end{equation}
In the following, we will use the notation
\begin{equation}
\begin{aligned}
g^{\ell L}_{\rm SM}=& \frac{e}{2\sw \cw}\left(1-2\sw^2\right) \,,\qquad &
g^{\ell R}_{\rm SM}=&  -\frac{e \sw}{\cw} \,,\\
g^{\nu}_{\rm SM}=&  - \frac{e}{2\sw\cw}\,,\qquad &
g^{\ell\nu}_{\rm SM} =&  - \frac{e}{\sqrt 2  \sw}\,.
\end{aligned}
\label{eq:glSM}
\end{equation}
\smallskip
Note that the $Z\ell\ell$ couplings, $g _{ij}^{\ell L}$ and $g _{ij}^{\ell R}$, are not modified at tree-level, while the interactions of the EW gauge bosons with neutrinos receive contributions proportional to $\Sij{ij}$ and can therefore be flavour off-diagonal. The interactions of the charged Goldstone bosons with neutrinos are modified in a similar way. All relevant Feynman rules are given in Table~\ref{tab:FRsExpanded}. 
\smallskip

\begin{table}[ht]
	\begin{center}
		\setlength{\tabcolsep}{12pt}
		\begin{tabular}{c c}
		Interaction &  Expanded Feynman rule\\
		\hline \hline\\[-4mm]
		$\overline{\lLR}_i\,W^-_\mu \nALi{j}$ & $-\dfrac{e}{\sqrt{2}\sw}\left(\delta_{ij}-\dfrac{\Sij{ij}}{2}\right)\gamma^{\mu}P_L$\\[3mm]
		$\overline{\lLR}_i\,W^-_\mu \nSLi{a}$ & $-\dfrac{e}{\sqrt{2}\sw}\left(\MD \MR^{-1}\right)_{ia}\gamma^{\mu}P_L$\\[3mm]
			\hline \\[-4mm]
			$\nALi{i} Z_\mu \nALi{j}$  & $-\dfrac{e}{2\sw\cw}\left(\delta_{ij}-\Sij{ij}\right)\,\gamma^{\mu}P_L$\\[3mm]
			$\nALi{i} Z_\mu \nSLi{a}$  & $-\dfrac{e}{2\sw\cw}\left(\MD \MR^{-1}\right)_{ia}\,\gamma^{\mu}P_L$\\[3mm]
			$\nSLi{a} Z_\mu \nSLi{b}$  & $-\dfrac{e}{2\sw\cw}\left(\MR^{-1}\MDdag \MD \MR^{-1}\right)_{ab}\,\gamma^{\mu}P_L$\\[3mm]
			\hline\\[-4mm]
			$\overline{\lLR}_i\varphi^- \,\nARi{j}$  & $0$\\[3mm]
			$\overline{\lLR}_i\varphi^- \,\nSRi{a}$  & $\dfrac{\sqrt{2}}{v} \MDij{ia} P_R $\\[3mm]
			\hline
		\end{tabular}
	\end{center}	
	\caption{Feynman rules at leading order in the seesaw expansion, neglecting charged lepton masses. The active (light) states are denoted as $\nA$, the sterile (heavy) states as $\nS$. $\MD$ is the $3\times n$ Dirac mass matrix defined in \Eq{eq:MD}, $\MR$ is the $n\times n$ real and diagonal Majorana mass matrix introduced in \Eq{eq:LN} and $\Sij{ij}$ the mass insertion defined in \Eq{eq:Sij}.\label{tab:FRsExpanded}}
\end{table}

\begin{figure}[ht]
\centering
	\hspace{5mm}
	\subfloat[\label{fig:Sij}]{%
	\includegraphics[scale=.7]{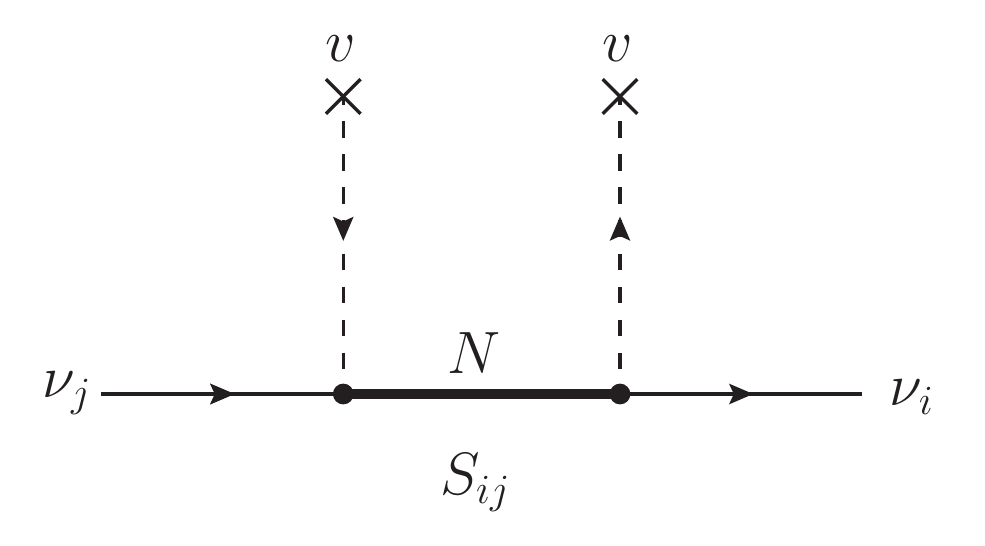}
	}
	\subfloat[\label{fig:Tij}]{%
	\includegraphics[scale=.7]{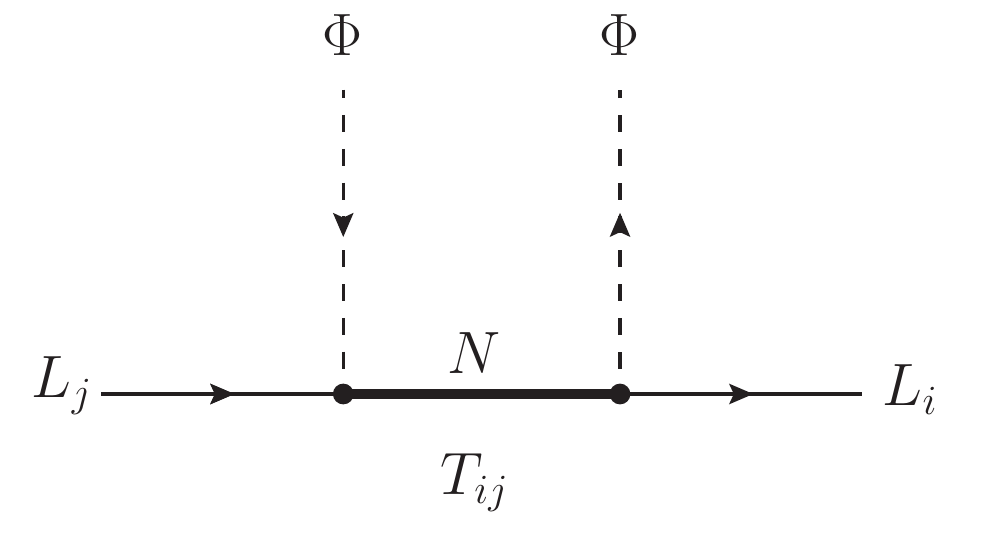}
	}
\caption{(a) Diagrammatic representation of the mass insertion $\Sij{ij}$ in the broken theory (see \Eq{eq:Sij}). $\nu_i$ and $\nu_j$ are SM-like left-handed neutrino gauge eigenstates, which can also interact with the SM gauge bosons, whereas $N$ is the right-handed neutrino, which is a singlet under the SM gauge group. The dashed lines indicate the insertions of the Higgs vev $v$.
(b) The related object, $\Tij{ij}$, defined in \Eq{eq:Tij}, which is relevant in the unbroken theory and enters the SMEFT matching relations. Here $L_i$ and $L_j$ are SM lepton doublets. $N$ is the right-handed neutrino.}\label{fig:MassInsertions}
\end{figure}

These (expanded) Feynman rules can be visualised in the Mass Insertion Approximation (MIA): Instead of working in the mass eigenbasis, one can remain in the interaction eigenbasis of \Eq{eq:Mvint}, and treat off-diagonal mass terms as perturbative interactions. This approach leads to the same amplitudes as those derived in the mass eigenbasis and afterwards expanded in the seesaw limit. Figures~\ref{fig:Sij} and \ref{fig:Tij} show how Eqs. \eqref{eq:Sij} and \eqref{eq:Tij} are represented or obtained diagrammatically.
\smallskip

\section{Matching onto the SMEFT}\label{sec:SMEFT}

\begin{table}[t]
	\begin{center}
		\setlength{\tabcolsep}{12pt}
		\begin{tabular}{c c}
		Interaction &  Feynman rule in the EFT\\
		\hline \hline\\[-4mm]
		$\overline{\lLR}_i\,W^-_\mu \nALi{j}$ & $- \dfrac{e}{\sqrt 2  \sw}\left( {\delta _{ij} + v^2C_{\varphi \ell,ij }^{(3)}}\right)\gamma^\mu P_L$ \\[3mm]
			\hline \\[-4mm]
			$\nALi{i} Z_\mu \nALi{j}$  & $- \dfrac{e}{2\sw\cw}\left( {{\delta _{ij}} + v^2\Xmij{ij}} \right)\gamma^\mu P_L$ \\[3mm] 
			\hline\\[-4mm]
			$\varphi^+\overline{\lLR}_i\nALi{j}$ & $\sqrt{2}v \,p_\mu^+ \, C_{\varphi \ell,ij }^{(3)}\, \gamma^\mu P_L$\\[3mm] 
			\hline\\[-4mm]
			$\varphi^-(p^-)\varphi^+(p^+) \nALibar{i}\nALi{j}$ & $\left(p_\mu^+-p_\mu^-\right) \Xpij{ij}\gamma^\mu P_L$\\[3mm] 
			\hline\\[-4mm]
			$\varphi^-(p^-)\varphi^+(p^+) \overline{\lLR}_i\lLR_j$ & $-\left(p_\mu^+-p_\mu^-\right)\Xmij{ij}\gamma^\mu P_L$\\[3mm] 
			\hline\\[-4mm]
			$\varphi^-(p^-)W^+ \nALibar{i}\nALi{j}$ & $-\dfrac{e}{\sw}v C_{\varphi \ell,ij }^{(1)}\gamma^\mu P_L$\\[3mm] 
			\hline\\[-4mm]
			$\varphi^-(p^-)W^+ \overline{\lLR}_i\lLR_j$ & $-\dfrac{e}{\sw}v C_{\varphi \ell,ij }^{(1)}P_L\gamma^\mu $\\[3mm] 
			\hline\\[-4mm]
		\end{tabular}
	\end{center}	
	\caption{Feynman rules for the relevant operators of the SMEFT for vanishing charged lepton masses. The tree-level results for $C_{\varphi\ell,ij}^{(3)}$ and $\Xmij{ij}$ are given in \Eqs{eq:C13tree} and \eqref{eq:XYtree}.\label{tab:FRsEFT}}
\end{table}

In this section we calculate the matching onto the SMEFT. These results could be used as initial conditions of a renormalisation group improved computation of charged lepton flavour violating observables.\footnote{Our results agree with Ref.~\cite{Zhang:2021jdf} (v3). We thank the authors for useful discussions. } For the derivation of our results, we made extensive use of the \texttt{Mathematica} packages \texttt{FeynRules}~\cite{Christensen:2008py}, \texttt{FeynArts}~\cite{Hahn:2000kx} and \texttt{Package-X}~\cite{Patel:2015tea,Patel:2016fam} in combination with \texttt{CollierLink}, a \texttt{Package-X} interface to the \texttt{Collier} library~\cite{Denner:2016kdg}.

\subsection{Conventions}
The SMEFT extends the SM Lagrangian $\mathcal{L}_{\rm SM}^{(4)}$ by higher dimensional operators, which are invariant under the full SM gauge group. Up to the dimension 6 level we write
\begin{align}
\mathcal{L}_{\rm SMEFT} = \mathcal{L}_{\rm SM}^{(4)}+  C^{(5)} \mathcal{O}^{(5)}  +  \sum_k C_k^{(6)}\mathcal{O}_k^{(6)}\, ,\label{eq:SMEFTLag}
\end{align}
where $\mathcal{O}^{(5)}$ is the dimension Weinberg operator
\begin{equation}
\mathcal{L}^{(5)}\equiv C^{(5)}_{ij} \mathcal{O}^{(5)}_{ij}+\hc\equiv  C^{(5)}_{ij} \left(\LLibar{i}^c \tilde{\Phi}^* \right)\left(\tilde{\Phi}^\dagger \LLi{j}\right)+\hc \,.
\label{eq:WeinbergOp}
\end{equation}
We will only consider the subset of dimension-six operators that can, at $\mathcal{O}\left(v^2/\MR^2\right)$ and for vanishing charged lepton masses, lead to direct contributions to lepton flavour violating observables. These are
\begin{align}
\sum_k C_k^{(6)}Q_k^{(6)}= &\;\Cuno \Ouno + \Ctre \Otre + \left(C_{eW}\mathcal{O}_{eW} +  C_{eB} \mathcal{O}_{eB} + \hc\right) \notag\\[-2mm]
&+ C_{\ell \ell} \mathcal{O}_{\ell \ell} + C_{\ell e} \mathcal{O}_{\ell e}+ C_{\ell q}^{(1)} \mathcal{O}_{\ell q}^{(1)}+ C_{\ell q}^{(3)} \mathcal{O}_{\ell q}^{(3)} + C_{\ell u} \mathcal{O}_{\ell u} + C_{\ell d} \mathcal{O}_{\ell d}\,,
\end{align}
which are defined as~\cite{Grzadkowski:2010es}
\begin{align}
\begin{aligned}
\Ounoij{ij}=&\left(\Phi^\dagger i\Dfb_\mu\Phi\right)\left(\LLibar{i}\gamma^\mu
  \LLi{j}\right)\, ,\\
\Otreij{ij}=&\left(\Phi^\dagger i\Dfb_\mu^I\Phi\right)\left(\LLibar{i}\sigma^I\gamma^\mu \LLi{j}\right)\, ,\\
\mathcal{O}_{eW,ij}=&\left(\LLibar{i}\sigma_{\mu\nu} \eRi{j}\right)\sigma^I\Phi W^{I\mu\nu}\,,\\
\mathcal{O}_{eB,ij}=&\left(\LLibar{i}\sigma_{\mu\nu} \eRi{j}\right)\Phi B^{\mu\nu}\,.\\
\left(\mathcal{O}_{\ell \ell}\right)_{ij,kl}=&\left(\LLibar{i}\gamma_\mu
  \LLi{j}\right)\left(\LLibar{k}\gamma_\mu \LLi{l}\right)\,,\\
\left(\mathcal{O}_{\ell e}\right)_{ij,kl}=&\left(\LLibar{i} \gamma_\mu \LLi{j}\right)\left(\eRibar{k}\gamma_\mu \eRi{l}\right)\,,\\
\left(\mathcal{O}_{\ell q}^{(1)}\right)_{ij,kl}\equiv &
\left(\LLibar{i} \gamma_\mu \LLi{j}\right)\left(\QLibar{k} \gamma^\mu \QLi{l}\right),\\
\left(\mathcal{O}_{\ell q}^{(3)}\right)_{ij,kl}\equiv &
\left(\LLibar{i} \gamma_\mu \sigma^I\LLi{j}\right)\left(\QLibar{k} \gamma^\mu \sigma^I\QLi{l}\right),\\
\left(\mathcal{O}_{\ell u}\right)_{ij,kl}\equiv &
\left(\LLibar{i} \gamma_\mu \LLi{j}\right)\left(\uRibar{k} \gamma^\mu \uRi{l} \right),\\
\left(\mathcal{O}_{\ell d}\right)_{ij,kl}\equiv&
\left(\LLibar{i} \gamma^\mu \LLi{j}\right)\left(\dRibar{k} \gamma_\mu \dRi{l} \right)\,.
\label{eq:2l2qOps}
\end{aligned}
\end{align}
Note that in all operators involving $SU(2)_L$ doublets, the $SU(2)_L$ indices are contracted within the fermion bilinears.  We follow the hypercharge conventions of Ref.~\cite{Grzadkowski:2010es}, given in Table~\ref{tab:HyperchargeConventions}.
\begin{table}[t]
\centering
\begin{tabular}{|c|c c c c c|c|}
\hline
    & $\LL$ & $\eR$ & $\QL$ & $u$ & $d$ & $\Phi$\\
\hline
hypercharge $Y$ & $-\frac{1}{2}$ & $-1$ & $\frac{1}{6}$ & $\frac{2}{3}$ & $-\frac{1}{3}$ & $\frac{1}{2}$\\
\hline
\end{tabular}
\caption{Hypercharges for the SM fermions and Higgs field.}\label{tab:HyperchargeConventions}
\end{table}
The covariant derivative is defined in Eq.~\eqref{eq:CoviD}. Using the short-hand notation $\Phi^\dagger \Db_\mu\Phi\equiv \left(D_\mu
\Phi\right)^\dagger \Phi$, we define the Hermitian derivative terms
\begin{equation}
\Phi^\dagger i\Dfb_\mu\Phi\equiv i\Phi^\dagger\left(D_\mu -\Db_\mu\right)\Phi\,,  \quad
\Phi^\dagger i\DfbImu\Phi\equiv i\Phi^\dagger\left(\sigma^I D_\mu -\Db_\mu\sigma^I\right)\Phi\,,
\label{eq:HermitD}
\end{equation}
which enter the operators $\Ouno$ and $\Otre$.
The field strength tensors $W_{\mu\nu}^I\equiv \partial_\mu W_\nu^I-\partial_\nu W_\mu^I-g_2\varepsilon^{IJK}W_\mu^J W_\mu^K$ and $B_{\mu\nu}\equiv \partial_\mu B_\nu-\partial_\nu B_\mu$, are associated to the $SU(2)_L$ and $U(1)_Y$ gauge fields $W^I$ and $B$, respectively. We follow the convention of summation over all flavour indices in the Lagrangian. For the operators involving four leptons, this means that we write the corresponding terms as follows
\begin{align}
C \mathcal{O}=\sum_{i,j,k,l}C_{ij,kl}\mathcal{O}_{ij,kl}, \qquad i,j,k,l\in\lbrace e,\,\mu,\,\tau\rbrace\,.
\end{align}
For operators whose fields are distinguishable, i.e.~$\mathcal{O}_{\ell e}$, $\mathcal{O}_{\ell q}^{(1)}$, $\mathcal{O}_{\ell q}^{(3)}$, $\mathcal{O}_{\ell u}$, $\mathcal{O}_{\ell d}$, and that thus cannot be fierzed into themselves, this summation convention has no impact. However, for $\mathcal{O}_{\ell\ell}$, which is invariant under the exchange of the two fermion bilinears, this leads to a factor 2 at the amplitude level (the contraction of $SU(2)_L$ indices is taken into account):
\begin{align}
C_{ij,kl}^{\ell\ell}\left(\bar L_i^a \gamma_\mu L_j^a\right)\left(\bar L_k^b \gamma^\mu L_l^b\right)
+C_{kl,ij}^{\ell\ell}\left(\bar L_k^a \gamma_\mu L_l^a\right)\left(\bar L_i^b \gamma^\mu L_j^b\right)\notag\\
+C_{kj,il}^{\ell\ell}\left(\bar L_k^a \gamma_\mu L_j^a\right)\left(\bar L_i^b \gamma^\mu L_l^b\right)
+C_{il,kj}^{\ell\ell}\left(\bar L_i^a \gamma_\mu L_l^a\right)\left(\bar L_k^b \gamma^\mu L_j^b\right)\notag\\
\to 2\left(C_{ij,kl}^{\ell\ell}\left(\bar L_i^a \gamma_\mu L_j^a\right)\left(\bar L_k^b \gamma^\mu L_l^b\right)+C_{ij,kl}^{\ell\ell}\left(\bar L_i^a \gamma_\mu L_j^b\right)\left(\bar L_k^b \gamma^\mu L_l^a\right)\right)\,.
\label{eq:M4LWithSummation}
\end{align}
Note that $L$ on the left-handed side corresponds to a field, while on the right-handed side it denotes a spinor. $a$ and $b$ denote $SU(2)_L$ indices.

\subsection{Tree Level Matching}\label{sec:TreeLevelMatchingSMEFT}

\label{sec:WeinbergOp}
\begin{figure}
    \centering
    \includegraphics[scale=.7]{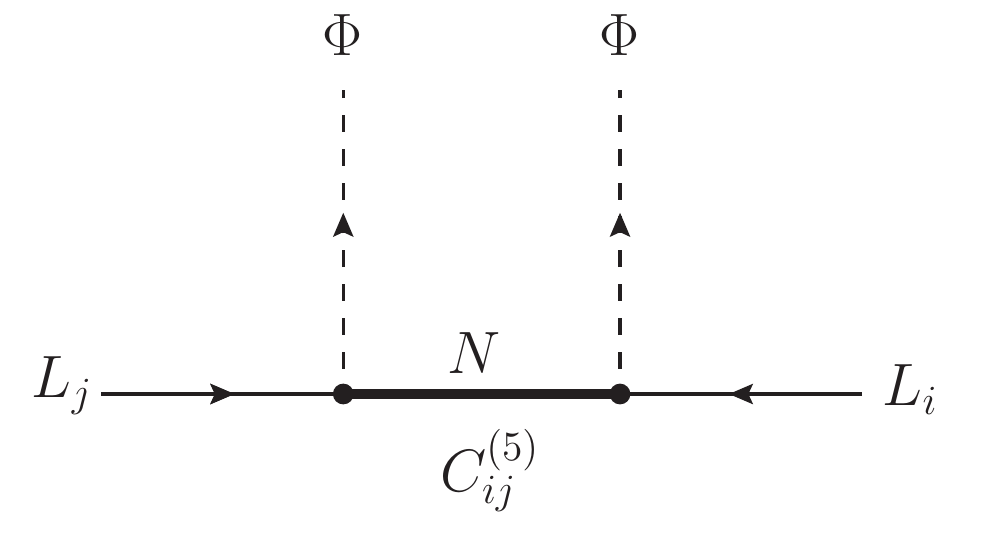}
    \caption{Feynman diagram in the type-I seesaw contributing to the Wilson coefficient of the Weinberg operator. In the inverse seesaw limit, the neutrino Yukawa matrices are chosen in such as way that this diagram vanishes and lepton number is conserved.}
    \label{fig:WeinbergOp}
\end{figure}
The Feynman diagram in Figure~\ref{fig:WeinbergOp} leads to the following Wilson coefficient of the Weinberg operator
\begin{align}
C^{(5)}=\frac{1}{2}\Yvconj \MR^{-1} \Yvdag\,.
\end{align}
After electroweak symmetry breaking, this can be expressed as $C^{(5)}=\frac{1}{v^2}\MDconj \MR^{-1} \MDdag$, which features, by definition, the same combination of Dirac- and Majorana matrices as the active neutrino block of the neutrino mass matrix. In the inverse seesaw limit, this combination of matrices is set to zero.
\smallskip

\begin{figure}[ht]
\center
	\subfloat[\label{fig:TreeDiag}]{%
		\includegraphics[width=0.43\textwidth]{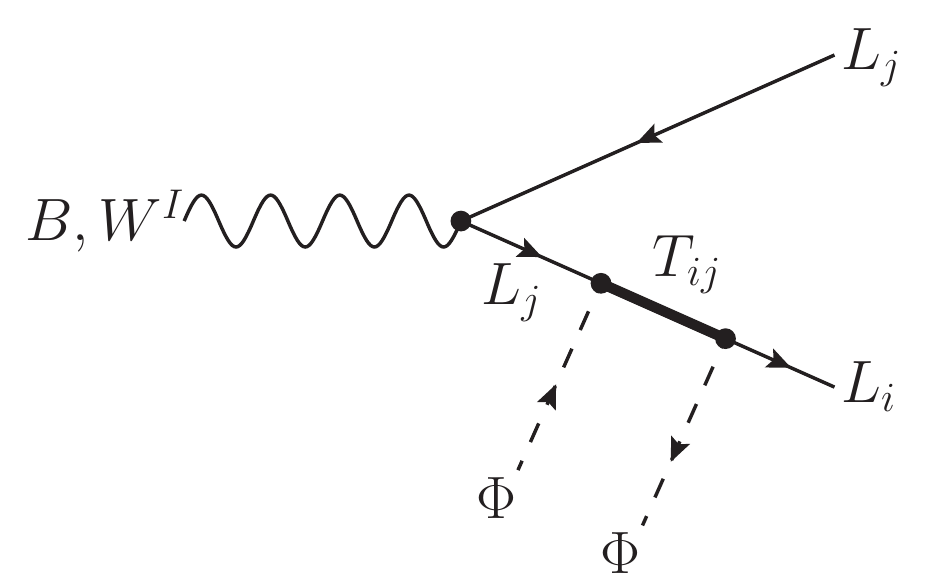}%
	}
	\hspace{5mm}
	\subfloat[\label{fig:TreeMatching}]{%
		\includegraphics[width=0.38\textwidth]{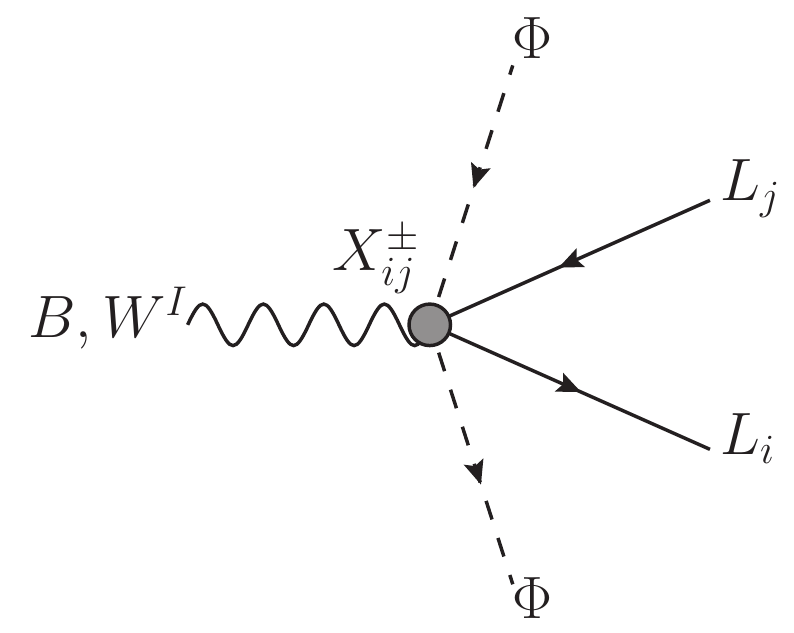}%
	}\\
	\caption{(a) Diagram giving rise to modifications of the lepton couplings to the $U(1)_Y$ and $SU(2)_L$ gauge bosons in the type-I seesaw. (b) Corresponding diagram in the effective theory. }
	\label{fig:MIAtree}
\end{figure}

At the dimension-6 level, the Wilson coefficients of the operators $\mathcal{O}_{\varphi \ell}^{(1)}$ and $\mathcal{O}_{\varphi \ell}^{(3)}$ receive tree-level contributions induced by the diagrams in Figure~\ref{fig:TreeDiag}:
\begin{align}
C_{\varphi \ell,ij }^{(3)} =-C_{\varphi \ell,ij }^{(1)}=-\frac{1}{4}\Tij{ij}\,,\label{eq:C13tree}
\end{align}
$\Tij{ij}$ is defined in Eq.~\eqref{eq:Tij}. The relation $\Ctre=-\Cuno$, which follows from the fact that only neutrino couplings, no charged lepton couplings, are modified,  motivates a change of basis from $\{\Ouno,\Otre\}$ to $\{\Op,\Om\}$,
\begin{align}
\Opij{ij}=&\,\frac{1}{2}\left(\mathcal{O}_{\varphi \ell,ij }^{(3)}+\mathcal{O}_{\varphi \ell,ij }^{(1)}\right)\,,
&
\hspace{-5mm}\Omij{ij}=&\,\frac{1}{2}\left(\mathcal{O}_{\varphi \ell,ij }^{(3)}-\mathcal{O}_{\varphi \ell,ij }^{(1)}\right)
\notag
\\[2mm]
\Xpij{ij} =&\, C_{\varphi\ell,ij}^{(3)} + C_{\varphi\ell,ij}^{(1)} \,,
&
\Xmij{ij} =&\, C_{\varphi\ell,ij}^{(3)} - C_{\varphi\ell,ij}^{(1)} \,,
\label{eq:DefXY}
\end{align}
with
\begin{align}
\Xpij{ij} =  0\,,\quad
\Xmij{ij}  =-\frac{1}{2}\Tij{ij}\,.
\label{eq:XYtree}
\end{align}
The corresponding diagrams in the full and effective theory are shown in Figure~\ref{fig:TreeDiag} and Figure~\ref{fig:TreeMatching}, respectively. 

\subsection{One-Loop Matching}\label{sec:OneLoopMatchingSMEFT}

\begin{figure}[ht]
\center
	\subfloat[\label{}]{%
		\includegraphics[width=0.46\textwidth]{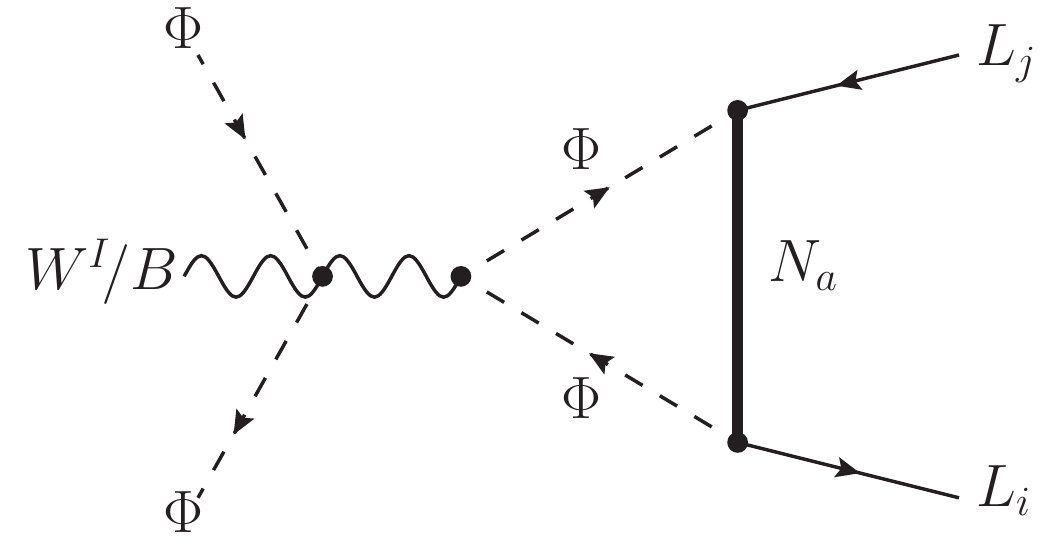}
	}
	\hspace{5mm}
	\subfloat[\label{}]{%
		\includegraphics[width=0.42\textwidth]{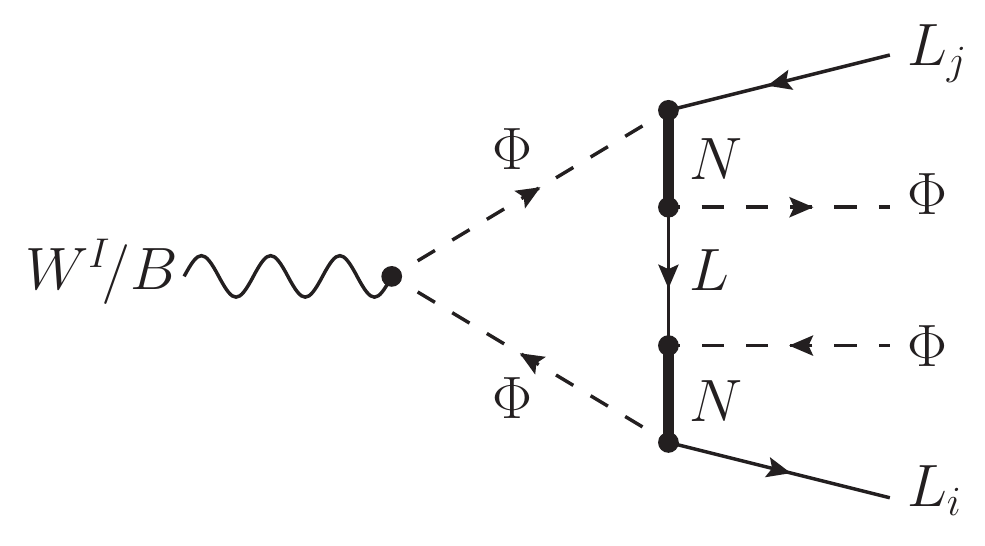}
	}\\
	\subfloat[\label{}]{%
		\includegraphics[width=0.46\textwidth]{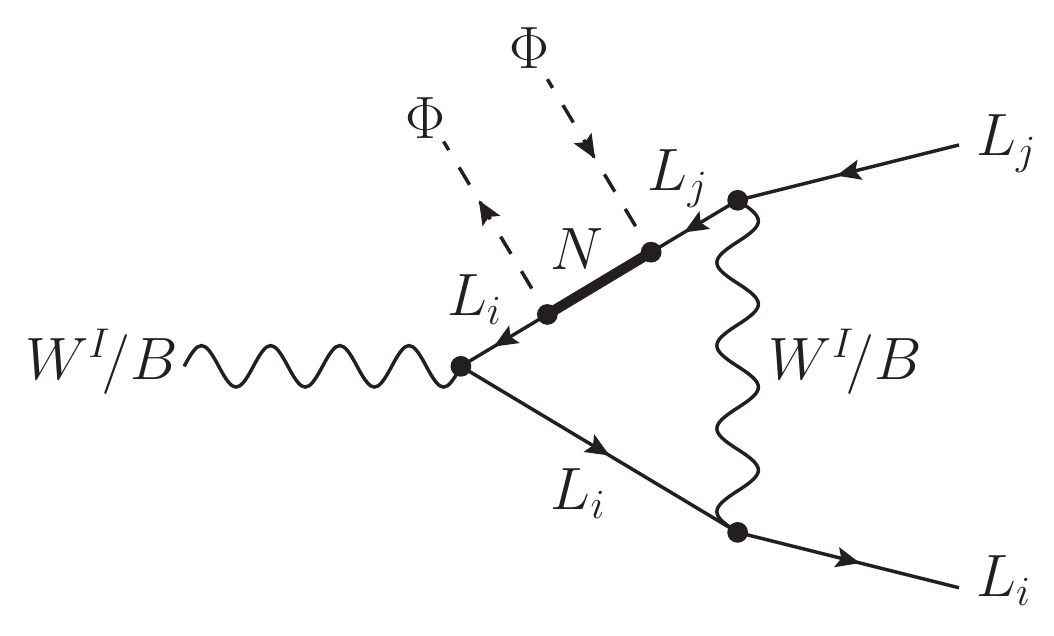}
	}
	\hspace{5mm}
	\subfloat[\label{}]{%
		\includegraphics[width=0.42\textwidth]{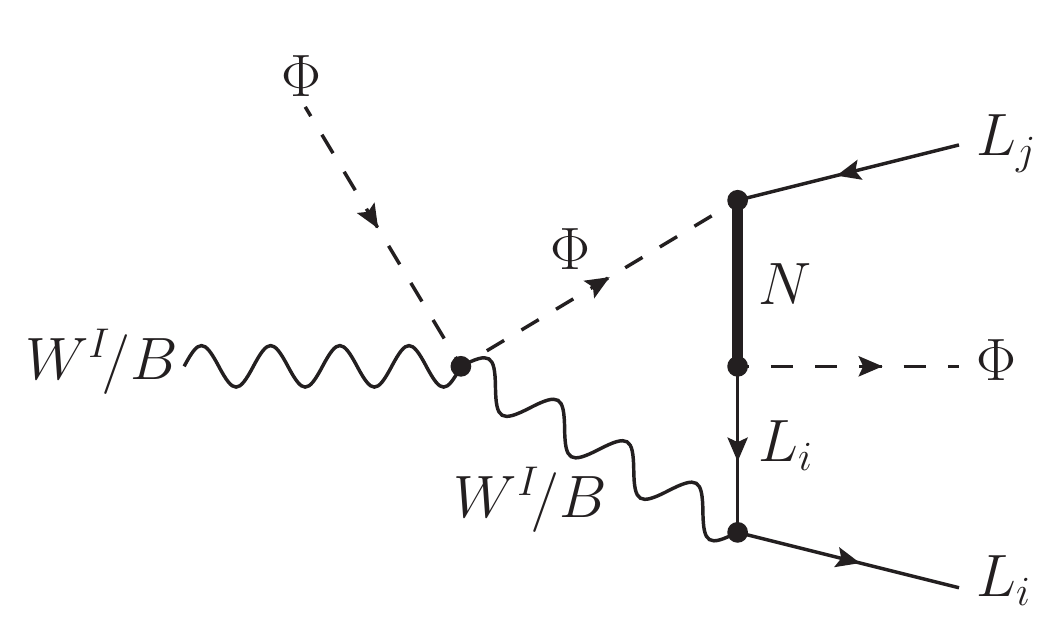}
	}\\
	\subfloat[\label{}]{%
		\includegraphics[width=0.46\textwidth]{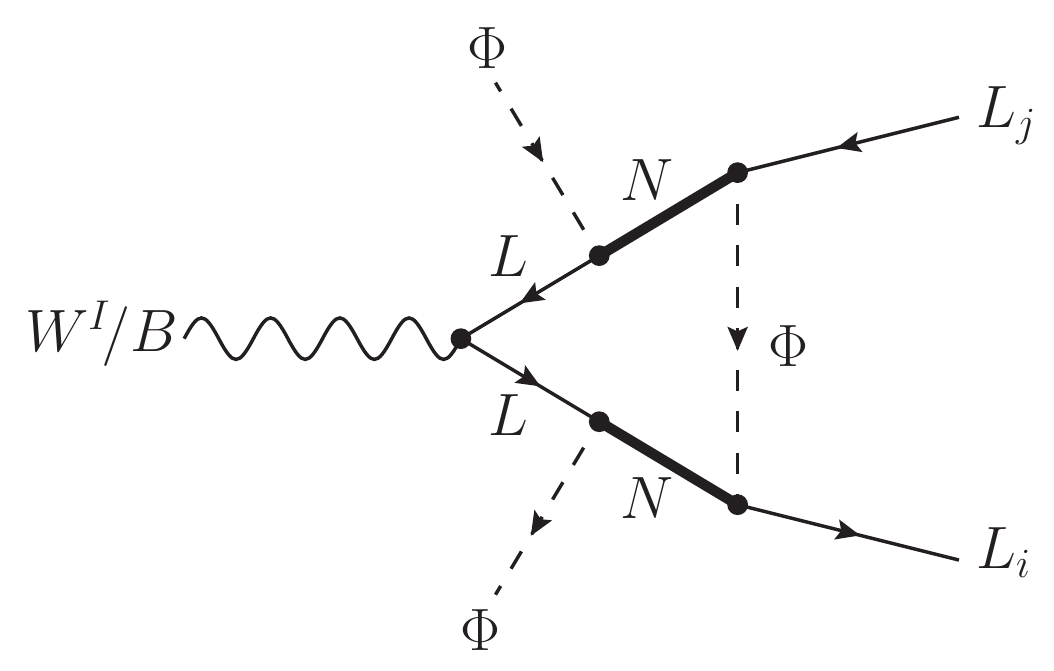}
	}
	\subfloat[\label{fig:WBll_Majorana}]{%
		\includegraphics[width=0.46\textwidth]{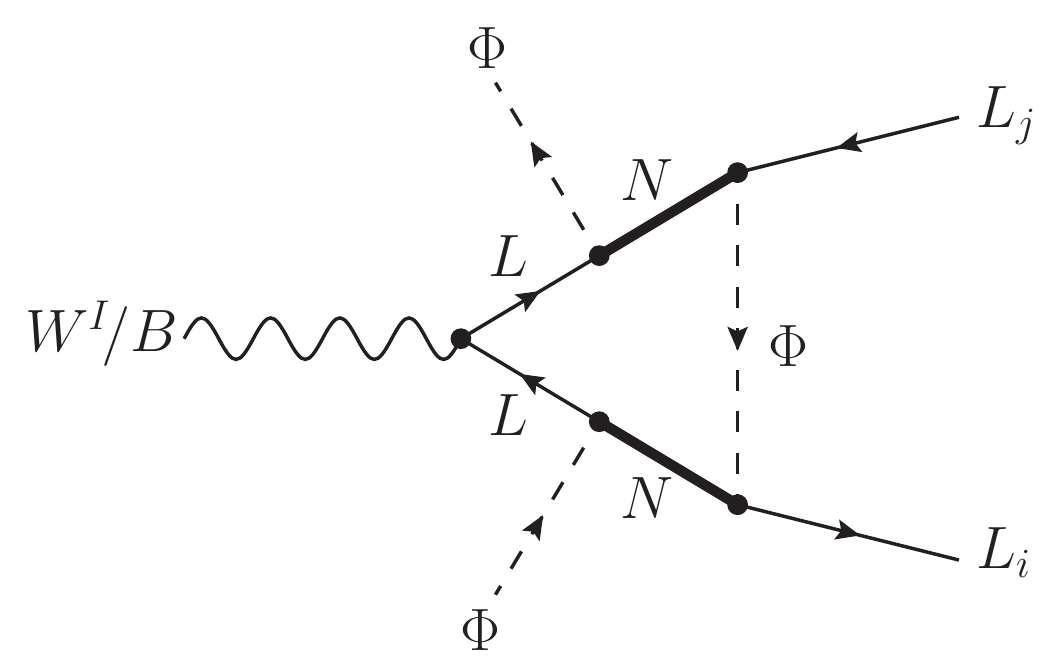}
	}
	\\	\subfloat[\label{}]{%
		\includegraphics[width=0.42\textwidth]{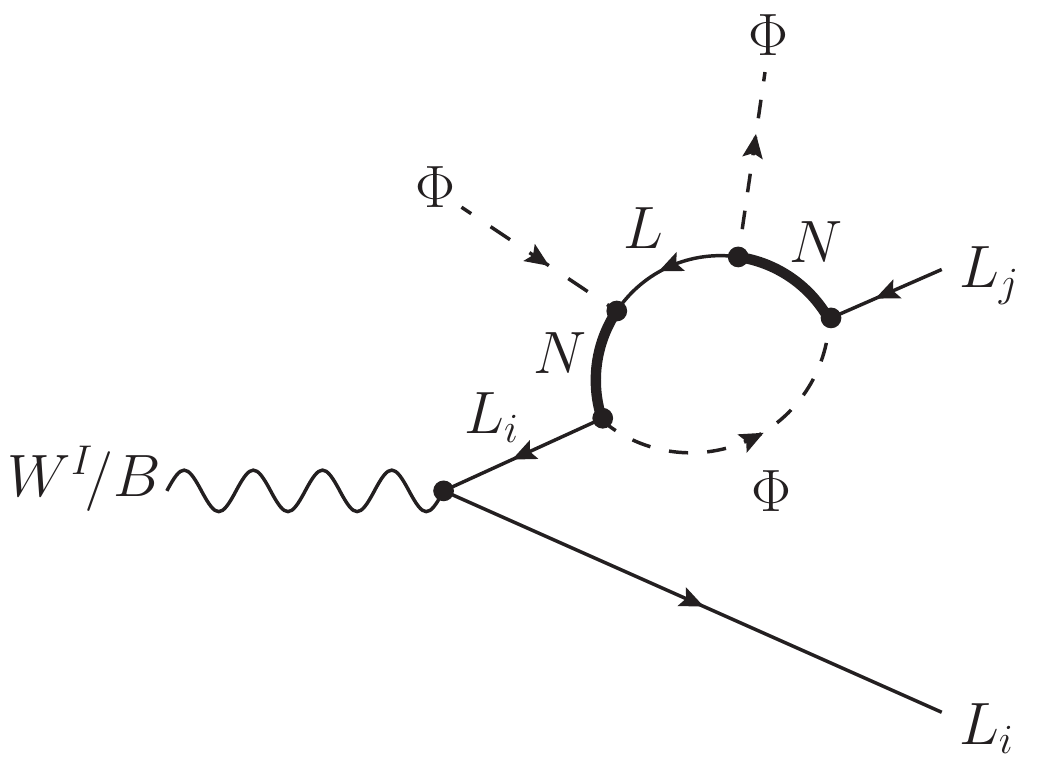}
		}
	\hspace{5mm}
	\subfloat[\label{}]{%
		\includegraphics[width=0.42\textwidth]{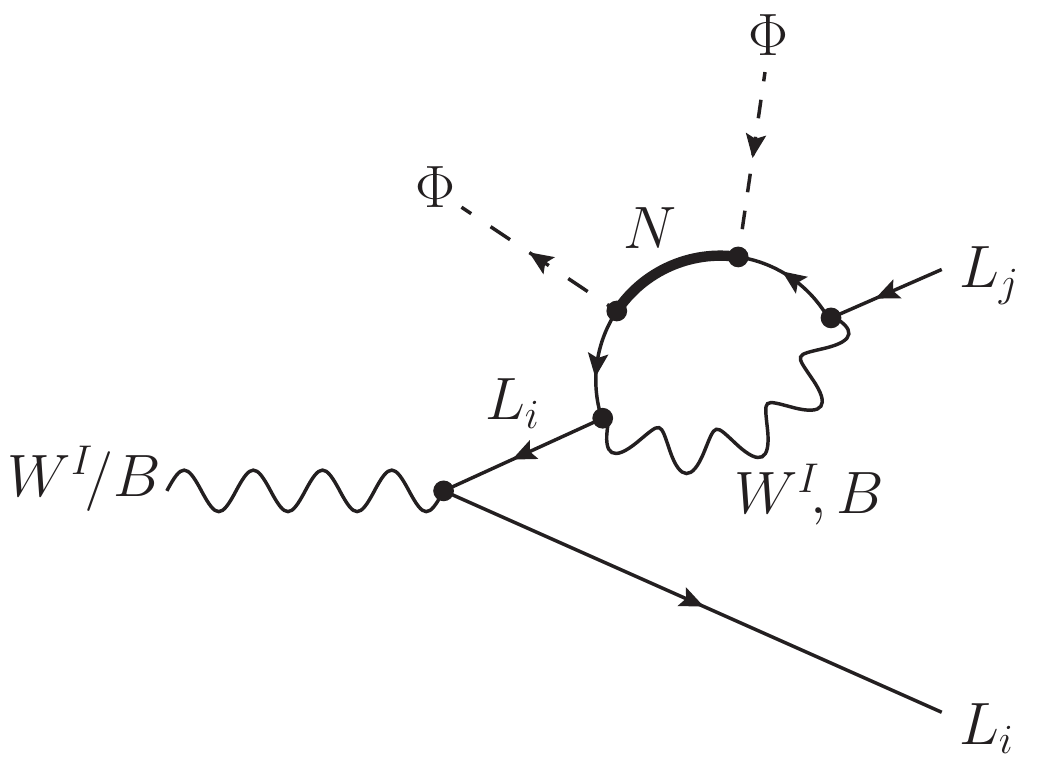}%
	}		
	\caption{Diagrams contribution to the 1-loop matching onto $\Ouno$ and $\Otre$ (see Figure~\ref{fig:TreeMatching}) or, equivalently, onto $\Op$ and $\Om$.}
	\label{fig:WBllhard}
\end{figure}

\subsubsection{Modified Gauge-Boson Couplings ($\mathcal{O}_{\varphi \ell}^{(1)}$ and $\mathcal{O}_{\varphi \ell}^{(3)}$)}\label{sec:OneLoopMatchingSMEFTO13}

Note that since $\Xmij{ij}\neq 0$ at tree-level, we will not calculate loop corrections to the corresponding operator, but rather focus on $\Xpij{ij}$, where finite corrections generate novel effects such as modified $Z\ell^+\ell^-$ couplings (after EW symmetry breaking). 
Indeed, at the one-loop level, the relation $C_{\varphi\ell,ij}^{(3)\alpha\beta}= -C_{\varphi\ell,ij}^{(1)}$ or, equivalently, $\Xp=0$, is broken by the contributions of the diagrams shown in Figure~\ref{fig:WBllhard}. Performing an on-shell matching, we find
\begin{align}
\Xpij{ij}
    =&\frac{ 1 }{2304 \pi ^2}
    \sum_{a=1}^n\Yvij{ia} \MRi{a}^{-2} \Yvijconj{ja}\;
     \left(g_1^2+17g_2^2\right)\left(11+6 \left(\frac{1}{\varepsilon}+\log \left(\frac{\mu ^2}{\MRi{a}^2}\right)\right)\right)\notag\\
	& -\frac{1}{32\pi^2}\sum _{a,b=1}^n \Yvij{ia}\left(\sum_{c=1}^3 \Yvijconj{ca}\Yvij{cb}\right)\Yvijconj{jb}\; \;\dfrac{1}{\MRi{a}^2-\MRi{b}^2}\log\left(\dfrac{\MRi{a}^2}{\MRi{b}^2}\right)\notag\\
	& -\frac{1}{128\pi^2}\sum _{a,b=1}^N \Yvij{ia}\MRi{a}^{-1}\left(\sum_{c=1}^3 \Yvij{ca}\Yvijconj{cb}\right)\MRi{b}^{-1}\Yvijconj{jb} 
	\frac{\MRi{a}^2+\MRi{b}^2}{\MRi{a}^2-\MRi{b}^2}\log\left(\frac{\MRi{a}^2}{\MRi{b}^2}\right)\,,
	\label{eq:XP1loop}
\end{align}
where the $1/\varepsilon$ pole is cancelled by the renormalisation of the EFT operator, leading to the corresponding renormalisation group evolution (RGE)~\cite{Jenkins:2013zja}. 

The third line of \eqref{eq:XP1loop} originates from diagram given in Figure~\ref{fig:WBll_Majorana}, i.e.~from penguins involving a combination of two $\Delta L=2$ interactions. This term is only relevant in presence of large mass splitting between the sterile neutrinos, since it involves the same Yukawa structure as the Wilson coefficient of the Weinberg operator and vanishes in the limit of degenerate heavy neutrino masses. See Appendices~\ref{sec:FullNeutrinoMatrix} and ~\ref{sec:Zll} for a discussion of the identities we used for the derivation of this result.

\subsubsection{Four-Lepton Operators ($\mathcal{O}_{\ell\ell}$ and $\mathcal{O}_{\ell e}$)}\label{sec:OneLoopMatchingFourLepton}

\begin{figure}[ht]
	\center
	\subfloat[]{%
		\includegraphics[width=.37\textwidth]{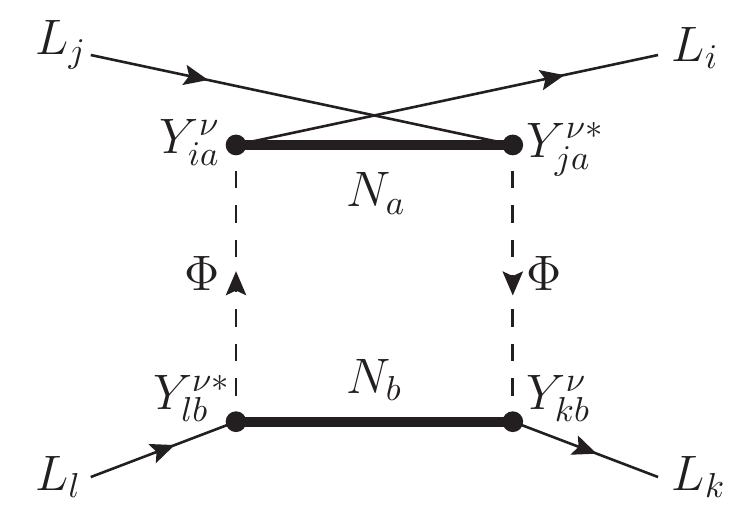}\label{fig:FFBox}%
	}
	\hspace{15mm}
	\subfloat[]{%
		\includegraphics[width=.37\textwidth]{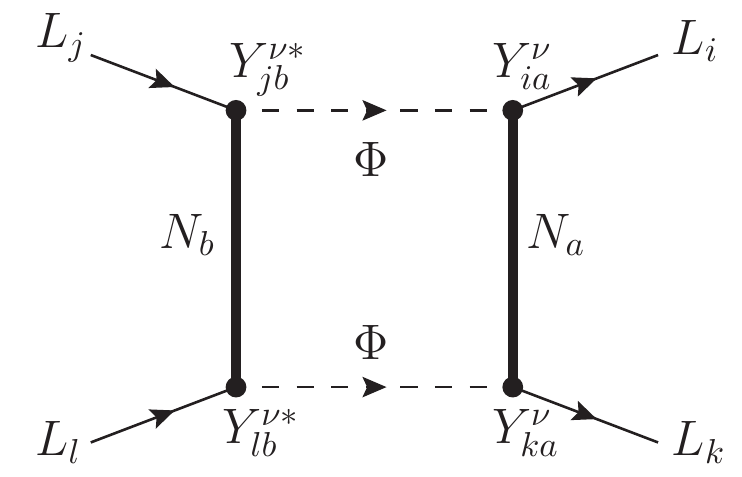}\label{fig:FFMajoranaBox}%
	}
	\\[4mm]
	\subfloat[]{%
		\includegraphics[width=.43\textwidth]{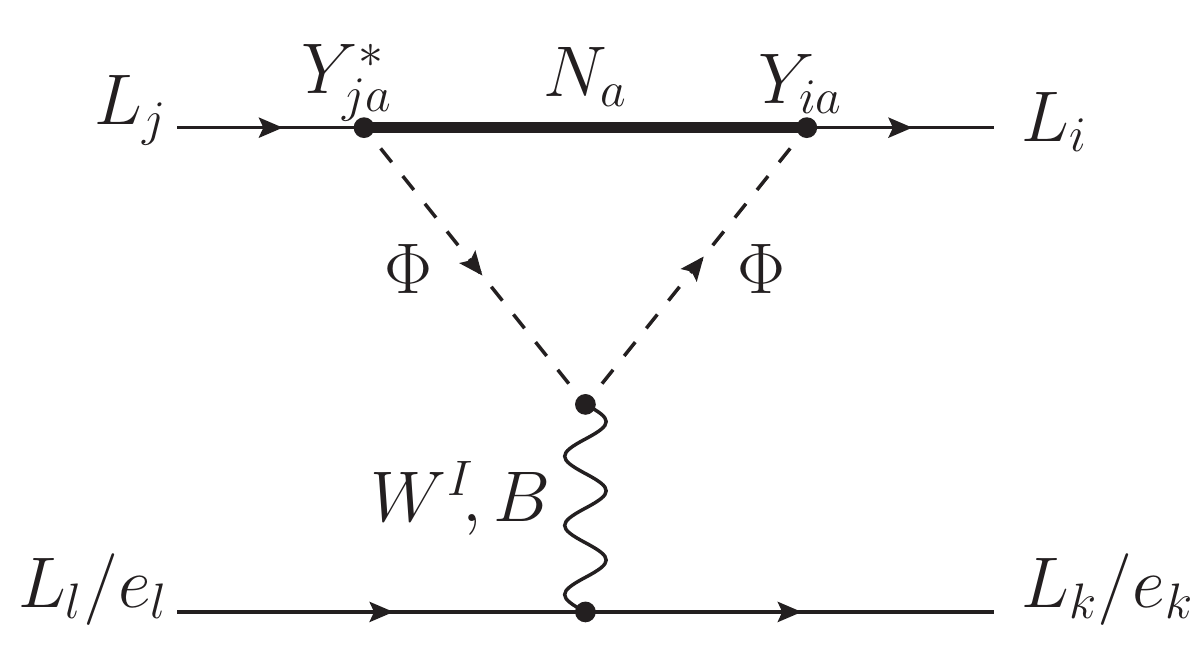}\label{fig:FFPenguin}%
	}
	\hspace{5mm}
	\subfloat[]{%
		\includegraphics[width=.4\textwidth]{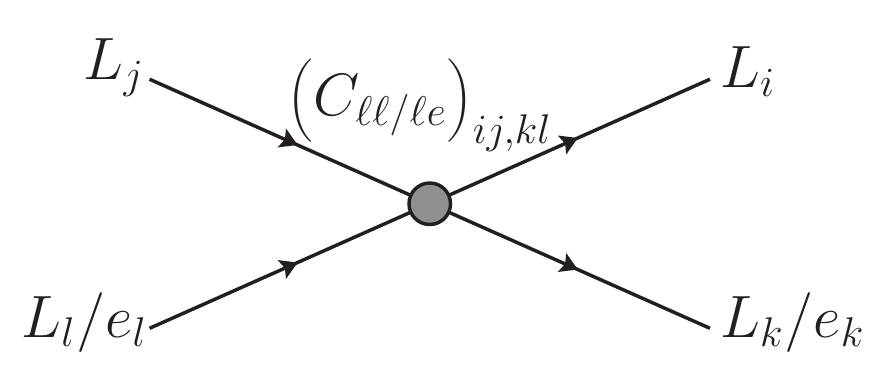}\label{fig:FFMatching}%
	}
   \caption{The double-Higgs boxes and the double-Higgs penguin are the only contributions to the loop-level matching onto $\mathcal{O}_{\ell\ell}$ and $\mathcal{O}_{\ell e}$.  }\label{fig:FourFermionMatching}	
\end{figure}

The four-lepton operators $\mathcal{O}_{\ell\ell}$ and $\mathcal{O}_{\ell e}$ receive contributions from off-shell $B$ and $W$ penguins and Higgs-neutrino boxes (see Figure~\ref{fig:FourFermionMatching}). The latter contribute only to $\mathcal{O}_{\ell\ell}$. We find the following ($\xi$-independent) Wilson coefficients:
\begin{align}
\left(C_{\ell \ell}\right)_{ij,kl}=
&\frac{g_1^2+g_2^2}{4608 \pi ^2}\sum _{a=1}^n \left(
\Yvij{ia}\MRi{a}^{-2}\Yvijconj{ja}\delta_{kl}
+\Yvij{ka}\MRi{a}^{-2}\Yvijconj{la}\delta_{ij}
\right)
\left(11+6 \left(\frac{1}{\varepsilon}+\log \left(\frac{\mu ^2}{\MRi{a}^2}\right)\right)\right)\notag\\
&+\frac{1}{128\pi^2}\sum _{a,b=1}^n 
\Yvij{ia}\Yvijconj{ja}\Yvij{kb}\Yvijconj{lb}
\dfrac{1}{\MRi{a}^2-\MRi{b}^2}\log\left(\dfrac{\MRi{a}^2}{\MRi{b}^2}\right)\notag\\
&+\frac{1}{128\pi^2}
\sum _{a=1}^N \left( \Yvij{ia}\MRi{a}^{-1}\Yvij{ka}\right)\left(\Yvijconj{jb}\MRi{b}^{-1}\Yvijconj{lb}\right)
\frac{\MRi{a}^2+\MRi{b}^2}{\MRi{a}^2-\MRi{b}^2}\log\left(\frac{\MRi{a}^2}{\MRi{b}^2}\right)
\label{eq:Cll}\\
\left(C_{\ell e}\right)_{ij,kl}=&\frac{g_1^2}{1152 \pi ^2}\sum _{a=1}^n \Yvij{ia}\MRi{a}^{-2}\Yvijconj{ja}\delta_{kl}
\left(11+6\left(\frac{1}{\varepsilon}+ \log \left(\frac{\mu ^2}{\MRi{a}^2}\right)\right)\right)\,.
\label{eq:Cle}
\end{align}
The third line of Eq.~(\ref{eq:Cll}) corresponds to the contribution of the diagram in Figure~\ref{fig:FFMajoranaBox}, which can only arise if Majorana particles are in the loop, since it features two lepton number violating interactions. Given that we are imposing the inverse seesaw condition of Eq.~\eqref{eq:InverseSeesawCond}, these diagrams only contribute in presence of sterile neutrino mass splitting.

\subsubsection{Two-Lepton-Two-Quark Operators ($\mathcal{O}_{\ell q}^{(1)},\,\mathcal{O}_{\ell q}^{(3)},\,\mathcal{O}_{\ell u}$ and $\mathcal{O}_{\ell d}$)}
Next we consider contributions to the two-lepton-two-quark operators $\mathcal{O}_{\ell q}^{(1)},\,\mathcal{O}_{\ell q}^{(3)},\,\mathcal{O}_{\ell u}$ and $\mathcal{O}_{\ell d}$, defined in \Eq{eq:2l2qOps}, which receive contributions from $W^I$ and $B$ penguins similar to the ones shown in Figure~\ref{fig:FFPenguin}. Here we find the Wilson coefficients
\begin{align}
C_{\ell q,ij}^{(1)}=&-\frac{g_1^2}{6912\pi^2}\sum_{a=1}^n\Yvij{ia}\MRi{a}^{-2}\Yvijconj{ja}\left(11+6\left(\frac{1}{\varepsilon}+\log\left(\frac{\mu^2}{\MRi{a}^2}\right)\right)\right)
\label{eq:Clq1}\\
C_{\ell q,ij}^{(3)}=&\;\;\;\;\frac{ g_2^2}{2304\pi^2}\sum_{a=1}^n\Yvij{ia}\MRi{a}^{-2}\Yvijconj{ja}\left(11+6\left(\frac{1}{\varepsilon}+\log\left(\frac{\mu^2}{\MRi{a}^2}\right)\right)\right)
\label{eq:Clq3}\\
C_{\ell u,ij}^{\mathrm{V}LR}=&-\frac{g_1^2}{1728\pi^2}\sum_{a=1}^n\Yvij{ia}\MRi{a}^{-2}\Yvijconj{ja}\left(11+6\left(\frac{1}{\varepsilon}+\log\left(\frac{\mu^2}{\MRi{a}^2}\right)\right)\right)
\label{eq:Clu}\\
C_{\ell d,ij}^{\mathrm{V}LR}=&\;\;\;\;\frac{g_1^2}{3456\pi^2}\sum_{a=1}^n\Yvij{ia}\MRi{a}^{-2}\Yvijconj{ja}\left(11+6\left(\frac{1}{\varepsilon}+\log\left(\frac{\mu^2}{\MRi{a}^2}\right)\right)\right)\,.
\label{eq:Cld}
\end{align}
Note that the box contributions vanish in the limit of zero quark Yukawa couplings. Only the box involving the top quark could be {sizeable}, which, however, is not relevant for charged lepton flavour violating processes, such as $\mu\to e$ conversion in nuclei.

\begin{figure}[ht]
	\center
	\subfloat[]{%
		\includegraphics[width=.37\textwidth]{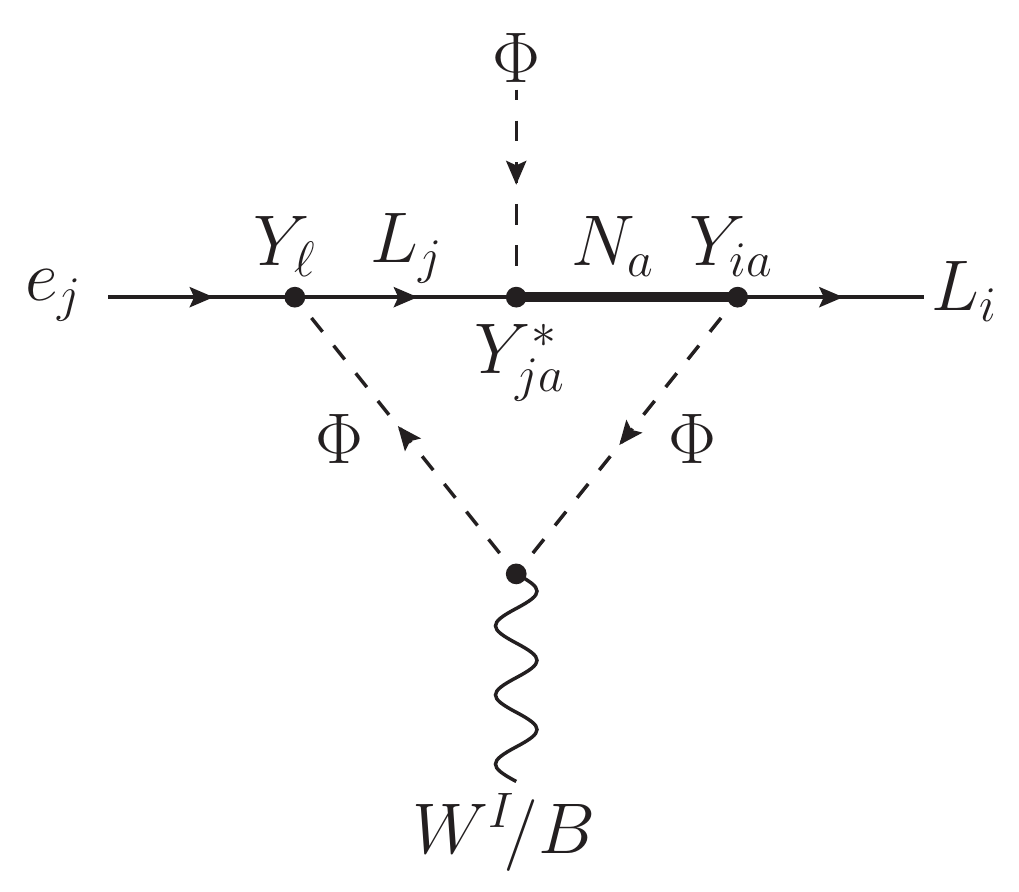}\label{}%
	}
	\hspace{5mm}
	\subfloat[]{%
		\includegraphics[width=.37\textwidth]{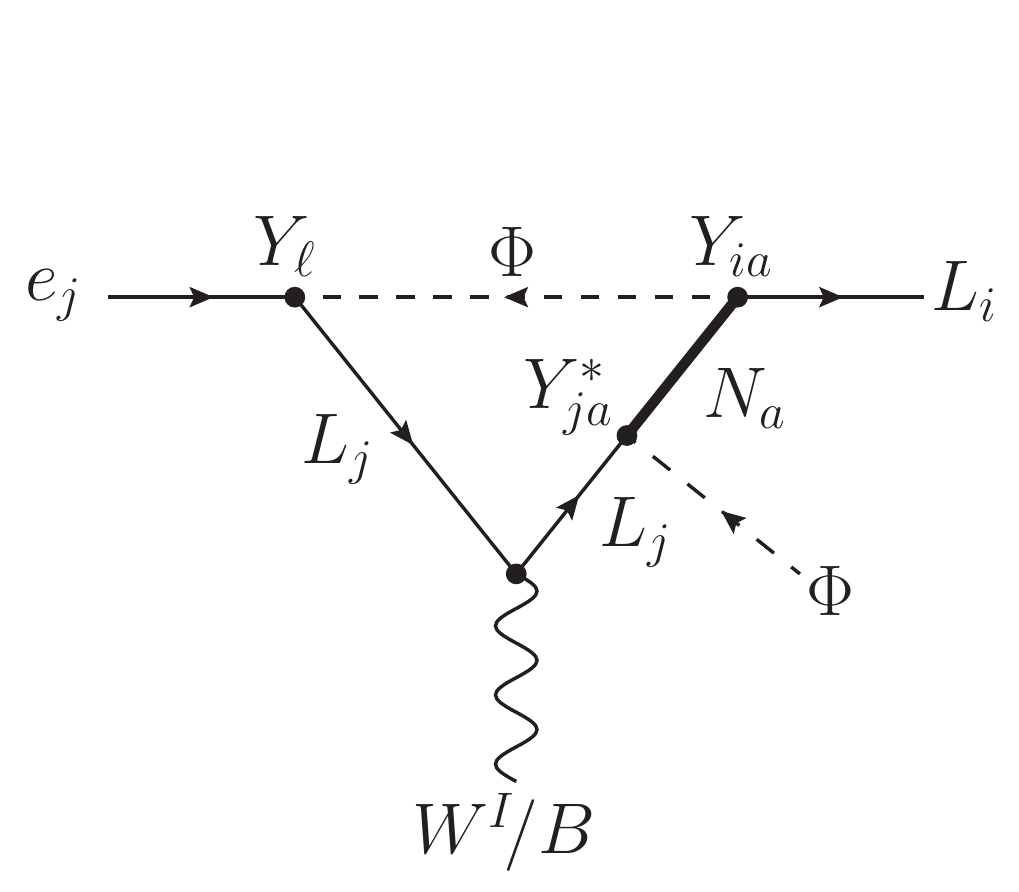}\label{}%
	}\\
	\subfloat[]{%
		\includegraphics[width=.37\textwidth]{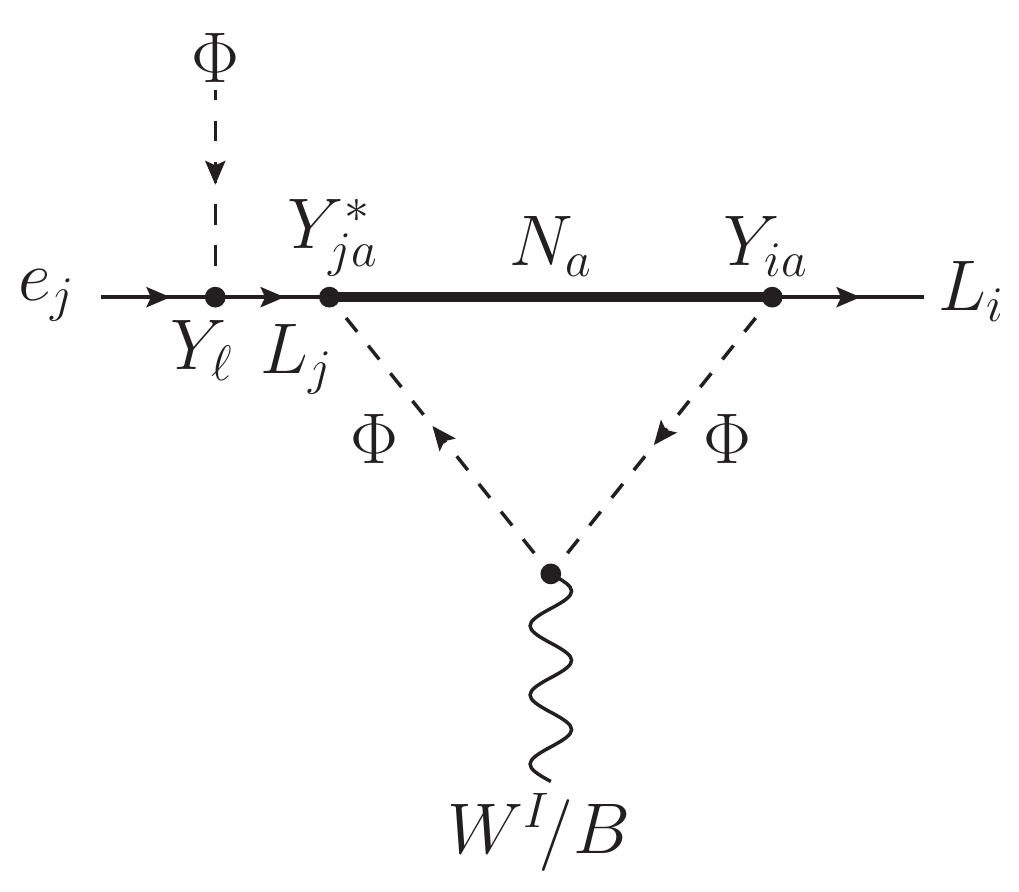}\label{}%
	}
	\subfloat[]{%
		\includegraphics[width=.37\textwidth]{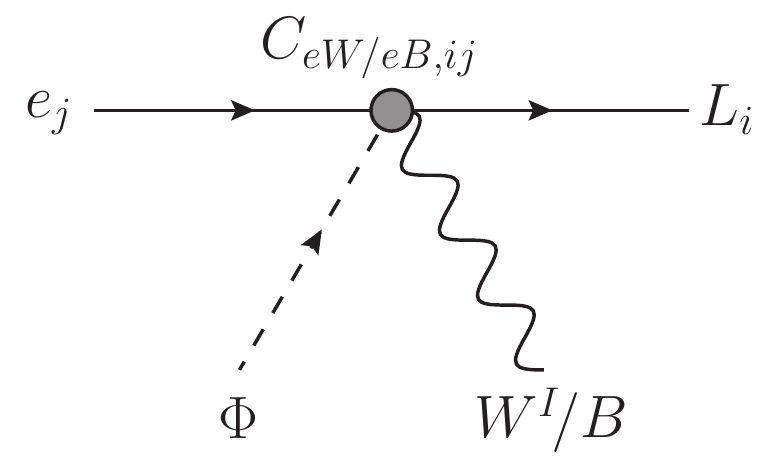}\label{}%
	}
   \caption{(a-c) Diagrams contributing to the matching onto the magnetic operators $\mathcal{O}_{eW}$ and $\mathcal{O}_{eB}$. (d) Diagram in the effective theory  originating from the magnetic operators. }\label{fig:MagnOpDiags}	
\end{figure}

\subsubsection{Magnetic Operators ($\mathcal{O}_{eW}$ and $\mathcal{O}_{eB}$)}\label{sec:OneLoopMatchingSMEFTmagn}
The diagrams contributing to the matching onto the magnetic operators are shown in Figure~\ref{fig:MagnOpDiags}. These result in
\begin{align}
C_{eW,ij}=&-\frac{5g_2}{384\pi^2} \Tij{ij} \Yli{j}\,,
\label{eq:CeW}\\
C_{eB,ij}=&-\frac{g_1}{384\pi^2} \Tij{ij} \Yli{j}\,,
\label{eq:CeB}
\end{align}
where $Y^\ell_j$ is the Yukawa coupling of the charged lepton $\ell_j$. These matching conditions agree with the results in Ref.~\cite{Zhang:2021tsq}. Note the absence of logarithms, which can be explained by the fact that, in the lepton number conserving limit, the type-I seesaw introduces purely left-handed, i.e.~chirality conserving, new physics, resulting in the absence of operator mixing in the EFT at the one-loop level.
\smallskip

\section{Observables}\label{sec:Flavour}

In the last section we calculated the matching of the type-I seesaw onto the SMEFT at the scale $\MR$. However, these results are not sufficient for a phenomenological analysis of charged lepton flavour violating observables. For this, the running from $M_R$ to the EW scale, the matching at this scale onto the Low-Energy Effective Field Theory (LEFT)~\cite{Dekens:2019ept}, as well as the RGE from the weak scale to the charged lepton scale~\cite{Crivellin:2017rmk,Jenkins:2017dyc} and the evaluation of the matrix elements at the tau or muon scale would be required. As these results are not fully available, in particular, because even the calculation of 2-loop effects would be necessary for a consistent treatment, we naively calculate in this section the relevant diagrams without scale separation (i.e.~without resumming the logarithms). Note that these formulae nonetheless include the (potential) leading logarithm as well as the finite scheme independent terms, which correspond to the sum of any hard and soft contributions to the amplitudes. We will then use these results in our phenomenological analysis in Section~\ref{sec:Pheno}. 

\subsection{Lepton Flavour Universality Tests}
\label{sec:LFU}

\begin{figure}
    \centering
    \includegraphics[scale=.77]{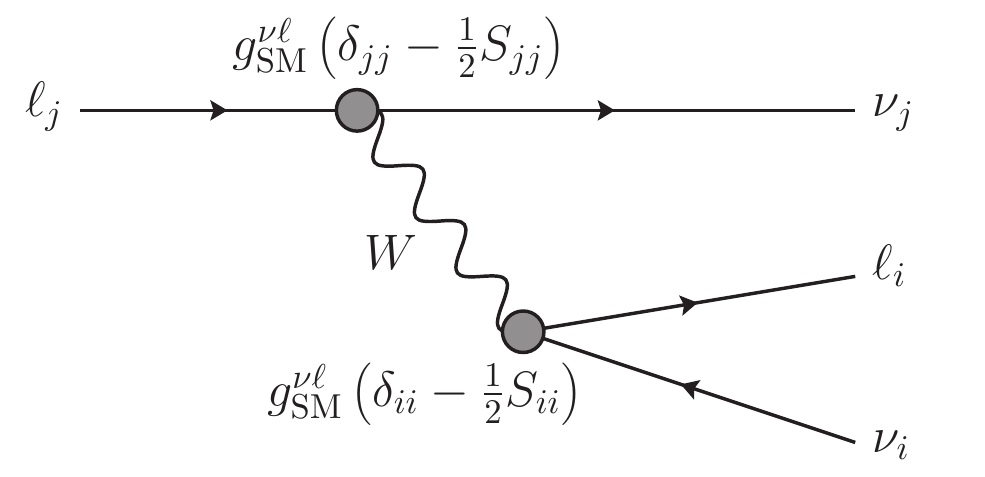}
    \caption{Tree-level diagrams generating $\ell\to\ell'\nu\nu$, in the type-I seesaw. $g_{\rm SM}^{\nu\ell}$ is the SM $W\ell\nu$ coupling defined in Eq.~\eqref{eq:glSM}and $S_{ii}$ the new physics modification of Eq.~\eqref{eq:Sij}.}
    \label{fig:l_lnunu}
\end{figure}

The $W\ell\nu$ couplings, which are modified at tree-level by the neutrino mixing, lead to effects in processes such as $\ell\to \ell^\prime\nu\nu$ (see Figure~\ref{fig:l_lnunu}), $\tau\to \pi\nu$, $\pi \to \ell\bar\nu$, $\tau\to K\nu$ or $K \to \ell\bar\nu$. For these decays, LFU ratios can be formed, which we compare to the HFLAV fit results~\cite{HFLAV:2019otj,HFLAV22} for the coupling fractions $\left(g_i/g_j\right)_\tau$, which are obtained using pure leptonic processes, and $\left(g_i/g_j\right)_P$, $P\in \{\pi,K\}$, which are defined as~\cite{HFLAV:2019otj,HFLAV22}
\begin{align}
\left\vert\frac{g_\tau}{g_\mu}\right\vert_P ^2 = \frac{\mathrm{Br}(\tau \to P\nu_\tau)}{\mathrm{Br}(P \to \mu\bar{\nu}_\mu)}\frac{ \tau_P}{\tau_\tau}\frac{2m_P m_\mu^2}{m_\tau^3}\left(\frac{1-m_\mu^2/m_P^2}{1-m_P^2/m_\tau^2}\right)^2 \frac{1}{1+\delta R_{\tau/P}}\,,
\end{align}
where $\delta R_{\tau/P}$ accounts for the radiative corrections to $\Gamma(\tau\to P\nu_\tau)/\Gamma (P\to \mu\bar{\nu}_\mu)$, $P\in \{\pi,K\}$,
which have been estimated as~\cite{Decker:1994dd,Decker:1994ea,Decker:1994kw,Marciano:1993sh,Pich:2013lsa}
\begin{align}
\delta R_{\tau /\pi} = (0.16\pm 0.14)\%, \qquad \delta R_{\tau/K} = (0.90\pm 0.22)\%\,.
\end{align}

We identify the new physics amplitude fractions directly with the current HFLAV fit results~\cite{HFLAV:2019otj,HFLAV22} for the coupling fractions $g_i/g_j$, $i,j\in\{e,\mu,\tau\}$.
\begin{align}
\begin{aligned}
 \frac{\mathcal{M}(\tau\to e\bar \nu\nu)}{\mathcal{M}(\mu\to e\bar \nu\nu)} \simeq
 & 1-\frac{1}{2}\left(\Sij{\tau\tau}-\Sij{\mu\mu}\right) \equiv 
 \left\vert\frac{g_\tau}{g_\mu}\right\vert_\tau = 1.0009(14)
\\
\frac{\mathcal{M}(\tau\to \mu\bar \nu\nu)}{\mathcal{M}(\mu\to e\bar \nu\nu)} \simeq
& 1-\frac{1}{2}\left(\Sij{\tau\tau}-\Sij{ee}\right)\equiv 
\left\vert\frac{g_\tau}{g_e}\right\vert_\tau = 1.0027(14)
\\
\frac{\mathcal{M}(\tau\to \mu\bar \nu\nu)}{\mathcal{M}(\tau\to e\bar \nu\nu)} \simeq
& 1-\frac{1}{2}\left(\Sij{\mu\mu}-\Sij{ee}\right)\equiv 
\left\vert\frac{g_\mu}{g_e}\right\vert_\tau = 1.0019(14)
\\
\frac{\mathcal{M}(\tau\to K \nu_\tau)}{\mathcal{M}(K\to\mu\bar\nu_\mu)} \simeq
& 1-\frac{1}{2}\left(\Sij{\tau\tau}-\Sij{\mu\mu}\right)\equiv
\left\vert\frac{g_\tau}{g_\mu}\right\vert_K = 0.9855(75)
\\
 \frac{\mathcal{M}(\tau\to \pi \nu_\tau)}{\mathcal{M}(\pi\to\mu\bar\nu_\mu)} =& 
 1-\frac{1}{2}\left(\Sij{\tau\tau}-\Sij{\mu\mu}\right)\equiv
\left\vert\frac{g_\tau}{g_\mu}\right\vert_\pi = 0.9959(38)
\\
\end{aligned}
\label{eq:geff_vals}
\end{align}
The HFLAV fit results come with the correlation matrix~\cite{HFLAV:2019otj,HFLAV22}
\begin{center}
\begingroup
\setlength{\tabcolsep}{5pt} 
\renewcommand{\arraystretch}{2}
\begin{tabular}{ c c }
\begingroup 
\setlength{\tabcolsep}{4pt}
 \begin{tabular}{ccccc}
     $\left\vert\dfrac{g_\tau}{g_\mu}\right\vert_\tau$  & $\left\vert\dfrac{g_\tau}{g_e}\right\vert_\tau$   & $\left\vert\dfrac{g_\mu}{g_e}\right\vert_\tau$  & $\left\vert\dfrac{g_\tau}{g_\mu}\right\vert_\pi$  & $\left\vert\dfrac{g_\tau}{g_\mu}\right\vert_K$ 
 \end{tabular}
\endgroup&
\\
$\begin{pmatrix}
    1\; & \;0.51\; & \;-0.49~\text{\footnotemark} \; & \; 0.16 \; & \; 0.12\\
    0.51\; & \;1\; & \;0.49 \; & \; 0.18 \; &\; 0.11 \\
    -0.49\; & \;0.49\; & \;1  \; & \; 0.01 \; & \; -0.01\\
    0.16 \; & \; 0.18 \; & \; 0.01 \; & \; 1 \; & \; 0.07 \\
    0.12 \; & \; 0.11 \; & \; -0.01 \; & \; 0.07 \; & \; 1\\
\end{pmatrix}$ &
\begin{tabular}{ccc}
     $\left\vert g_\tau /g_\mu\right\vert_\tau$  \\[1mm] $\left\vert g_\tau/g_e\right\vert_\tau$   \\[1mm]  $\left\vert g_\mu / g_e\right\vert_\tau$ 
     \\[1mm] $\left\vert g_\tau /g_\mu\right\vert_\pi$\\[1mm] $\left\vert g_\tau /g_\mu\right\vert_K$
 \end{tabular} 
\end{tabular}
\endgroup
\end{center}

\footnotetext{The HFLAV collaboration reports -0.50 due to numerical limitations, however, we use $-0.49$ in order to have a positive semi-definite correlation matrix.}
Belle II, which will produce approximately ten times more tauons than Belle or BaBar, is expected to improve the measurements of $\tau\to \mu\nu\bar{\nu}$ and $\tau\to e\nu\bar{\nu}$~\cite{Belle-II:2018jsg}. 

Further LFU ratios that are relevant in this context are~\cite{Pich:2013lsa}
\begin{align}
\left\vert\frac{g_\mu}{g_e}\right\vert_{P\to e/\mu}^2 = 
\frac{\mathrm{Br}(P^-\to \mu \bar{\nu}_\mu(\gamma))}{\mathrm{Br}(P^-\to e \bar{\nu}_e (\gamma))}\frac{m_e^2}{m_\mu^2}\left(\frac{1-m_e^2/m_P^2}{1-m_\mu^2/m_P^2}\right)^2 (1+\delta R_{P\to e/\mu})\,,
\end{align}
where $\delta R_{P\to e/\mu}$ denotes the radiative corrections, including a summation of the leading QED logarithms $\alpha^n \log(m_\mu /m_e)$~\cite{Marciano:1993sh,Finkemeier:1995gi}, and a two-loop calculation of $\mathcal{O}(e^2p^4)$ effects within chiral perturbation theory~\cite{Cirigliano:2007xi}. Comparing the SM predictions~\cite{Cirigliano:2007xi} with the experimental results~\cite{NA62:2012lny,KLOE:2009urs} for $K\to e/\nu$, for $\pi\to e/\nu$), one obtains~\cite{Pich:2013lsa}
\begin{align}
\dfrac{\mathcal{M}(K\to \mu\nu)}{\mathcal{M}(K\to e\nu)}\simeq
1-\frac{1}{2}\left(\Sij{\mu\mu}-\Sij{ee}\right)
\equiv \left\vert\dfrac{g_\mu}{g_e}\right\vert_{K\to e/\mu}
=0.9978(20)\,.
\end{align}
This measurement will also be performed by J-PARC E36~\cite{Shimizu:2018jgs}.
Comparing the SM predictions~\cite{Cirigliano:2007xi} with the experimental results for $\pi\to e/\nu$~\cite{PiENu:2015seu}, one finds~\cite{Pich:2013lsa}
\begin{align}
\dfrac{\mathcal{M}(\pi \to \mu\nu)}{\mathcal{M}(\pi\to e\nu)}\simeq
1-\frac{1}{2}\left(\Sij{\mu\mu}-\Sij{ee}\right)
\equiv \left\vert\dfrac{g_\mu}{g_e}\right\vert_{\pi\to e/\mu}
= 1.0010(09)\,.
\end{align}
The PEN experiment expects to improve the sensitivity to $\pi\to \mu\nu/\pi\to e\nu$ by more than a factor three~\cite{PEN:2018kgj}.

Decays of the form $K\to \pi\ell\bar{\nu}_\ell$~\cite{Cirigliano:2011ny} are not helicity suppressed and are used for the determination of the Cabibbo angle. Comparing the values for the CKM element $V_{us}$ from  $K\to \pi\mu\bar{\nu}_\mu$ with the $V_{us}$ from  $K\to \pi e\bar{\nu}_e$ allows for a further test of LFU: ~\cite{Cirigliano:2011ny,FlaviaNetWorkingGrouponKaonDecays:2010lot,Pich:2013lsa}, 
\begin{align}
\dfrac{\mathcal{M}(K\to \pi\mu\bar{\nu})}{\mathcal{M}(K\to \pi e\bar{\nu})}
 \simeq 1-\frac{1}{2}\left(\Sij{\mu\mu}-\Sij{ee}\right)
 \equiv \left\vert\dfrac{g_\mu}{g_e}\right\vert _{K\to \pi (\mu/e)}  =   1.0010(25) \,.
\end{align}

LFU can also be tested directly in leptonic $W$ boson decays, however, these channels are statistically limited~\cite{ALEPH:2013dgf,Filipuzzi:2012mg,Pich:2013lsa,ParticleDataGroup:2020ssz} and the resulting bounds are not competitive with the ones from tau, kaon and pion decays.
However, future $e^+e^-$ colliders such as the International Linear Collider (ILC)~\cite{Baer:2013cma}, the Compact Linear Collider (CLIC)~\cite{CLIC:2018fvx} or the Future Circular Collider (FCC-ee)~\cite{FCC:2018byv,FCC:2018evy} could improve these bounds. 

\subsection{$Z\to \nu\nu$}
Also $Z\to\nu_i \bar{\nu}_j$ receives corrections at tree-level in presence of neutrino mixing. The corresponding amplitude 
\begin{align}
\mathcal{M}(Z\to \nu_j \bar{\nu}_i) = -\frac{e}{2\sw\cw}\left(\delta_{ij}-S_{ij}\right)\,\bar{\nu}_i \gamma^\mu P_L \nu_j Z_\mu 
\label{eq:Zvv}
\end{align}
affects the effective number of light neutrino species~\cite{ALEPH:1989kcj,ALEPH:2005ab}
\begin{align}
N_\nu^\mathrm{eff.} = 2.9840 \pm 0.0082 \,.
\label{eq:Nv}
\end{align}
Considering Eq.~\eqref{eq:Zvv}, we can approximate $N_\nu \sim 3 -2\sum_{i} S_{ii}$ in the type-I seesaw model if we neglect effects that do not interfere with the SM contribution.

\subsection{$Z\to \ell\ell^\prime$}
\begin{figure}
	\center
	\subfloat[\label{fig:Gdiags}]{%
		\includegraphics[width=0.4\textwidth]{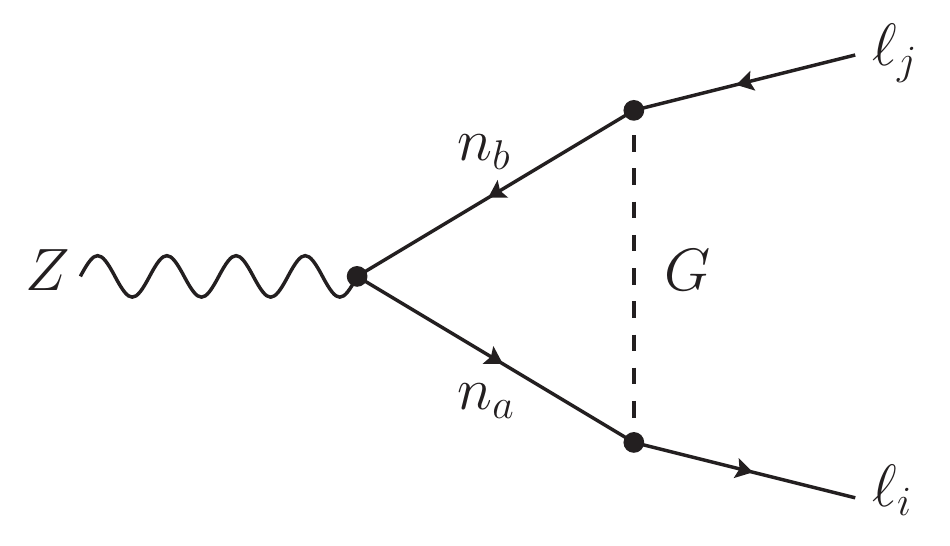}%
	}
	\hspace{5mm}
	\subfloat[\label{fig:Wdiags}]{%
		\includegraphics[width=0.4\textwidth]{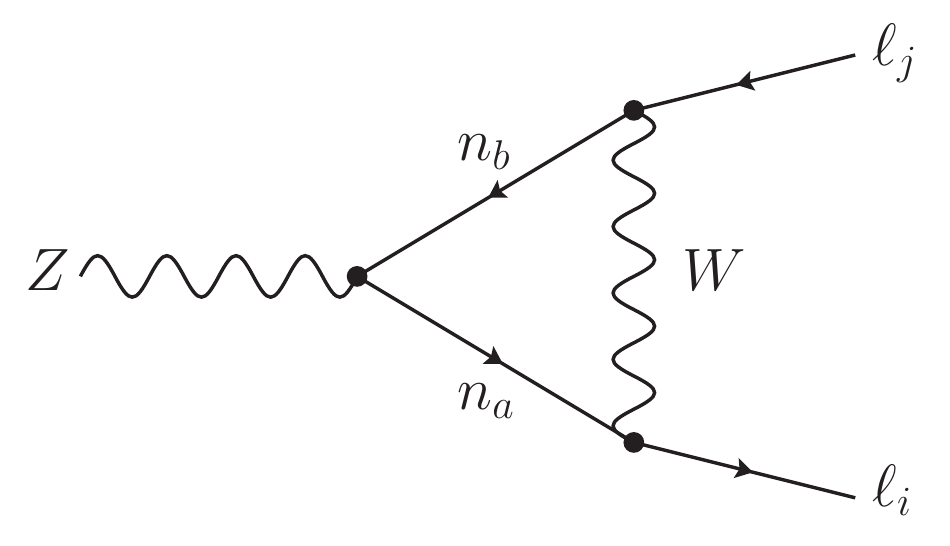}%
	}\\
	\subfloat[\label{fig:GGdiags}]{%
		\includegraphics[width=0.4\textwidth]{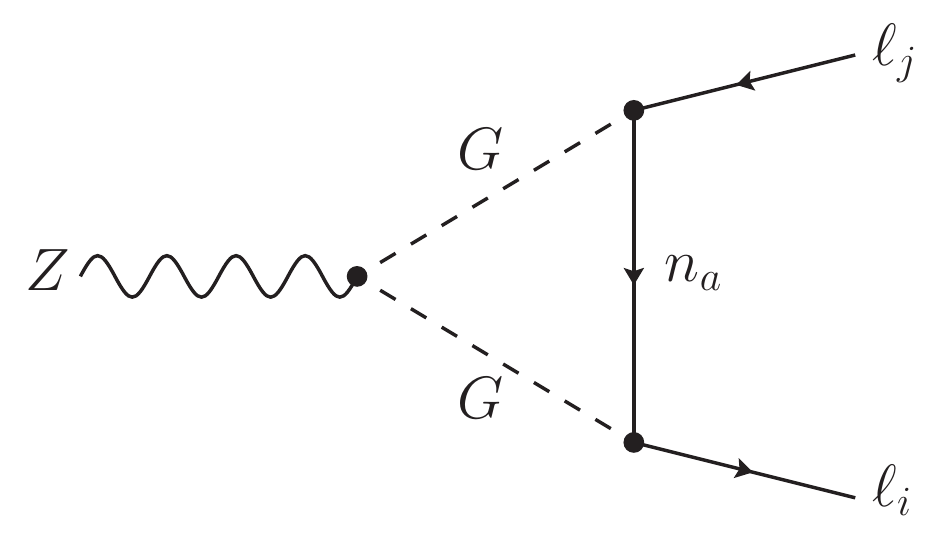}%
	}
	\subfloat[\label{fig:WWdiags}]{%
		\includegraphics[width=0.4\textwidth]{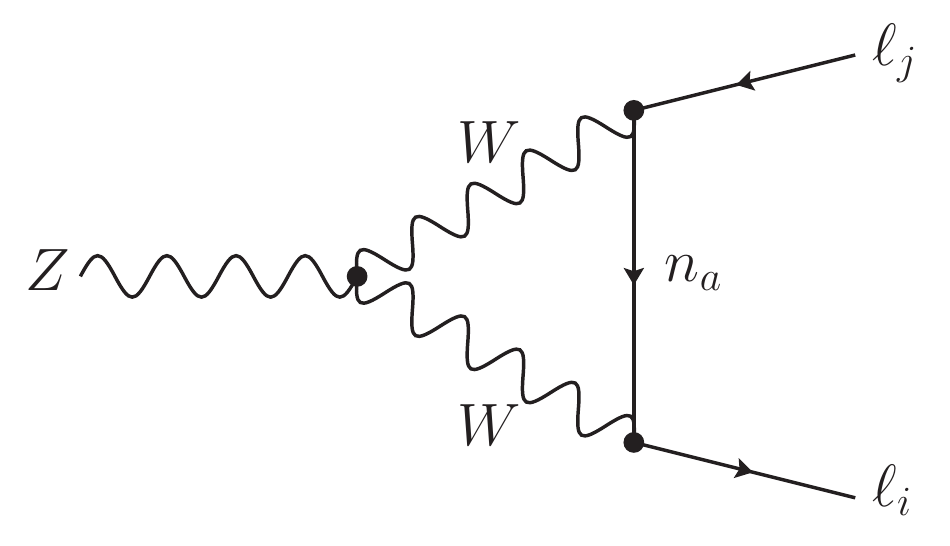}%
	}
	\hspace{5mm}
   \caption{Diagrams contributing to the $Z\ell\ell'$ coupling at the one-loop level. Here $n_a$ and $n_b$ correspond to neutrino mass eigenstates. The formulae in the main text are obtained by expanding the  amplitudes corresponding to these diagrams in $v^2/M_R^2$, i.e.~they are given in the seesaw limit. The unexpanded results are given in Appendix~\ref{sec:FullResults}.}\label{fig:Zlldiags}	
\end{figure}

$Z$ decays into charged leptons $Z\to\bar{\ell}_i\ell_j$ only receive corrections at loop-level in the type-I seesaw model. Expanding the diagrams shown in Figure~\ref{fig:Zlldiags} in $v^2/M_R^2$, we find 
\begin{align}
\mathcal{M}(Z\to \lLRibar{i}\lLRi{j})=-\frac{e^3}{16\pi^2\cw\sw^3}\;\overline{\mathcal{Z}}^{\mathrm{V}L}_{ij}\!(q_Z^2)\;
\lLRibar{i}(p_i)\slashed{Z}P_L \lLRi{j}(p_j)\,,
\label{eq:Zll}
\end{align}
with
\begin{align}
\overline{\mathcal{Z}}^{\mathrm{V}L}_{ij}(M_Z^2)=&\;
\sum _{a=1}^n \MDij{ia} \MRi{a}^{-2} \MDijconj{ja} \left(f(\cw^2)+g(\cw^2)\log\left(\frac{\MW^2}{\MRi{a}^2}\right)\right)\notag\\
&\quad+\frac{1}{4\MW^2}\sum _{a,b=1}^n \MDij{ia}\left(\sum_{c=1}^3 \MDijconj{ca}\MDij{cb}\right)\MDij{jb}\; \; h\left(\MRi{a}^2,\MRi{b}^2\right)\notag\\
&\quad +\frac{1}{8\MW^2}\sum_{a,b=1}^n\MDij{ia}\MRi{a}^{-1}\left(\sum_{k=1}^3\MDij{ka}\MDijconj{kb}\right)\MRi{b}^{-1}\MDijconj{jb}\; \;k\left(\MRi{a}^2,\MRi{b}^2\right)\,,
\label{eq:ZllseesawMOR}
\end{align}
where
\begin{align}
f(x)=& \;\frac{13}{6}+\frac{5}{144\; x} \left(1-2 x\right)+\frac{11}{12} \,x-x^2-\frac{1}{3} \,\pi ^2 \left(1+x\right)^2\notag\\&
+\frac{1}{2} \log \left(x\right) \left(3+2 \, x+\left(1+x\right)^2 \log\left(x\right)\right)\notag\\&
+\left(1+x\right)^2 \mathrm{Li}_2\left(1+x\right)\notag  + i \pi  \left(\frac{3}{2}+x+\left(1+x\right)^2 \log \left(x\right)\right)\notag\\&
  -4 \,  x^2\left(2+x\right) \arctan^2\left(\left(4  \,x-1\right)^{-1/2}\right)\notag\\&
  +\sqrt{4 \, x-1} \left(-\frac{1}{12 \,x}-\frac{3}{2}+\frac{7}{3}\, x +2\, x^2\right) \arctan\left(\left(4 \, x-1\right)^{-1/2}\right),\\[5mm]
g(x)=& \;-\frac{2}{3}\left(1+\dfrac{1}{16\,x}\right)
\label{eq:ZllseesawMORLoopFcts}\\
h\left(\MRi{a}^2,\MRi{b}^2\right)=&
\dfrac{1}{\MRi{a}^2-\MRi{b}^2}\log\left(\dfrac{\MRi{a}^2}{\MRi{b}^2}\right)\,,
\label{eq:hFct}\\
k\left(\MRi{a}^2,\MRi{b}^2\right)=&\frac{1}{2}\frac{\MRi{a}^2+\MRi{b}^2}{\MRi{a}^2-\MRi{b}^2}\log\left(\frac{\MRi{a}^2}{\MRi{b}^2}\right)
\label{eq:kFct}
\,.
\end{align}
Note that while the first two terms of Eq.~\eqref{eq:ZllseesawMOR} agree with Ref.~\cite{Herrero:2018luu}, the third term, which is only present for Majorana neutrinos, was not included there.

The branching ratio of $Z\to \ell_i\ell_j$, for $i\neq j$ is given by
\begin{align}
\mathrm{Br}(Z\to\ell_i^\pm \ell_j^\mp)\equiv &\mathrm{Br}(Z\to\ell_i^+\ell_j^-) +\mathrm{Br}(Z\to\ell_i^-\ell_j^+)\notag  \\
=& \frac{1}{24\pi}\frac{\MZ}{\Gamma_Z} \left(\frac{e^3}{16\pi^2 \cw \sw^3}\right)^2\left(\left\vert \overline{\mathcal{Z}}_{ij}(M_Z^2)\right\vert ^2 + \left\vert \overline{\mathcal{Z}}_{ji}(M_Z^2)\right\vert ^2\right)
\end{align}
We compare this result with the ATLAS and LEP measurements given in Table~\ref{tab:Zll_constraints} and to the future sensitivities in Table~\ref{tab:Zll_sensitivities}.
\begin{table}[t]
\centering
\begin{tabular}{|r l l|}
\hline
$\mathrm{Br}\left[Z\to e^\pm \mu^\mp\right]$    & $<7.5\times 10^{-7}\; (95\%\,{\rm CL})$   & ATLAS: \cite{ATLAS:2014vur}\\
 $\mathrm{Br}\left[Z\to e^\pm\tau^\mp\right]$   & $<5.0\times 10^{-6}\; (95\%\,{\rm CL})$   & ATLAS: \cite{ATLAS:2021bdj}\\
 $\mathrm{Br}\left[Z\to \mu^\pm\tau^\mp\right]$ & $<6.5\times 10^{-6}\; (95\%\,{\rm CL})$   & ATLAS:  \cite{ATLAS:2021bdj}\\
\hline
\end{tabular}
\caption{Bounds on $\mathrm{Br}(Z\to \ell\ell')$.}\label{tab:Zll_constraints}
\end{table}

\begin{table}[t]
\centering
\begin{tabular}{|r l l|}
\hline
$\mathrm{Br}\left[Z\to e^\pm \mu^\mp\right]$    & $10^{-7} - 10^{-8} $  & FCC-ee: \cite{FCC:2018evy,FCCeeZllpSensitivities}\\
$\mathrm{Br}\left[Z\to e^\pm \mu^\mp\right]$    & $3.0 \times 10^{-9} $  & CEPC: \cite{CEPCStudyGroup:2018ghi}\\
$\mathrm{Br}\left[Z\to e^\pm\tau^\mp\right]$    & $10^{-9} $            & FCC-ee: \cite{FCC:2018evy,FCCeeZllpSensitivities} \\
$\mathrm{Br}\left[Z\to e^\pm\tau^\mp\right]$    & $2.0 \times 10^{-8} $  & CEPC: \cite{CEPCStudyGroup:2018ghi} \\
$\mathrm{Br}\left[Z\to \mu^\pm\tau^\mp\right]$  & $10^{-9} $            & FCC-ee: \cite{FCC:2018evy,FCCeeZllpSensitivities} \\
$\mathrm{Br}\left[Z\to \mu^\pm\tau^\mp\right]$  & $2.0 \times 10^{-8} $ & CEPC: \cite{CEPCStudyGroup:2018ghi} \\
\hline
\end{tabular}
\caption{Future sensitivities to $\mathrm{Br}(Z\to \ell\ell')$}\label{tab:Zll_sensitivities}
\end{table}

\subsection{$\ell\to\ell'\gamma$}
Defining the effective Lagrangian in broken $SU(2)_L$, 
\begin{equation}
\mathcal{L}_{\rm eff}=
\mathcal{A}_{ij}^\mathrm{M}\lLRibar{i}\sigma_{\mu \nu} P_R \lLRi{j} F^{\mu \nu}+\hc\,,
\end{equation}
where $F^{\mu \nu}=\partial^\mu A^\nu-\partial^\nu A^\mu$ is the electromagnetic field strength tensor, we find
\begin{align}
\mathcal{A}^\mathrm{M}_{ij}=&\frac{e^3  m_{\ell_j}}{128\pi^2\sw^2\MW^2}\sum_{a=4}^n U_{ia} U_{ja}^*\approx\frac{e^3 m_{\ell_j}}{128\pi^2\sw^2 M_W^2}S_{ij}\,,\label{eq:AM}
\end{align}
in the seesaw limit, $v\ll \MR$, with $\Sij{ij}$ as defined in \Eq{eq:Sij}. The absence of a logarithm involving a sterile neutrino mass can be understood as a consequence of the purely left-handed new physics effect in this model. This avoids chiral enhancement, and implies that the anomalous magnetic moments do not yield relevant bounds, and that electric dipole moments are absent~\cite{Crivellin:2018qmi}.

\begin{table}[ht]
\centering
\begin{tabular}{|r l l|}
\hline
$\mathrm{Br}\left[\mu\to e\gamma\right]$      & $<4.2\times 10^{-13}$  & MEG: \cite{MEG:2016leq} \\
$\mathrm{Br}\left[\tau\to e\gamma\right]$     & $<3.3\times 10^{-8} $  & BaBar: \cite{BaBar:2009hkt}\\
$\mathrm{Br}\left[\tau\to \mu\gamma\right]$   & $<4.2\times 10^{-8} $  & Belle: \cite{Belle:2021ysv} \\
\hline
\end{tabular}
\caption{Current experimental upper bounds on $\mathrm{Br}(\ell \to \ell' \gamma)$.}\label{tab:llpgamma_constraints}
\end{table}

\begin{table}[ht]
\centering
\begin{tabular}{|r l l|}
\hline
$\mathrm{Br}\left[\mu\to e\gamma\right]$      & $6 \times 10^{-14}$      & MEG II: \cite{MEGII:2018kmf} \\
$\mathrm{Br}\left[\tau\to \mu\gamma\right]$   & $2.7\times 10^{-8}$      & Belle II: \cite{Belle-II:2018jsg} \\
\hline
\end{tabular}
\caption{Future sensitivities to $\mathrm{Br}(\ell \to \ell' \gamma)$.}\label{tab:llpgamma_sensitivities}
\end{table}

\subsection{$\ell \to 3\ell'$ and $\ell \to \ell' \ell''\ell''$}

Processes of the type $\ell \to 3\ell'$ receive contributions from photon and $Z$ penguins, as well as from box diagrams. In the seesaw limit, the sum of the off-shell photon penguins, the off-shell $Z$ penguins and boxes, leads to
\begin{align}
{\rm Br}\left[\tau^\mp\to e^\mp\mu^\pm\mu^\mp\right]=
&-\frac{e^2\, m_\tau^3}{48\, \pi ^3\,\Gamma_\tau} 
\left(\left|\mathcal{A}_{e\tau}^\mathrm{M}\right|^2+\left|\mathcal{A}_{\tau e}^\mathrm{M}\right|^2\right) 
\left(3+\log\left(\frac{m_\mu^2}{m_\tau^2}\right)\right)\notag\\
&+\frac{m_\tau^5}{1536\, \pi ^3 \,\Gamma_\tau }
\Bigg(\left. \left| \mathcal{F}^{\mathrm{V}LL}_{e\tau, \mu\mu} + \mathcal{F}^{\mathrm{V}LL}_{\mu\tau, e\mu}\right|^2 +\left|\mathcal{F}^{\mathrm{V}LR}_{e\tau, \mu\mu} \right|^2 + \left|\mathcal{F}^{\mathrm{V}LR}_{\mu \tau ,e\mu}\right|^2\right.\Bigg)
\notag\\
&+\frac{e \,m_\tau{}^4}{192\, \pi ^3 \,\Gamma_\tau}\Big({\rm Re}\left[\mathcal{A}_{e\tau}^{\mathrm{M}*}\left(\mathcal{F}^{\mathrm{V}LL}_{e\tau,\mu\mu}+\mathcal{F}^{\mathrm{V}LL}_{\mu\tau,e\mu}+\mathcal{F}^{\mathrm{V}LR}_{e\tau, \mu\mu}+\mathcal{F}^{\mathrm{V}LR}_{\mu \tau, e\mu}\Big)\right]
\right)\,,
\notag \\[4mm]
{\rm Br}\left[\tau \to 3\mu\right]=&-\frac{e^2\, m_\tau^3 }{192\, \pi ^3\,\Gamma_\tau}
\left(|\mathcal{A}_{\mu\tau}^\mathrm{M}|^2+|\mathcal{A}_{\tau\mu}^\mathrm{M}|^2\right)
\left(11+4 \log \left(\frac{m_\mu^2}{m_\tau^2}\right)\right)\notag\\
&+\frac{m_\tau^5 }{1536 \,\pi ^3\, \Gamma_\tau}
\left(2 \left|\mathcal{F}_{\mu\tau, \mu\mu}^{\mathrm{V}LL}\right|^2+\left|\mathcal{F}_{\mu\tau, \mu\mu}^{\mathrm{V}LR}\right|^2
\right)\notag\\
&+\frac{e\, m_\tau^4}{192\, \pi ^3 \,\Gamma_\tau }
\Bigg(\left.{\rm Re}\left[\mathcal{A}_{\mu\tau}^{\mathrm{M}*} \left(2\, \mathcal{F}_{\mu\tau, \mu\mu}^{\mathrm{V}LL}+\mathcal{F}_{\mu\tau,\mu\mu}^{\mathrm{V}LR}\right)\right]\right.\Bigg)\,,
\notag \\[4mm]
{\rm Br}\left[\tau^\mp\to \mu^\mp e^\pm \mu^\mp\right]=&\frac{m_\tau^5}{1536\,\pi^3\Gamma_\tau}
\left(2 \,\left|\mathcal{F}^{\mathrm{V}LL}_{\mu\tau,\mu e}\right|^2 + \left|\mathcal{F}^{\mathrm{V}LR}_{\mu\tau,\mu e}\right|^2\right)\,,
\label{eq:l3lBranchingRatios}
\end{align}
with
\begin{align}
\mathcal{F}_{ij,kl}^{\mathrm{V}LL}=&
\frac{e^4}{384\pi^2\sw^4\MW^2}\Bigg\lbrace 
\notag
\sum_{a=1}^n \MDij{ia}\MRi{a}^{-2}\MDijconj{ja} \delta_{kl} \left(9-37\sw^2+\left(-9+16\sw^2\right)\log\left(\frac{\MRi{a}^2}{\MW^2}\right)\right)\notag\\
&\qquad+\frac{3}{\MW^2}\sum_{a,b=1}^n \Bigg(\left(-1+2\sw^2\right)\left(\MDij{ia}\left(\sum_{c=1}^3\MDijconj{ca}\MDij{cb}\right)\MDijconj{jb}\right) \delta_{kl}\notag\\
&\qquad\qquad\qquad\qquad+\frac{1}{2}\left(\MDij{ia}\MDijconj{ja}\right)\left(\MDij{kb}\MDijconj{lb}\right)\Bigg)h\!\left(\MRi{a}^2,\MRi{b}^2\right)\notag\\
&\qquad+\frac{3}{2\MW^2}\sum_{a,b=1}^n \Bigg(\left(-1+2\sw^2\right)\left(\MDij{ia}\left(\sum_{c=1}^3\MDijconj{ca}\MDij{cb}\right)\MDijconj{jb}\right) \delta_{kl}\notag\\
&\qquad\qquad\qquad\qquad+2\left(\MDij{ia}\MDijconj{ja}\right)\left(\MDij{kb}\MDijconj{lb}\right)\Bigg)k\!\left(\MRi{a}^2,\MRi{b}^2\right)
\Bigg\rbrace\,,
\label{eq:FVLL}\\
\mathcal{F}_{ij,kl}^{\mathrm{V}LR}=&
\frac{e^4}{384\pi^2\sw^2\MW^2}\Bigg\lbrace 
\notag
\sum_{a=1}^n \MDij{ia}\MRi{a}^{-2}\MDijconj{ja} \delta_{kl} \left(-37+16\log\left(\frac{\MRi{a}^2}{\MW^2}\right)\right)\notag\\
&\qquad + \frac{6}{\MW^2}\sum_{a,b=1}^n \left(\MDij{ia}\left(\sum_{c=1}^3\MDijconj{ca}\MDij{cb}\right)\MDijconj{jb}\right)\delta_{kl}\, h\!\left(\MRi{a}^2,\MRi{b}^2\right)\Bigg\rbrace
\,,
\label{eq:FVLR}
\end{align}
with the functions $h$ and $k$ given in Eq.~\eqref{eq:kFct} and Eq.~\Eq{eq:hFct}, respectively. The corresponding formula for ${\rm Br}\left[\tau \to 3e\right]$ and ${\rm Br}\left[\mu \to 3e\right]$ can be obtained by the appropriate replacement of flavour indices.
\smallskip

\begin{table}[t]
\centering
\begin{tabular}{|r l l|}
\hline
$\mathrm{Br}\left[\mu^- \rightarrow e^- e^+ e^-\right], \; 90\%~\mathrm{C.L.}$        & $\leq 1.0 \times 10^{-12} $   & SINDRUM: \cite{SINDRUM:1987nra}\\
$\mathrm{Br}\left[\tau^- \rightarrow e^- e^+ e^-\right], \; 90\%~\mathrm{C.L.}$       & $\leq 2.7 \times 10^{-8}  $   & Belle: \cite{Hayasaka:2010np}\\
$\mathrm{Br}\left[\tau^- \rightarrow e^- \mu^+ \mu^-\right], \; 90\%~\mathrm{C.L.}$   & $\leq 2.7 \times 10^{-8}  $   & Belle: \cite{Hayasaka:2010np}\\
$\mathrm{Br}\left[\tau^- \rightarrow \mu^- e^+ \mu^-\right], \; 90\%~\mathrm{C.L.}$   & $\leq 1.7 \times 10^{-8}  $   & Belle: \cite{Hayasaka:2010np}\\
$\mathrm{Br}\left[\tau^- \rightarrow \mu^- e^+ e^-\right], \; 90\%~\mathrm{C.L.}$     & $\leq 1.8 \times 10^{-8}  $   & Belle: \cite{Hayasaka:2010np}\\
$\mathrm{Br}\left[\tau^- \rightarrow e^- \mu^+ e^-\right], \; 90\%~\mathrm{C.L.}$     & $\leq 1.5 \times 10^{-8}  $   & Belle: \cite{Hayasaka:2010np}\\
$\mathrm{Br}\left[\tau^- \rightarrow \mu^- \mu^+ \mu^-\right], \; 90\%~\mathrm{C.L.}$ & $\leq 2.1 \times 10^{-8}  $   & Belle: \cite{Hayasaka:2010np}\\
\hline
\end{tabular}
\caption{Current upper bounds on lepton flavour violating decays of the type $\ell\to3\ell^\prime$.}\label{tab:l3l_bounds}
\end{table}

\begin{table}[ht]
\centering
\begin{tabular}{|r l l|}
\hline
$\mathrm{Br}\left[\mu\to 3e\right]$      & $10^{-15}$                  & Mu3e, phase I: \cite{Blondel:2013ia,Perrevoort:2016nuv} \\
$\mathrm{Br}\left[\mu\to 3e\right]$      & $10^{-16}$                         & Mu3e, phase II: \cite{Blondel:2013ia,Perrevoort:2016nuv} \\
$\mathrm{Br}\left[\tau\to 3e\right]$     & $10^{-9} $                         & Belle-II: \cite{Belle-II:2018jsg}\\
$\mathrm{Br}\left[\tau\to 3\mu\right]$   & $\mathcal{O}\left(10^{-9}\right)$  & CMS, ATLAS, LHCb at HL-LHC \cite{Cerri:2018ypt} \\
$\mathrm{Br}\left[\tau\to 3\mu\right]$   & $3.3 \times 10^{-10}$              & Belle-II: \cite{Belle-II:2018jsg}  \\
\hline
\end{tabular}
\caption{Future sensitivities to $\mathrm{Br}(\ell \to 3\ell')$.}\label{tab:l3l_sensitivities}
\end{table}

\subsection{$\mu \to e$ Conversion in Nuclei}
Next we consider $\mu \to e$ conversion in nuclei and define
\begin{align}
\mathcal{L}_{\text{eff}}=\!\!\!\sum_{q=u,d}\!\left(\mathcal{F}_{e\mu,qq}^{LL}\,\mathcal{O}_{e\mu,qq}^{LL}\,+\,\mathcal{F}_{e\mu,qq}^{LR}\,\mathcal{O}_{e\mu,qq}^{LR}\right)+\left(L\leftrightarrow R\right)+\mathrm{h.c.}\,,
\end{align}
with
\begin{align}
\begin{split}
\mathcal{O}^{\mathrm{V}LL}_{ij,qq}=&\left(\lLRibar{i} \gamma_\mu P_L \lLRi{j}\right)\left(\qLRbar \gamma_\mu P_L \qLR\right)\,,\\
\mathcal{O}^{\mathrm{V}LR}_{ij,qq}=&\left(\lLRibar{i} \gamma_\mu P_L \lLRi{j}\right)\left(\qLRbar \gamma_\mu P_R \qLR\right)\,.
\end{split}
\end{align}
This process receives contributions from photon and $Z$ penguins, as well as from box diagrams, resulting in
\begin{align}
\mathcal{F}_{ij,uu}^{\mathrm{V}LL}=&\frac{e^4}{384\pi^2\sw^4\MW^2}
\Bigg\lbrace
\sum_{a=1}^n\MDij{ia}\MRi{a}^{-2}\MDij{ja}^* {\frac{1}{3} \left(27+74 \sw^2+\left(27-32 \sw^2\right) \log \left(\frac{\MRi{a}^2}{\MW^2}\right)\right)}\notag\\
&\qquad\qquad\qquad+\frac{3-4\sw^2}{\MW^2}\sum_{a,b=1}^n \MDij{ia}\left(\sum_{c=1}^3\MDijconj{ca}\MDij{cb}\right)\MDijconj{jb} h(\MRi{a}^2,\MRi{b}^2)
\Bigg\rbrace\notag\\[5mm]
\mathcal{F}_{ij,uu}^{\mathrm{V}LR}=&\frac{e^4}{384\pi^2\sw^2\MW^2}\Bigg\lbrace\sum_{a=1}^n\MDij{ia}\MRi{a}^{-2}\MDij{ja}^*\frac{2}{3}\left(37-16\log\left(\frac{\MRi{a}^2}{\MW^2}\right)\right)\notag\\
&\qquad\qquad\qquad-\frac{4}{\MW^2}\sum_{a,b=1}^n \MDij{ia}\left(\sum_{c=1}^3\MDijconj{ca}\MDij{cb}\right)\MDijconj{jb} h(\MRi{a}^2,\MRi{b}^2)
\Bigg\rbrace\notag\\[5mm]
\mathcal{F}_{ij,dd}^{\mathrm{V}LL}=&\frac{e^4}{384\pi^2\sw^4\MW^2}
\Bigg\lbrace
\sum_{a=1}^n\MDij{ia}\MRi{a}^{-2}\MDij{ja}^*{\frac{1}{3} \left(27-37 \sw^2+\left(-27+16 \sw^2\right) \log \left(\frac{\MRi{a}^2}{\MW^2}\right)\right)}\notag\\
&\qquad\qquad\qquad-\frac{3-2\sw^2}{\MW^2}\sum_{a,b=1}^n \MDij{ia}\left(\sum_{c=1}^3\MDijconj{ca}\MDij{cb}\right)\MDijconj{jb} h(\MRi{a}^2,\MRi{b}^2)
\Bigg\rbrace\notag\\[5mm]
\mathcal{F}_{ij,dd}^{\mathrm{V}LR}=&\frac{e^4}{384\pi^2\sw^4\MW^2}
\Bigg\lbrace
\sum_{a=1}^n\MDij{ia}\MRi{a}^{-2}\MDij{ja}^*\frac{\sw^2}{3}\left(-37+16\log\left(\frac{\MRi{a}^2}{\MW^2}\right)\right)\notag\\
&\qquad\qquad\qquad+\frac{2\sw^2}{\MW^2}\sum_{a,b=1}^n \MDij{ia}\left(\sum_{c=1}^3\MDijconj{ca}\MDij{cb}\right)\MDijconj{jb} h(\MRi{a}^2,\MRi{b}^2)\Bigg\rbrace\notag\\[5mm]
\mathcal{F}_{ij,qq}^{\mathrm{V}RL}=&0,\qquad \mathcal{F}_{ij,qq}^{\mathrm{V}RR}=0, 
\label{eq:mueconvFormFactors}
\end{align}
with $q=u,d$ and $h$ as given in \Eq{eq:hFct}. Here we neglected the quark masses and CKM-suppressed effects in the box diagrams by using $\left(V^\mathrm{CKM}V^\mathrm{CKM\dagger}\right)_{kl}=\delta_{kl}$.

Together with $\mathcal{A}^\mathrm{M}_{ij}$ from the magnetic photon penguin (see \Eq{eq:AM}), the  transition rate $\Gamma_{\mu\to e}^N\equiv \Gamma(\mu N\to eN)$ follows as
\begin{align}
\Gamma^{\mu\to e}_N =&\frac{m_\mu^5}{4}\Bigg \lbrace \Bigg \vert \frac{\mathcal{A}^\mathrm{M}_{e\mu}}{m_\mu}D_N+4\sum_{q=u,d}\left(\mathcal{F}_{e\mu,qq}^{LL}+\mathcal{F}_{e\mu,qq}^{LR}\right)\left(f_{Vp}^{(q)}V_N^p+ f_{Vn}^{(q)}V_N^n\right) 
\Bigg\vert^2
+ \Bigg \vert \frac{\mathcal{A}^\mathrm{M}_{\mu e}}{m_\mu}D_N\Bigg \vert^2\Bigg \rbrace \,.
\end{align}
For the overlap integrals between the muon and electron wave functions and the nucleon densities $D_{\rm Au}$, $V_{\text{Au}}^p$ and $V_{\text{Au}}$, we use the numerical values~\cite{Kitano:2002mt}
\begin{align}
\hspace{20mm} D_{\rm Au}&=0.189,\quad\quad
& V_{\text{Au}}^p & = 0.0974\,,\quad\quad
& V_{\text{Au}}^n & = 0.146\,,\hspace{20mm}
\\
D_{\rm Al}&= 0.0362 ,\quad \quad
& V_{\text{Al}}^p & = 0.0161\,, \quad \quad
& V_{\text{Al}}^n & = 0.0173\,.\hspace{20mm}
\end{align}
The nucleon form factors are given by
\begin{align}
f_{Vp}^{(u)}=2,\quad f_{Vn}^{(u)}=1,\quad f_{Vp}^{(d)}=1,\quad f_{Vn}^{(d)}=2\,.
\end{align}
The $\mu\to e$ conversion rate is defined as the $\mu\to e$ transition rate divided by the $\mu$ capture rate, which depends on the nature of the target $N$
\begin{align}
\mathrm{Cr}\left[\mu \to e,\,N\right]=\frac{\Gamma^{\mu\to e}_N}{\Gamma^{\rm capt}_N}\,.
\end{align}
For the capture rates for gold and aluminium, we use the values~\cite{Suzuki:1987jf}
\begin{align}
\Gamma_{\rm Au}^{\rm capt}=8.7\times 10^{-18}~\mathrm{GeV}\,,
\qquad \Gamma_{\text{Al}}^{\rm capt}=4.6\times 10^{-19}~\mathrm{GeV}\,.
\end{align}
The current best experimental limit on $\mu\to e$ conversion comes from SINDRUM II~\cite{SINDRUMII:2006dvw} (see Table~\ref{tab:mueconv_bound}). The COMET and Mu2e collaborations will be probing $\mathrm{Cr}\left[\mu \to e, \mathrm{Al}\right]$ and expect to improve the upper limit on $\mu \to e$ conversion by three orders of magnitude in the coming years~\cite{Baldini:2018uhj} (see Table~\ref{tab:mueconv_sensitivities}).

\begin{table}[ht]
\centering
\begin{tabular}{|r l l|}
\hline
$\mathrm{Cr}\left[\mu\to e,~\mathrm{Au}\right]$      & $<7.0\times 10^{-13}$    &  SINDRUM II:\cite{SINDRUMII:2006dvw} \\
\hline
\end{tabular}
\caption{Current experimental upper bound on $\mu\to e$ conversion in Gold.}\label{tab:mueconv_bound}
\end{table}

\begin{table}[ht]
\centering
\begin{tabular}{|r l l|}
\hline
$\mathrm{Cr}\left[\mu\to e,~\mathrm{Al}\right]$      & $2.6\times 10^{-17}$    &  COMET: \cite{COMET:2009qeh} \\
$\mathrm{Cr}\left[\mu\to e,~\mathrm{Al}\right]$      & $2.87\times 10^{-17}$   &  Mu2e: \cite{Mu2e:2014fns} \\
\hline
\end{tabular}
\caption{Future sensitivities to $\mu\to e$ conversion in Aluminium.}\label{tab:mueconv_sensitivities}
\end{table}

\section{Phenomenology}\label{sec:Pheno}

In this section we study the phenomenology of charged lepton flavour violating processes in the symmetry protected type-I seesaw, taking into account the constraints from $Z\to\nu\nu$ and tests of lepton flavour universality from pion, kaon and tau decays.\footnote{Even though also beta decays can be used as a probe of lepton flavour universality~\cite{Crivellin:2020lzu}, we do not include them here, since the Cabibbo angle anomaly points towards an enhanced $W-\mu-\nu$ coupling~\cite{Crivellin:2020ebi,Kirk:2020wdk}, which cannot be achieved in our model and increases the tensions in the EW fit~\cite{Crivellin:2020lzu}. Furthermore, such a modification would further increase the tension within the EW fit via its effect in the determination of the Fermi constant~\cite{Crivellin:2021njn}.}  For this we use the expressions for the processes obtained in Sec.~\ref{sec:Flavour}, and the structure of the neutrino Yukawa couplings given in Eq.~\eqref{eq:WeinberglessLimit}. In addition, we assume the case of three right-handed neutrinos with degenerate masses.\footnote{Note that the phenomenological analysis would be the same if we were to supplement the SM with two mass degenerate sterile neutrinos, given that Eqs.~\eqref{eq:TijExplicit}, \eqref{eq:Y4Explicit} and \eqref{eq:Y2Y2Explicit2} are obtained by a simple rescaling of Eqs.~\eqref{eq:TijExplicit}, \eqref{eq:Y4Explicit} and \eqref{eq:Y2Y2Explicit2}.}
\smallskip

Let us start by showing the dependence of the $\mu\to e$ processes $\mu\to 3e$, $\mu\to e\gamma$, $\mu\to e$ conversion in nuclei, and $Z\to e\mu$ on the right-handed neutrino mass. For this we fix the ratio $\lambda_i=\MR/(10^6~\mathrm{GeV})$, which involves the (approximately) degenerate sterile neutrino mass $\MR$, such that $\Tij{ij}$ (see Eq.~\eqref{eq:Tij}) becomes independent of $\MR$. Here and in the following, the complex number $z$ in Eq.~\eqref{eq:WeinberglessLimit} is fixed to $1$, however, as can be seen from Eqs.~\eqref{eq:TijExplicit},~\eqref{eq:Y4Explicit} and \eqref{eq:Y2Y2Explicit}, varying $|z|$ and $\mathrm{Arg}(z)$ would not add anything to our discussion. 
\smallskip

\begin{figure}
    \centering
    \includegraphics[scale=0.6]{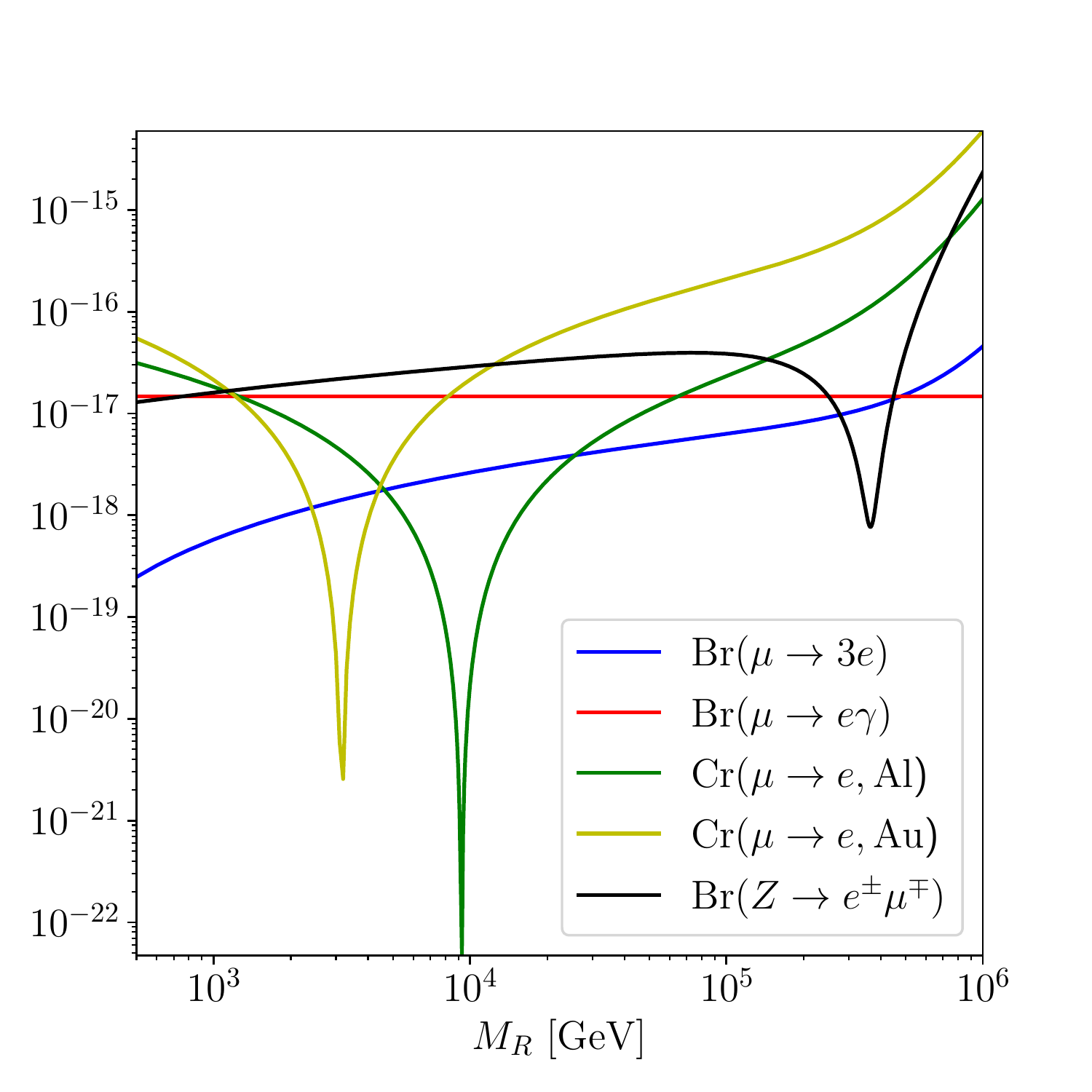}
    \caption{Branching ratios and conversion rates of processes involving $\mu\to e$ transitions for degenerate sterile neutrinos of mass $\MR$, and with the Yukawa coupling structure given in Eq.~\eqref{eq:WeinberglessLimit}, corresponding to the symmetry protected type-I seesaw. For this plot we set $\lambda_i=\MR/(10^6~\mathrm{GeV})$, $i\in \{e,\mu,\tau\}$, such that $T_{ij}$ is constant. Therefore, also $\mathrm{Br}(\mu\to e\gamma)$, which is proportional to $T_{ij}$, is independent of $\MR$, whereas the amplitudes entering $\mathrm{Br}(\mu\to 3e)$, $\mathrm{Cr}(\mu\to e)$ and $\mathrm{Br}(Z\to e\mu)$ involve terms with four Yukawa couplings in addition. Furthermore, $\mathrm{Cr}(\mu\to e, \mathrm{Au})$ and $\mathrm{Cr}(\mu\to e, \mathrm{Al})$ feature target dependent blind spots. }
    \label{fig:obs_vs_MR}
\end{figure}

As we can see from Fig.~\ref{fig:obs_vs_MR}, the $\mu\to e$ conversion rates show sharp dips for specific values of the sterile neutrino masses, whose positions depend on the target nucleus.\footnote{This behaviour was already observed in Ref.~\cite{Alonso:2012ji} and is due to a cancellation between $u$-quark and $d$-quark contributions which enter the $\mu\to e$ conversion rate with opposite sign.} For masses around these blind spots, $\mu\to e$ conversion leads to less stringent bounds on the neutrino Yukawa couplings, such that e.g.~the bounds from $\mu\to 3e$ can be competitive. Note that Br$(Z\to e^\pm \mu^\mp)$ not only displays a blind spot, but that this branching ratio is generally inaccessible to current experiments, and even at future $Z$ factories (taking into account the current limits from the other $\mu\to e$ processes). 
\smallskip

Let us now consider correlations between the different charged lepton flavour violating processes. For this we again assume three mass degenerate right-handed neutrinos with masses of either $\MR = 10^3$~GeV, $\MR = 10^4$~GeV or $\MR = 10^5$~GeV, whereas the couplings $\lambda_i$, $i\in\{e,\mu,\tau\}$ are logarithmically sampled within the range $\left[10^{-6},1\right]$. The resulting rates are compared to the current experimental bounds and future sensitivities given in Tables~\ref{tab:Zll_constraints},~\ref{tab:llpgamma_constraints},~\ref{tab:l3l_bounds}~\ref{tab:mueconv_bound} and in Tables~\ref{tab:Zll_sensitivities},~\ref{tab:llpgamma_sensitivities},~\ref{tab:l3l_sensitivities},~\ref{tab:mueconv_sensitivities}, respectively. In all plots we disregarded all points in parameter space that are disfavoured by the combined $\chi^2$ function of $Z\to \nu\nu$ and tests of lepton flavour universality of the charged current (at the 95\% CL), or that are excluded by any of the current upper bounds on charged lepton flavour violating observables that are not plotted on the axes.
\smallskip

\begin{figure}
\centering
\begin{subfigure}[b]{\textwidth}
\centering
\includegraphics[width=.95\textwidth]{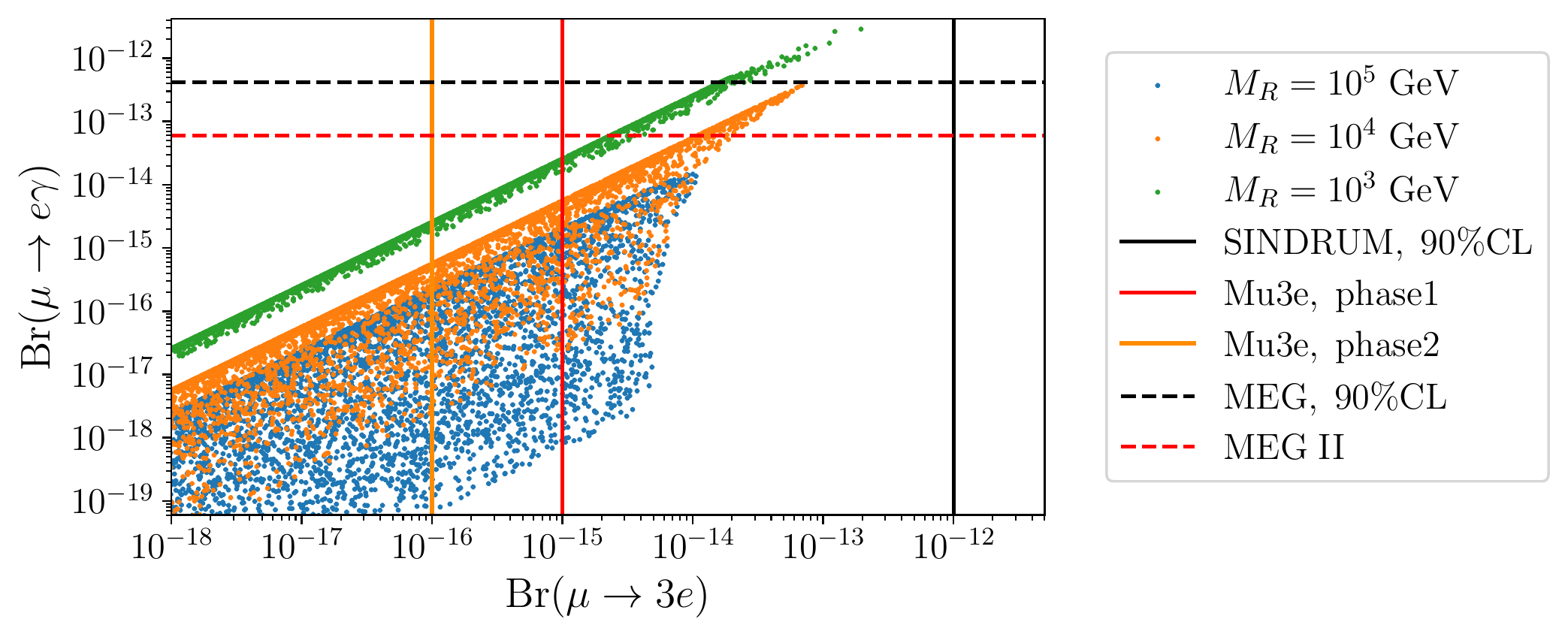}
\caption{\label{fig:mu3e_vs_muegamma}}
\end{subfigure}
\\[10mm]
\begin{subfigure}[b]{\textwidth}
\centering
\includegraphics[width=.95\textwidth]{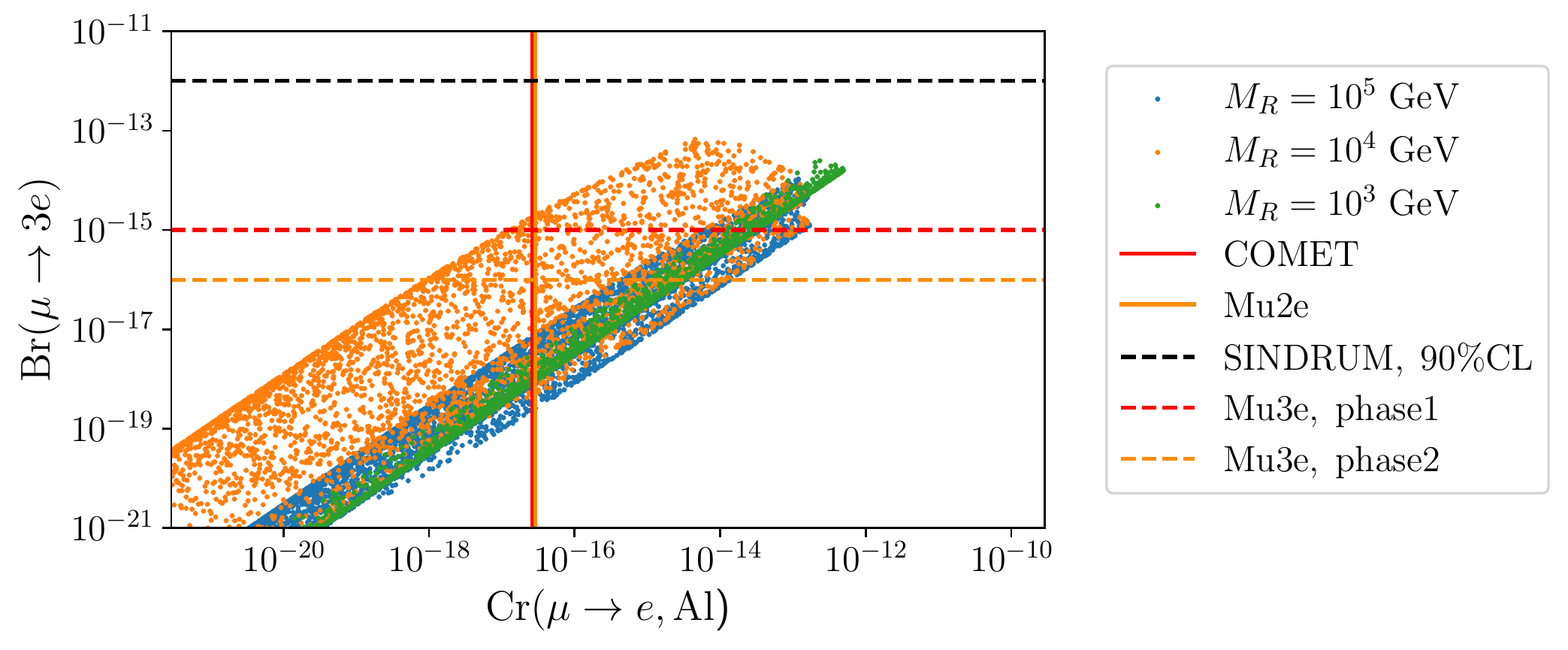}
\caption{\label{fig:mueconvAl_vs_mu3e}}
\end{subfigure}
\\[10mm]
\begin{subfigure}[b]{\textwidth}
\centering
\includegraphics[width=.95\textwidth]{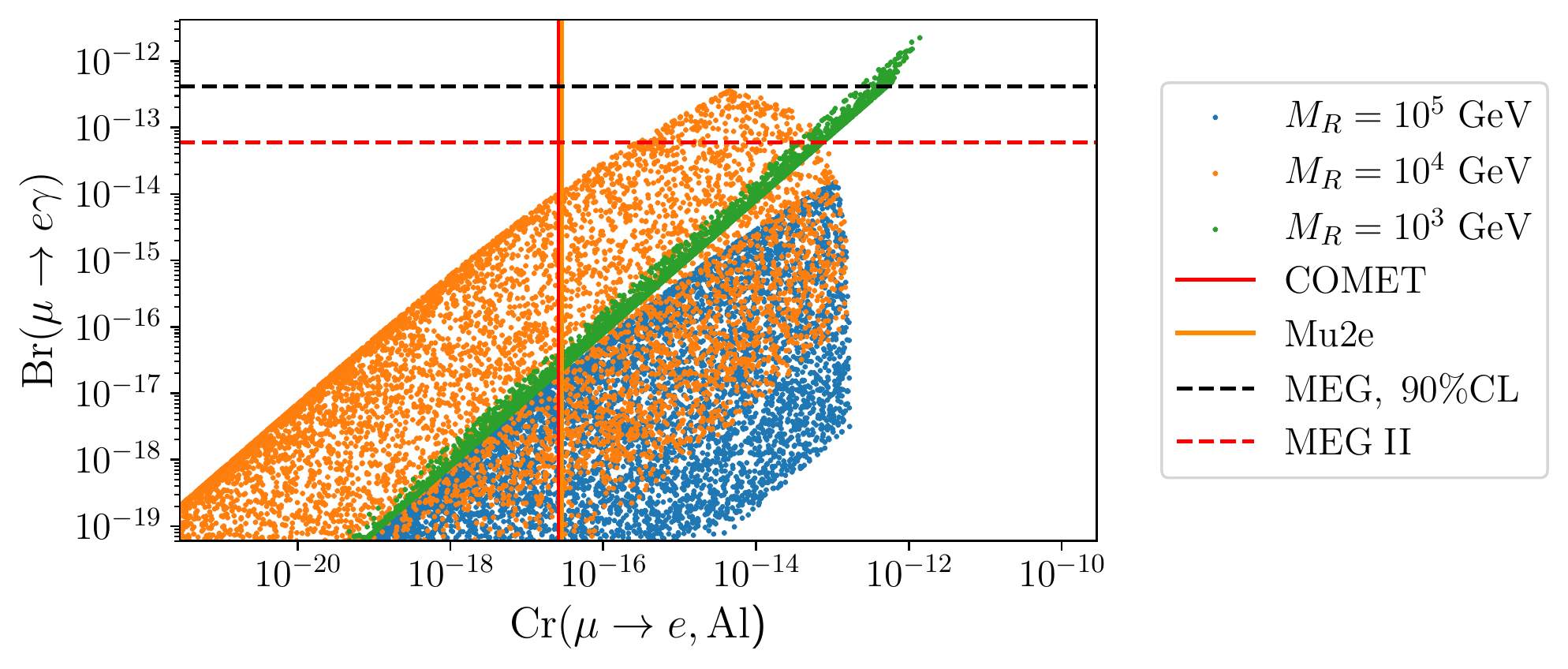}
\caption{\label{fig:mueconvAl_vs_muegamma}}
\end{subfigure}
\caption{Correlations between the processes $\mu \to 3e$, $\mu \to e \gamma$ and $\mu\to e$ conversion in aluminium nuclei. The current experimental upper limits are indicated by the black lines while the future sensitivities are shown in red and orange. 
}\label{fig:mu-e_pheno}
\end{figure}

In Figure~\ref{fig:mu-e_pheno} we show the three possible combinations of the $\mu\to e$ observables $\mathrm{Br}\left(\mu \to 3e\right)$, $\mathrm{Br}\left(\mu\to e\gamma\right)$ and $\mathrm{Cr}(\mu \to e,~ \mathrm{Al})$. Here we do not consider $\mathrm{Br}(Z\to \mu e)$, since it does not give relevant bounds, even once future prospects are taken into account (see Fig.~\ref{fig:obs_vs_MR}). We see that, apart from in the regions of parameter space around the blind spots shown in Fig.~\ref{fig:obs_vs_MR}, $\mu \to e$ conversion in nuclei is currently more constraining than $\mu\to 3e$, which is phase space suppressed. Furthermore, as Fig.~\ref{fig:obs_vs_MR} shows, the branching ratio of $\mu\to e\gamma$ is larger than that of $\mu\to 3e$ in the range of sterile neutrino masses considered here. Consequently, $\mu\to 3e$ can only lead to more stringent bounds than $\mu\to e\gamma$ if the corresponding experimental limit is more precise. This is currently not the case, however, the future Mu3e limits on the branching ratio will be lower than that of MEG II.
\smallskip

The spread of the points in Fig.~\ref{fig:mu3e_vs_muegamma} can be explained by the terms proportional to four powers of the neutrino Yukawa couplings. Indeed, if one disregarded these $Y^4$ terms, the predicted points in parameter space would converge to the upper left boundary of the current region, and we would obtain direct correlations between the two observables $\mathrm{Br}(\mu \to 3e)$ and $\mathrm{Br}(\mu \to e\gamma)$. In particular, the smaller the sterile neutrino mass, the stronger the bounds on the couplings and the smaller the $Y^4$ terms w.r.t.~the quadratic ones. Consequently, larger values of $\mathrm{Br}(\mu\to 3e)$ and $\mathrm{Br}(\mu\to e\gamma)$ can be attained by $\mathcal{O}(10^3)$~GeV and $\mathcal{O}(10^4)$~GeV sterile neutrinos than by $\mathcal{O}(10^5)$~GeV sterile neutrinos. For $\mathcal{O}(10^3)$~GeV and $\mathcal{O}(10^4)$~GeV sterile neutrinos, even the current bound on $\mathrm{Br}(\mu\to e\gamma)$ is constraining.
\smallskip

\begin{figure}
\centering
\begin{subfigure}[b]{\textwidth}
\centering
\includegraphics[width=.95\textwidth]{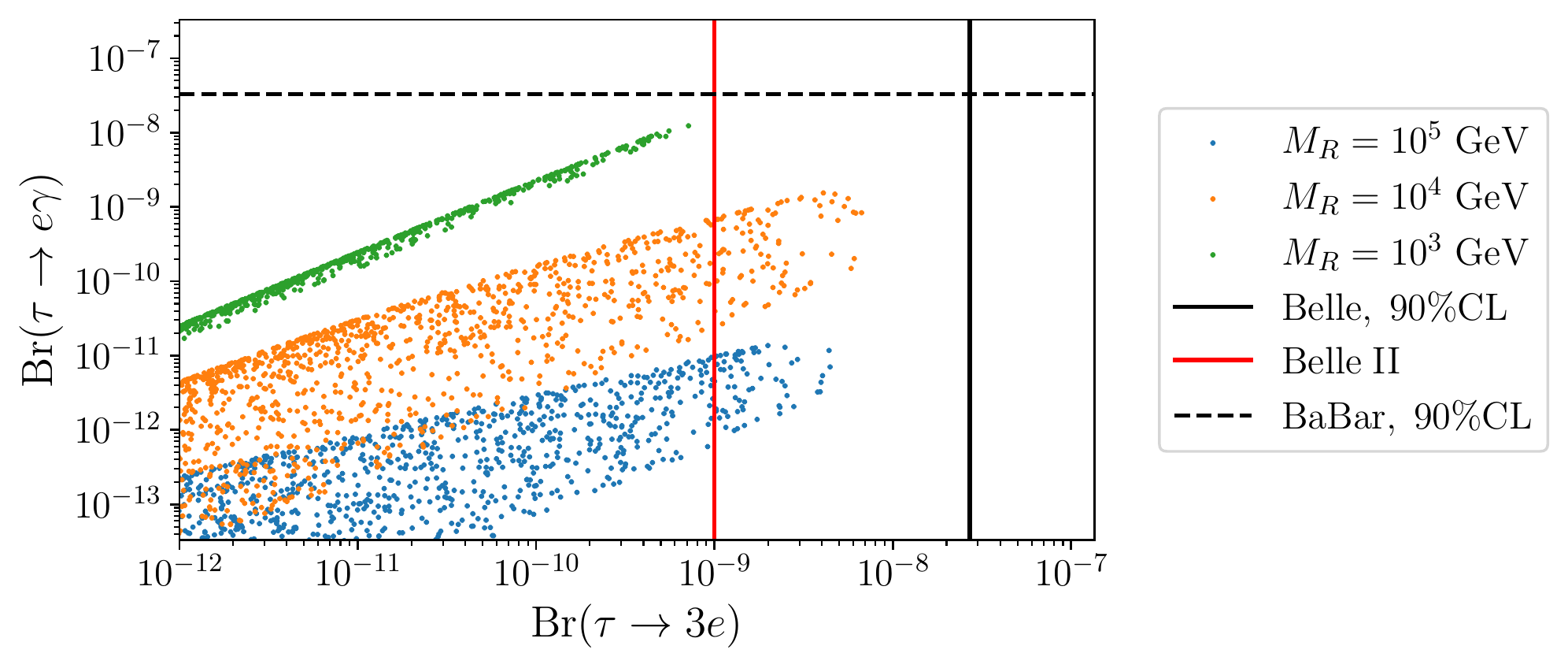}
\caption{\label{fig:tau3e_vs_tauegamma}}
\end{subfigure}
\\[10mm]
\begin{subfigure}[b]{\textwidth}
\centering
\includegraphics[width=.95\textwidth]{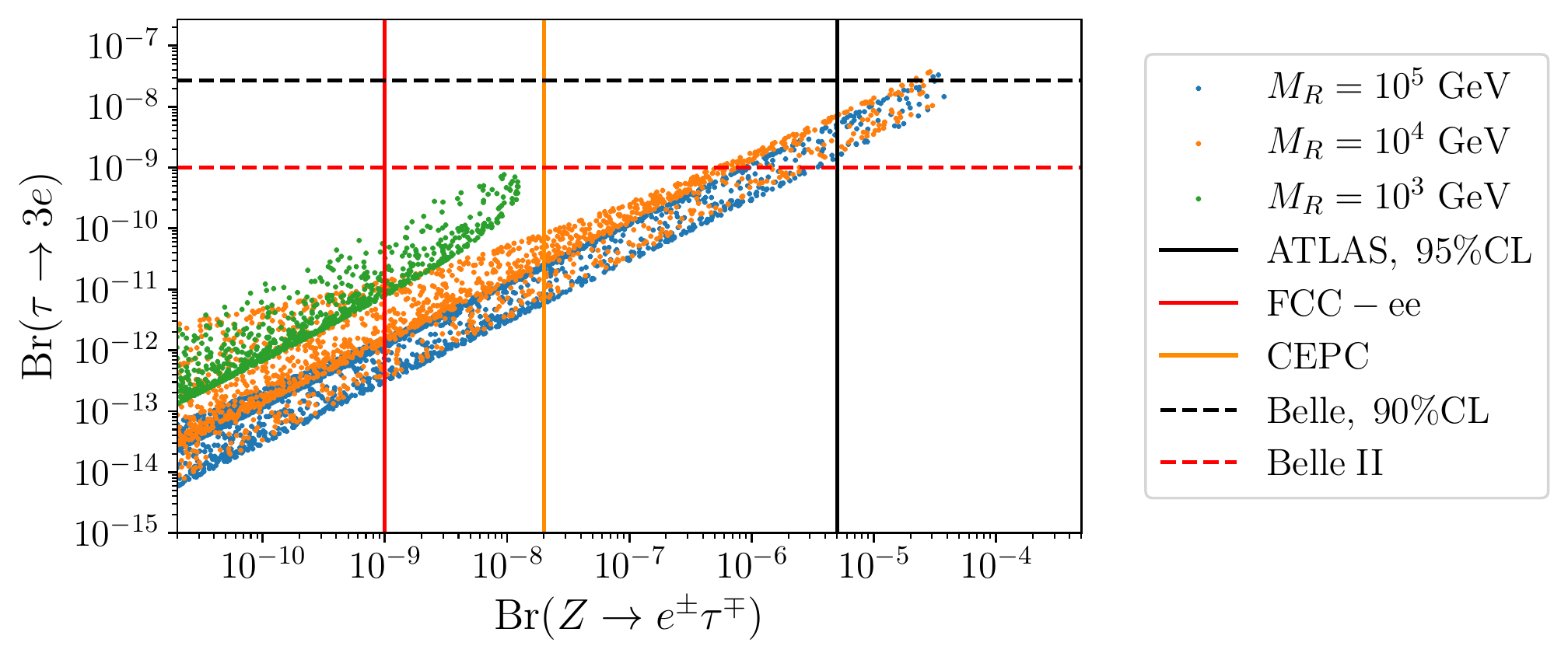}
\caption{\label{fig:Zetau_vs_tau3e}}
\end{subfigure}
\\[10mm]
\begin{subfigure}[b]{\textwidth}
\centering
\includegraphics[width=.95\textwidth]{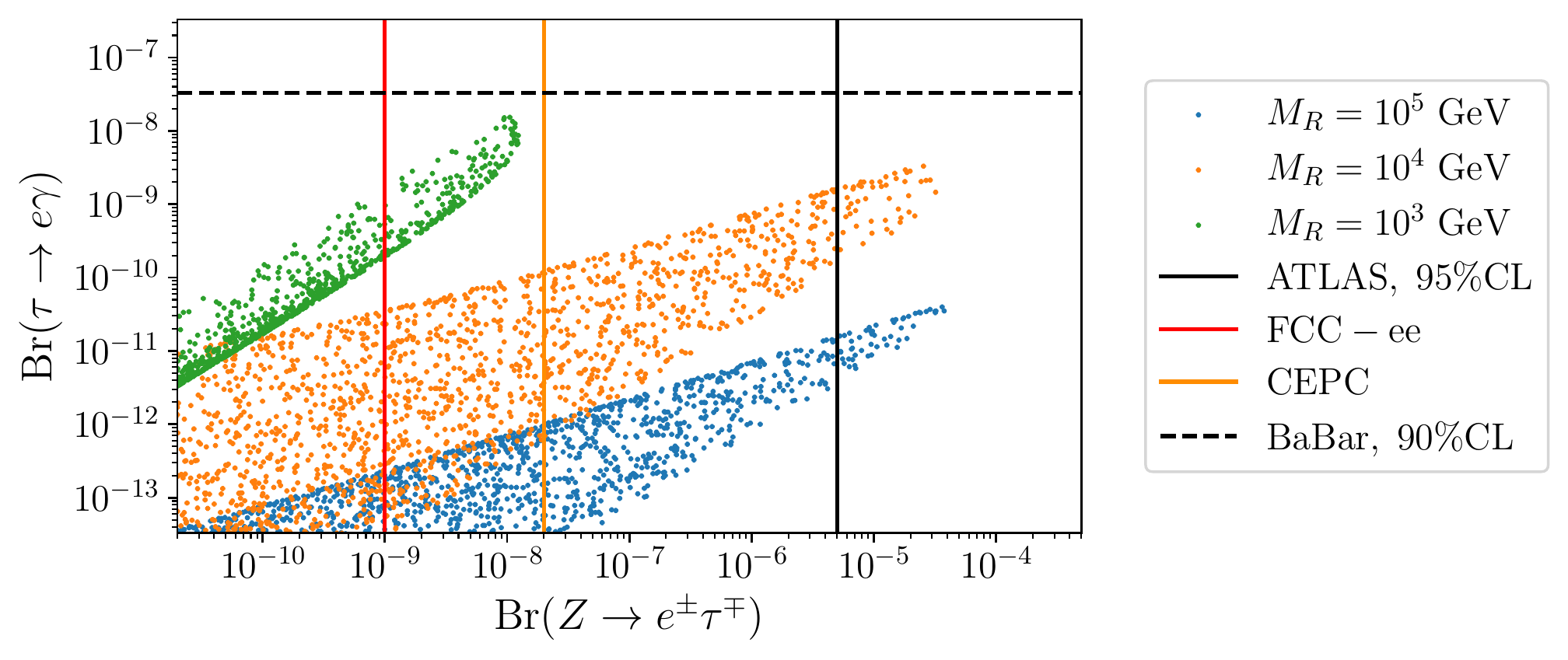}
\caption{\label{fig:Zetau_vs_tauegamma}}
\end{subfigure}
\caption{Correlations between the processes $\tau \to 3e$, $\tau \to e \gamma$ and $Z\to e\tau$. The current experimental upper limits are indicated by the black lines while the future sensitivities are shown in red and orange. 
}\label{fig:tau-e_pheno}
\end{figure}

\begin{figure}
\centering
\begin{subfigure}[b]{\textwidth}
\centering
\includegraphics[width=.95\textwidth]{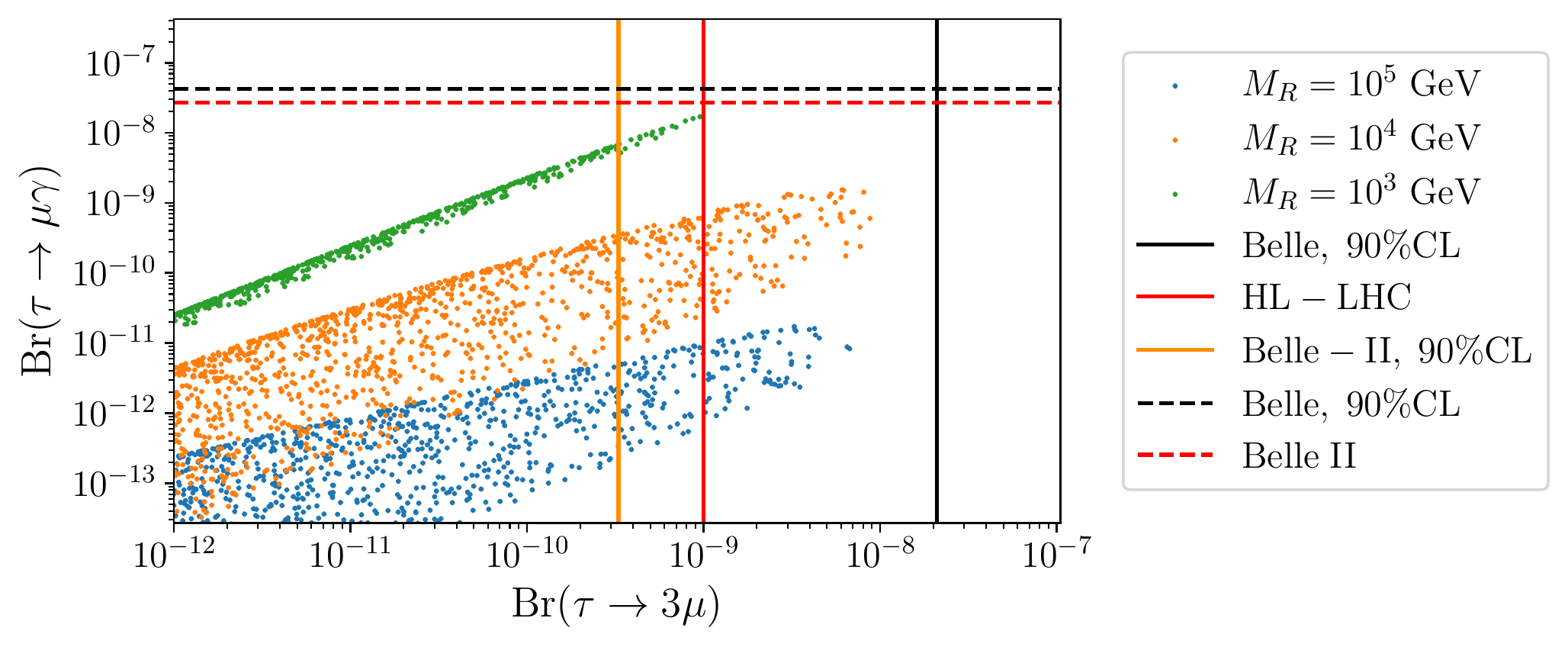}
\caption{\label{fig:tau3mu_vs_taumugamma}}
\end{subfigure}
\\[10mm]
\begin{subfigure}[b]{\textwidth}
\centering
\includegraphics[width=.95\textwidth]{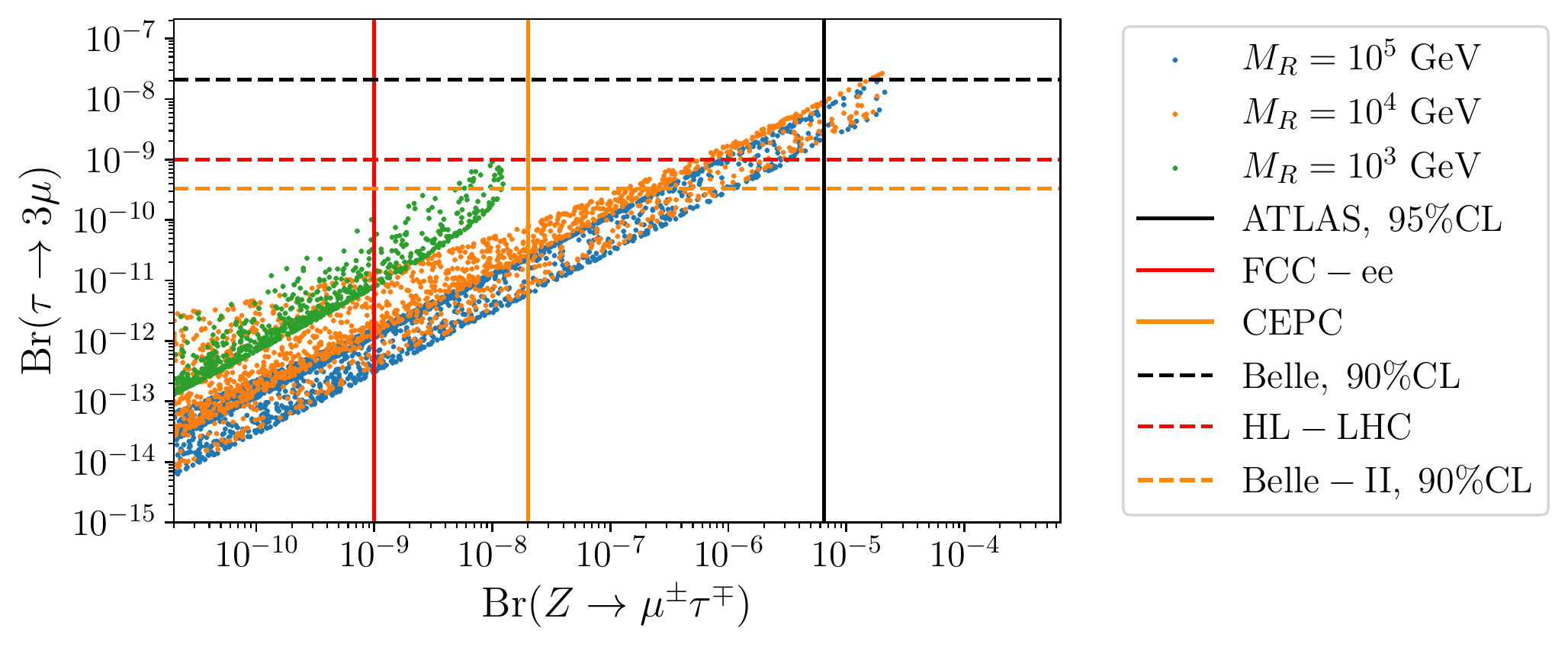}
\caption{\label{fig:Zmutau_vs_tau3mu}}
\end{subfigure}
\\[10mm]
\begin{subfigure}[b]{\textwidth}
\centering
\includegraphics[width=.95\textwidth]{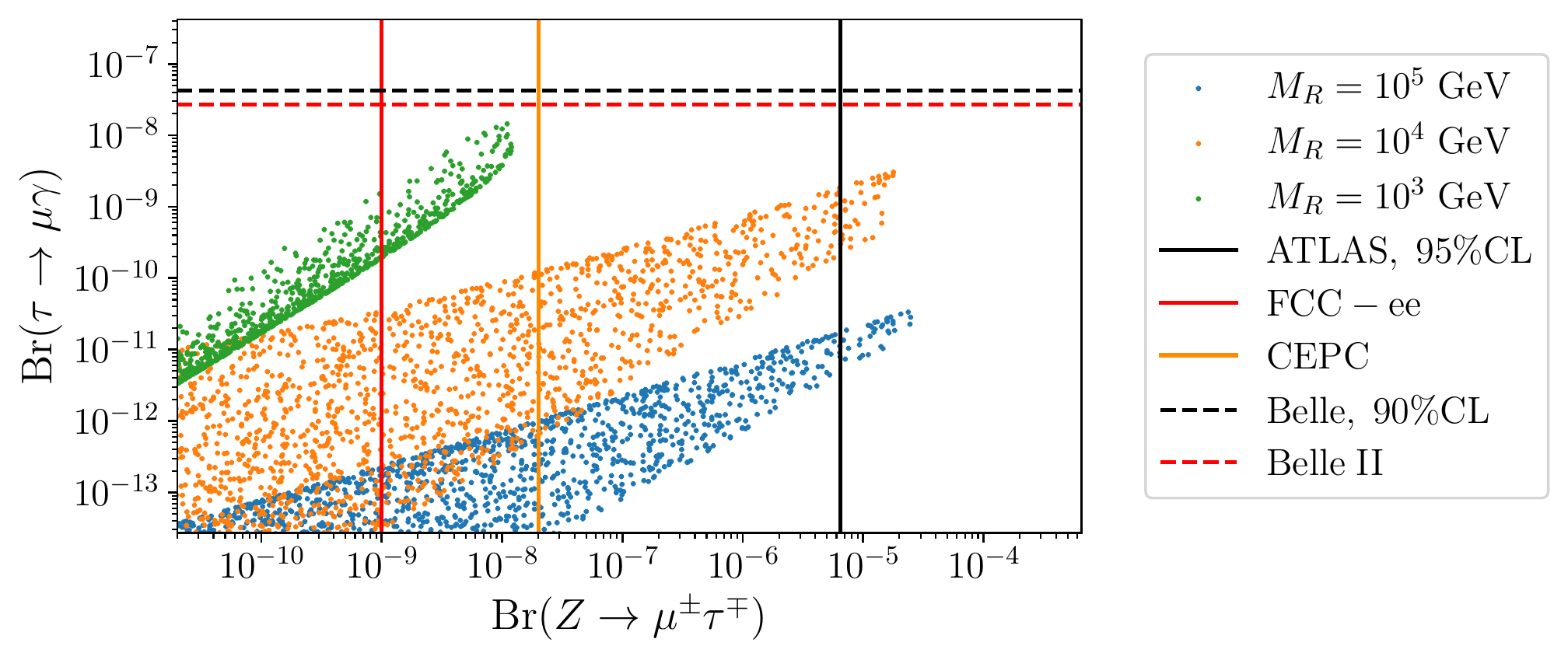}
\caption{\label{fig:Zmutau_vs_taumugamma}}
\end{subfigure}
\caption{Correlations between the lepton flavour violating processes $\tau \to 3\mu$, $\tau \to \mu \gamma$ and $Z\to \mu\tau$. The current experimental upper limits are indicated by the black lines while the future sensitivities are shown in red and in orange. 
}\label{fig:tau-mu_pheno}
\end{figure}

Fig.~\ref{fig:tau-e_pheno} (Fig.~\ref{fig:tau-mu_pheno}) depicts the $\tau\to e(\mu)$ observables $\mathrm{Br}\left(\tau \to e(\mu)\gamma\right)$, $\mathrm{Br}\left(\tau\to 3e(\mu)\right)$ and $\mathrm{Br}(Z\to e(\mu)\tau)$. Fig.~\ref{fig:tau3e_vs_tauegamma} implies that Belle~II will probe a part of the parameter space of sterile neutrinos with masses $\sim \mathcal{O}(10^4)-\mathcal{O}(10^5)$~GeV, via $\mathrm{Br}(\tau\to 3e)$, however, $\mathrm{Br}(Z\to e\tau)$, seems to be more sensitive to sterile neutrinos with masses $\sim \mathcal{O}(10^5)$~GeV (see also Figs.~\ref{fig:Zetau_vs_tau3e} and~\ref{fig:Zetau_vs_tauegamma}), since it can exclude a part of the parameter space lying below the current experimental bound on $\mathrm{Br}(\tau \to 3e)$. Furthermore, FCC-ee promises a substantial improvement in sensitivity to the $Z\to e\tau$ channel. The spread of the points can be understood in the same way as for $\mu\to e$ transitions.
\smallskip

Similar results are obtained for $\tau\to \mu$ processes, as can be seen in Figure~\ref{fig:tau-mu_pheno}. Among them, $Z\to \mu \tau$ is the observable most sensitive to sterile neutrinos with masses in the range of $\sim \mathcal{O}(10^5)$~GeV, however, future searches for $\tau\to 3\mu$ will also be able to probe the parameter space for $\sim \mathcal{O}(10^4)-\mathcal{O}(10^5)$~GeV sterile neutrinos.
Note that the largest possible values for the branching ratios $\mathrm{Br}(\tau\to e\mu e)$ and $\mathrm{Br}(\tau\to \mu e \mu)$, which feature two flavour changes, lie approximately eight orders of magnitude below their current bounds. For this reason, we do not study these two observables in more detail. We expect a similar behaviour for muonium-antimuonium oscillations, since also these feature two flavour transitions.
\smallskip

\section{Conclusions}\label{sec:Conclusions}

The type-I seesaw is a natural mechanism to generate the observed smallness of the active neutrino masses. In general, this requires the corresponding neutrino Yukawa couplings to be tiny for TeV scale right-handed neutrinos. However, the Wilson coefficient of the Weinberg operator can be protected from a non-zero contribution by a symmetry, as in the inverse seesaw model. We refer to this setup as the symmetry protected type-I seesaw, and to the corresponding limit as the inverse seesaw limit. In the inverse seesaw limit, the neutrino Yukawa couplings can be sizeable (for TeV scale sterile neutrions), such that observable effects in tests of lepton flavour (universality) violation are possible.  

Within this setup, we performed a complete and comprehensive analysis of charged lepton flavour violation. In particular, we calculated the matching of the type-I seesaw on the SMEFT at the dim-6 level, as well as the 1-loop contributions to the processes
\begin{itemize}
    \item $Z\to\ell\ell^\prime$
        \item $\ell\to\ell^\prime\gamma$
        \item $\ell\to3\ell^\prime$
        \item $\mu\to e$ conversion in nuclei
\end{itemize}
in the seesaw limit, i.e.~at leading order in $v^2/M_R^2$. 
In the appendix we also provide the corresponding expressions in a general $R_\xi$ gauge using exact diagonalisation of the neutrino mass matrix. 

In our phenomenological analysis, we correlated $\mu\to e\gamma$ to $\mu\to 3e$ and $\mu \to e$ conversion, as well as $\tau \to e (\mu)\gamma$ to $\tau\to 3e (\mu)$ and $Z\to \tau e (\mu)$. 
Taking into account the bounds from lepton flavour universality violation in tau, kaon and pion decays, and the limit on Br$(Z\to\nu\nu)$, we found that, while for sterile neutrino masses of the order of 1~TeV, the correlation between any two $\ell\to\ell^\prime$ processes is direct, i.e. showing a linear correlation, the allowed parameter space significantly broadens for heavier right-handed neutrinos. The reason for this behaviour is that for heavier right-handed neutrino masses, the neutrino Yukawa couplings $Y^\nu$ can be larger, while still respecting experimental bounds, such that the $(Y^\nu)^4$ effects in $\ell\to\ell^\prime$ and $Z\to\ell\ell^\prime$ can be (relatively) more important. In particular, we observed that
\begin{itemize}
    \item Mu3e and future $\mu\to e$ conversion experiments have the capability of covering a large portion of the so-far unexplored (i.e.~unconstrained) parameter space.
    \item The lepton flavour violating $Z$ decays, $Z\to e\tau$ and $Z\to \mu \tau$ can have sizeable branching ratios and could be observed at future $e^+e^-$ colliders such as FCC-ee or CEPC.
\end{itemize}

While we performed the phenomenological analysis without resummation of potentially large logarithms between the right-handed neutrino scale and the EW scale, the formulae for the matching on the SMEFT can in the future be used for an automated computation. However, for this both a (at least partial) two-loop renormalisation group evolution, as well as the inclusion of the (finite) loop contributions to the relevant observables within LEFT, i.e.~the contributions of the operators at the low scale to the matrix elements of the processes, would be necessary.

\acknowledgments
{We would like to thank Peter Stoffer and Zhang Di for useful discussions. This work is supported by the Swiss National Science Foundation, under Project No.~PP00P21\_76884.}

\medskip

\appendix

\section{Neutrino Mixing Matrix}\label{sec:FullNeutrinoMatrix}
For the derivation of the results for the amplitudes with exact diagonalisation of the neutrino mass matrix (see Appendix~\ref{sec:FullResults}), it is useful (as was already noticed in Ref.~\cite{Schechter:1980gr}) to define the $3\times (3+n)$ mixing matrix
\begin{align}
\U\equiv & \OlLdag  \begin{pmatrix}
\id{3} & \niente{3}
\end{pmatrix}
\Ov  \,,\notag\\
U_{is}\equiv &\sum_{j=1}^3 V_{ij}^{\ell L *} V_{js}\,,
\label{eq:U}
\end{align}
where $i$ and $j$ are lepton flavour indices that run from 1 to 3, whereas $s$ is a neutrino index which runs from 1 to $3+n$. $\Ov$ is the $(3+n)\times (3+n)$ neutrino mixing matrix introduced in \Eqs{eq:diagonalisation}-\eqref{eq:Mdiag} and $\OlL$ is one of the two $3\times 3$ matrices $\OlL$ and $\OlR$ that diagonalise the charged lepton Yukawa as in
\begin{align}
Y^{\ell,{\rm diag}}=\OlLdag \Yl \OlR\,.
\end{align} 
Note that $\Uscript$ is a semi-unitary matrix, since $\U \Udag=\id{3}$, but $\Udag \U\neq\id{3+n}$. At leading order in $v/M_R$, $\Uscript$ is given by
\begin{align}
\U\approx&\begin{pmatrix}
\id{3}-\frac{1}{2}\MD \MR^{-2} \MDdag & \;\;\MD \MR^{-1}
\end{pmatrix}\,,\label{eq:Uexp}
\end{align}
which corresponds to the upper $3\times (3+n)$ block of the seesaw-expanded neutrino mixing matrix $V$, given in \Eq{eq:Vexp}.

The Feynman rules for the type-I seesaw can be expressed in terms of the neutrino mixing matrix $\Uscript$. These are listed in Table~\ref{tab:FRsExact}. Expanding them in powers of $v/\MR$, we recover the Feynman rules given in Table~\ref{tab:FRsExpanded}.

\begin{table}[ht]
	\begin{center}
		\setlength{\tabcolsep}{12pt}
		\begin{tabular}{c c}
		Interaction & Exact Feynman rule\\
		\hline \hline\\[-4mm]
		$\overline{\lLR}_i\,W^-_\mu \nLi{s}$ & $-\dfrac{e}{\sqrt{2}\sw}\gamma^\mu \U_{is} P_L $\\[3mm]
			\hline \\[-4mm]
			$\nLi{s} Z_\mu \nLi{t}$ & $-\dfrac{e}{2 \sw \cw}\,\left(\Udag\gamma^\mu \U\right)_{st} P_L$ \\[3mm]
			\hline\\[-4mm]
			$\overline{\lLR}_i\varphi^- \,\nLi{s}$ & $-\dfrac{\sqrt{2}}{v}\left(m_\ell^\mathrm{diag} \U\right)_{is} P_L$ \\[3mm]
			\hline\\[-4mm]
			$\overline{\lLR}_i\varphi^- \,\nRi{s}$ & $\dfrac{\sqrt{2}}{v}\left(\U \Mdiag\right)_{is} P_R$ \\[3mm]
			\hline
		\end{tabular}
	\end{center}	
	\caption{Feynman rules for Majorana neutrinos in the mass eigenbasis, without applying the seesaw approximation. The lepton flavour index $i$ runs from 1 to 3, the neutrino indices $s$ and $t$ run from 1 to $3+n$. $m_\ell^\mathrm{diag}$ is the diagonal mass matrix of the charged leptons.\label{tab:FRsExact}}
\end{table}	

In the derivation of the results given in Appendix~\ref{sec:FullResults} we use
\begin{align}
\U\Mdiag \Utransp&=
\OlLdag
\begin{pmatrix}
\id{3} & \niente{3}
\end{pmatrix}
\Ov
\Mdiag
\left(\OlLdag
\begin{pmatrix}
\id{3} & \niente{3}
\end{pmatrix}
\Ov \right)^T\nonumber\\
&=\begin{pmatrix}
\id{3} & \niente{3}
\end{pmatrix}
\begin{pmatrix}
\niente{3} & M_D\\
M_D^T & M_R
\end{pmatrix}
\begin{pmatrix}
\id{3}\\
\niente{3}
\end{pmatrix}=\niente{3}\,,\label{eq:UMUT}
\end{align}
relating the mixing matrix $\U$, defined in \Eq{eq:U}, with the active and sterile neutrino masses. This identity is simply an extension of \Eq{eq:Mdiag}.

\section{Contributions to charged lepton flavour violating processes within the EFT}
\label{sec:soft}

By definition, the Wilson coefficients, obtained from a matching at a high scale, contain only the hard part of the corresponding amplitudes of the full theory (i.e.~the SM with right-handed neutrinos in our case). However, when calculating physical processes at fixed order, as is done in Sec.~\ref{sec:Flavour}, the full amplitudes, i.e.~the sum of the hard and the soft part of the amplitudes, enters. Therefore, the SMEFT matching of Sec.~\ref{sec:OneLoopMatchingSMEFT} would be insufficient to  calculate physical processes because the soft part of the amplitudes, corresponding to the loop-contributions to the respective processes within the EFT, would be missing. In this section we obtain these soft parts of the amplitudes by calculating the loop diagrams with the insertions of the modified tree-level couplings of neutrinos with $Z$ and $W$ bosons resulting from \Eq{eq:C13tree}.

\subsection{$\ell\to \ell' \gamma$}

\begin{figure}[ht]
\centering
	\captionsetup[sub]{justification=centering}{%
		\includegraphics[scale=.77]{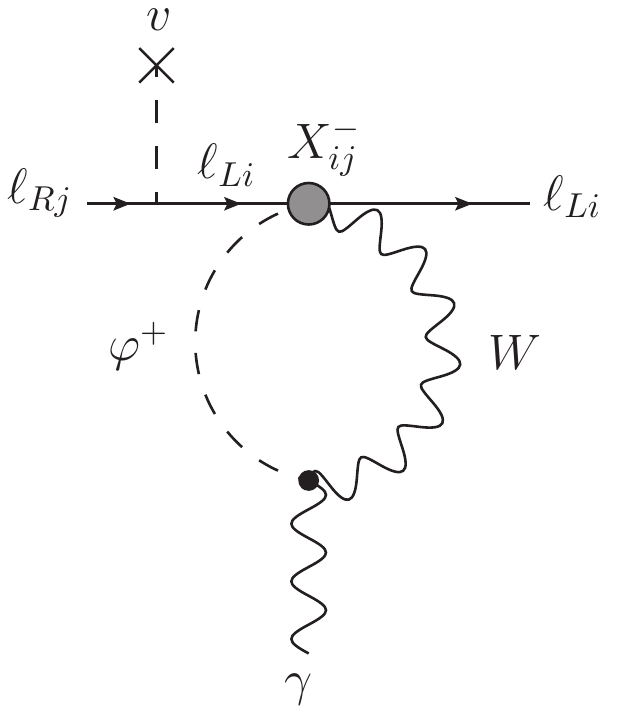}%
	}
	\hspace{5mm}
	\captionsetup[sub]{justification=centering}
{%
		\includegraphics[scale=.6]{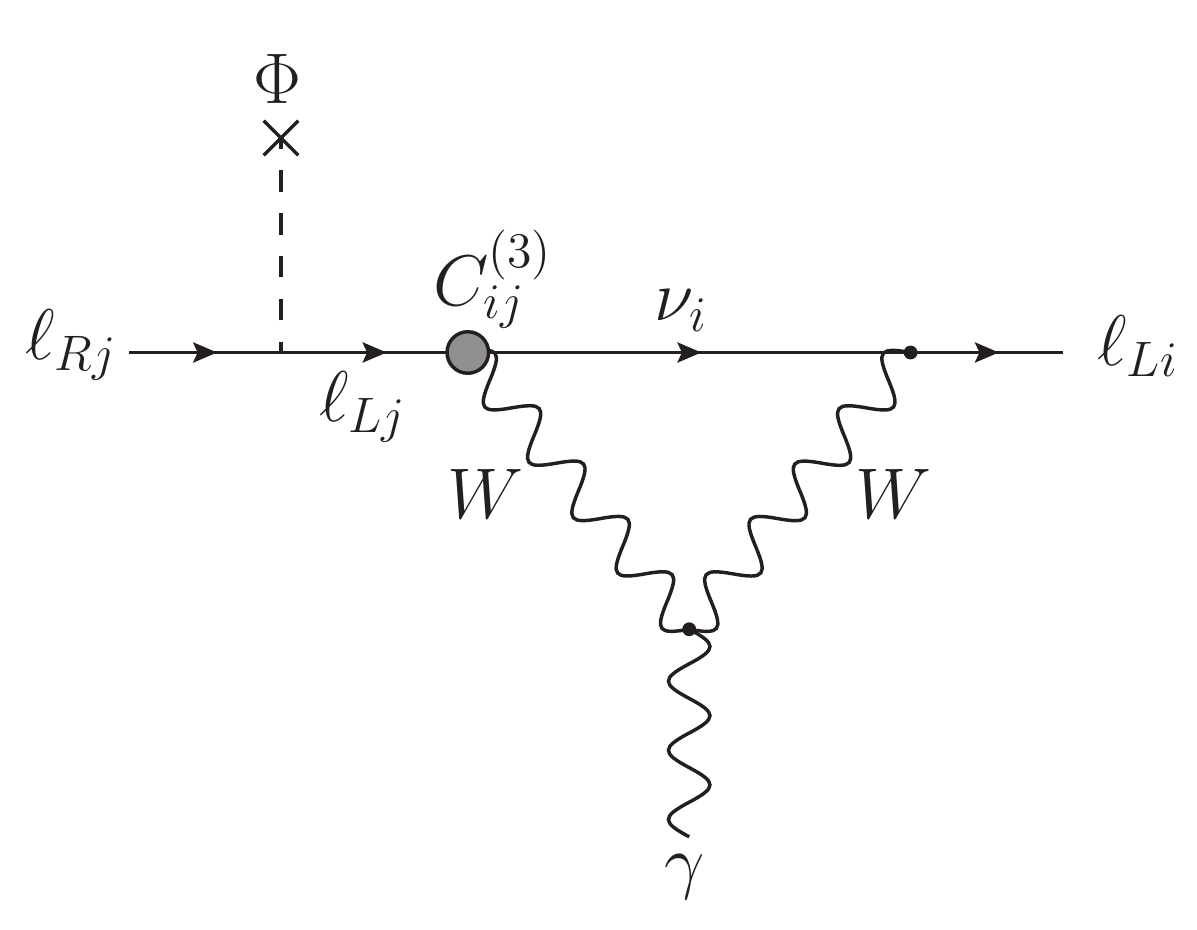}%
		}
	\caption{Diagrams with the insertions of the SMEFT operator modified at tree-level, contributing to  $\ell\to\ell'\gamma$.}\label{fig:MagnPhotonPenguinsSoft}	
\end{figure}

Defining
\begin{equation}
\mathcal{L}_{\rm eff}=
a_{ij}^\mathrm{M}\bar{\ell}_{L i}\sigma_{\mu \nu} \ell_{Rj} F^{\mu \nu}+\hc\,,
\end{equation}
where $F^{\mu \nu}=\partial^\mu A^\nu-\partial^\nu A^\mu$ is the electromagnetic field strength tensor, 
we find the contributions to $\ell\to\ell'\gamma$, shown in Figure~\ref{fig:MagnPhotonPenguinsSoft}, to be given by
\begin{align}
a_{ij}^\mathrm{M}=&
-\frac{5e^3 m_{\ell_j}}{384\pi^2\sw^2 M_W^2}\Sij{ij}\,.
\label{eq:aM}
\end{align}
Together with the contribution to the Wilson coefficients from the one-loop matching of the full theory onto SMEFT, reported in Eqs.~\eqref{eq:CeB} and \eqref{eq:CeW}, this combines to the full $\ell\to\ell'\gamma$ amplitude given in Eq.~\eqref{eq:AM}.

\subsection{Four Lepton Amplitudes}
The amplitudes of $\ell_j\ell_l\to \ell_i\ell_k$ processes, can be decomposed as
\begin{align}
\mathcal{M}= &
\left(a_{ij,kl}^{\mathrm{V}LL} +\tilde{z}_{ij,kl}^{\mathrm{V}LL} + d_{ij,kl}^{\mathrm{V}LL}\right)\left(\overline{\ell}_i \gamma_\mu P_L \ell_j\right)\left(\overline{\ell}_k\gamma_\mu P_L\ell_l\right)\notag\\
&+\left(a_{ij,kl}^{\mathrm{V}LR} +\tilde{z}_{ij,kl}^{\mathrm{V}LR} \right)\left(\overline{\ell}_i \gamma_\mu P_L \ell_j\right)\left(\overline{\ell}_k\gamma_\mu P_R\ell_l\right)\,,
\end{align}
where $a_{ij,kl}^{\mathrm{V}AB}$ denotes the off-shell photon penguin contributions, $\tilde{z}_{ij,kl}^{\mathrm{V}AB}$ the $Z$ penguin, and $d_{ij,kl}^{\mathrm{V}AB}$ the box contributions.

\begin{figure}[ht]
\centering
{%
		\includegraphics[scale=.77]{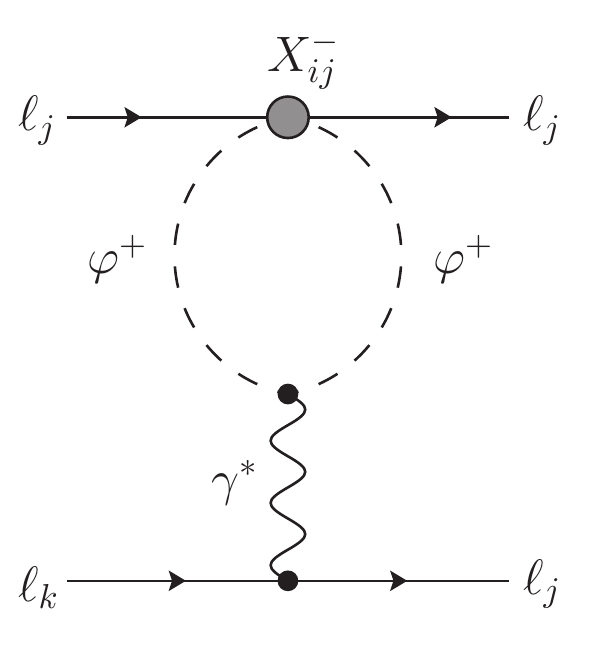}\label{fig:2GPhotonPenguin}%
	}
	\hspace{5mm}
	\captionsetup[sub]{justification=centering}
{%
		\includegraphics[scale=.77]{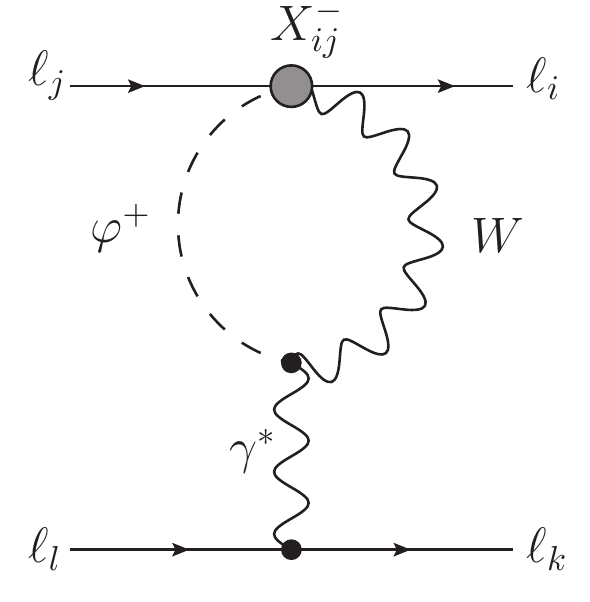}\label{fig:GPWPhotonPenguin}%
		}
	\hspace{5mm}{%
		\includegraphics[scale=.55]{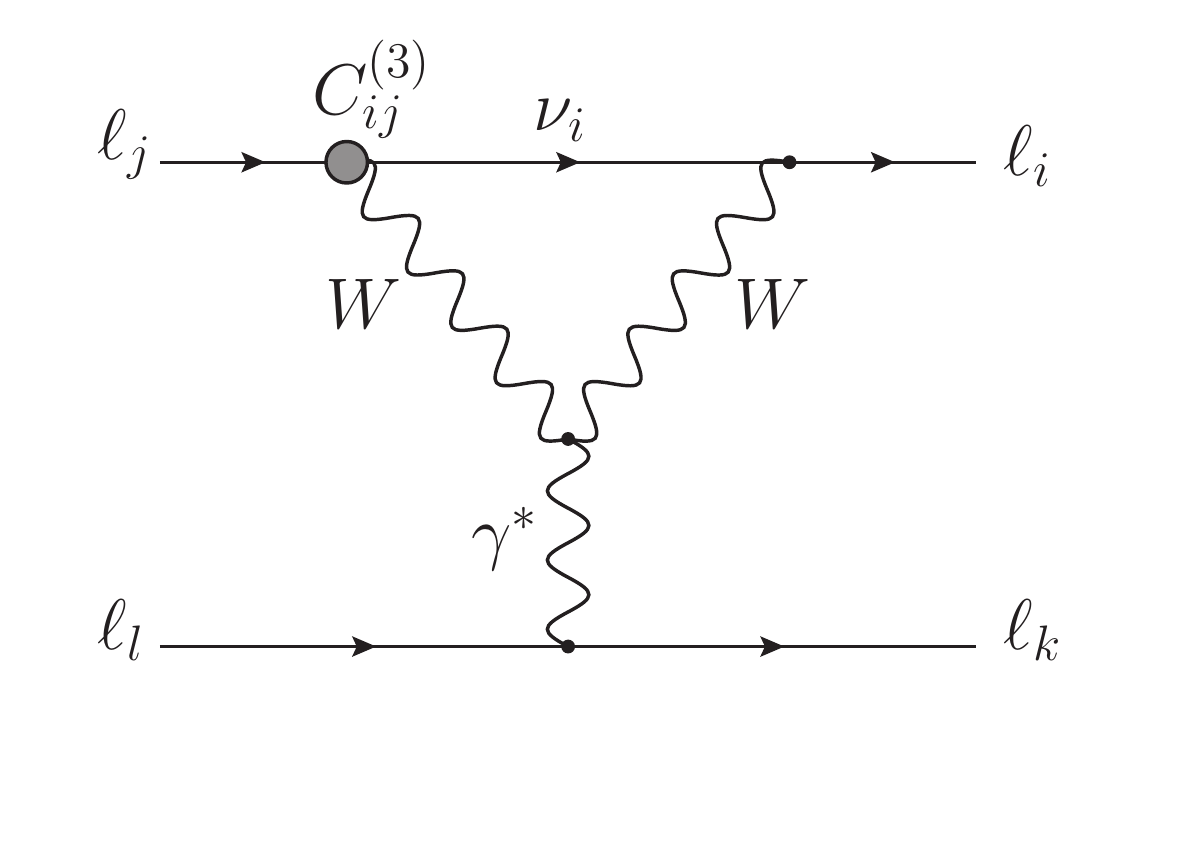}\label{fig:2WPhotonPenguin}%
		}
	\caption{Off-shell photon penguins contributing to four-lepton processes}\label{fig:PhotonPenguinSoft}	
\end{figure}

There are three classes of diagrams with an off-shell photon exchange, as shown in Figure~\ref{fig:PhotonPenguinSoft}):
\begin{itemize}
\item Diagrams with two Goldstone bosons in the loop
\item Diagrams with a Goldstone and W boson in the loop 
\item Diagrams with two $W$ bosons and a light neutrino in the loop
\end{itemize}
Due to the vectorial nature of the photon coupling, $a_{ij,kl}^{\mathrm{V}LL}=a_{ij,kl}^{\mathrm{V}LR}$ and we find 
\begin{align}
a_{ij,kl}^{\mathrm{V}LL}=a_{ij,kl}^{\mathrm{V}LR}=-e Q_ea_{ij}^\mathrm{V}\delta_{kl}=e\,, a_{ij}^\mathrm{V}\delta_{kl}
\label{eq:aV}
\end{align}
with $a_{ij}^\mathrm{V}$ given in $R_\xi$-gauge by
\begin{align}
a_{ij}^\mathrm{V}= -\frac{e^3}{576\pi^2\sw^2\MW^2}
\sum _{a=1}^n M_{D,ia}\MRi{a}^{-2}M^*_{D,ja}
\frac{1}{2}\left(95-9\xi+54\frac{\xi\log\xi}{1-\xi}+6\log\left(\frac{\mu^2}{\MW^2}\right)\right)\,.
\end{align}
Together with the result of the one-loop matching, the Z-penguin and box contributions, this leads to the results in Eqs.~\eqref{eq:FVLL} and \eqref{eq:FVLR}.
\smallskip

\begin{figure}[ht]
	\center	
	\subfloat[]{%
		\includegraphics[width=0.4\textwidth]{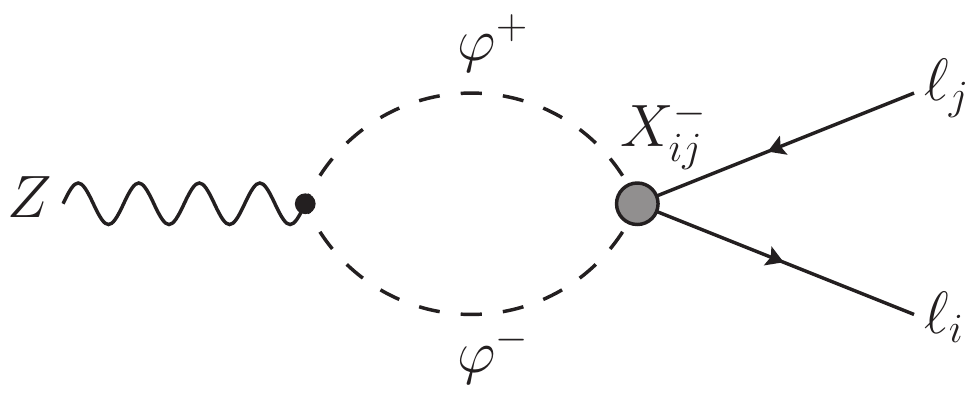}\label{fig:ZllGGDiagSoft}%
	}
	\hspace{5mm}
	\subfloat[]{%
		\includegraphics[width=0.4\textwidth]{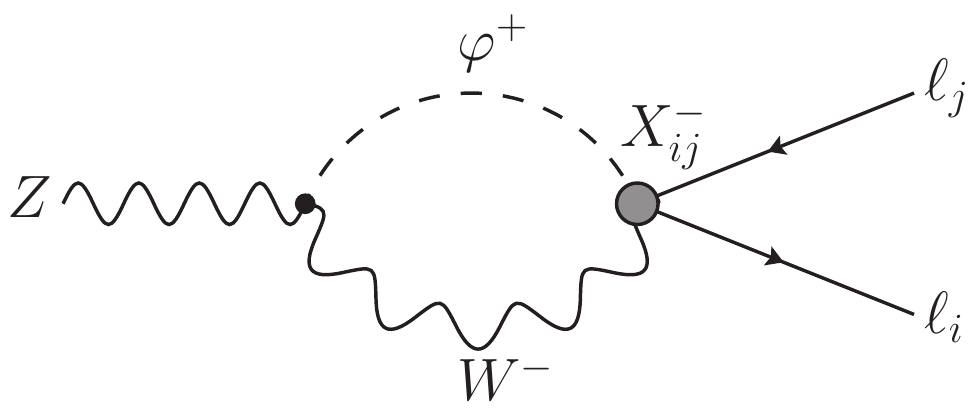}%
	}
	\hspace{5mm}
	\subfloat[]{%
		\includegraphics[width=0.4\textwidth]{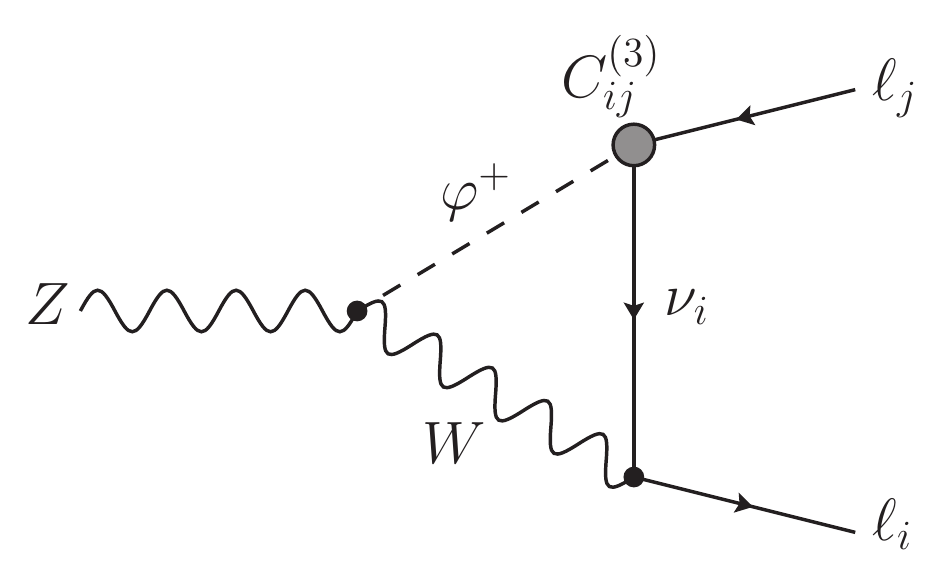}%
	}\\
	\subfloat[]{%
		\includegraphics[width=0.4\textwidth]{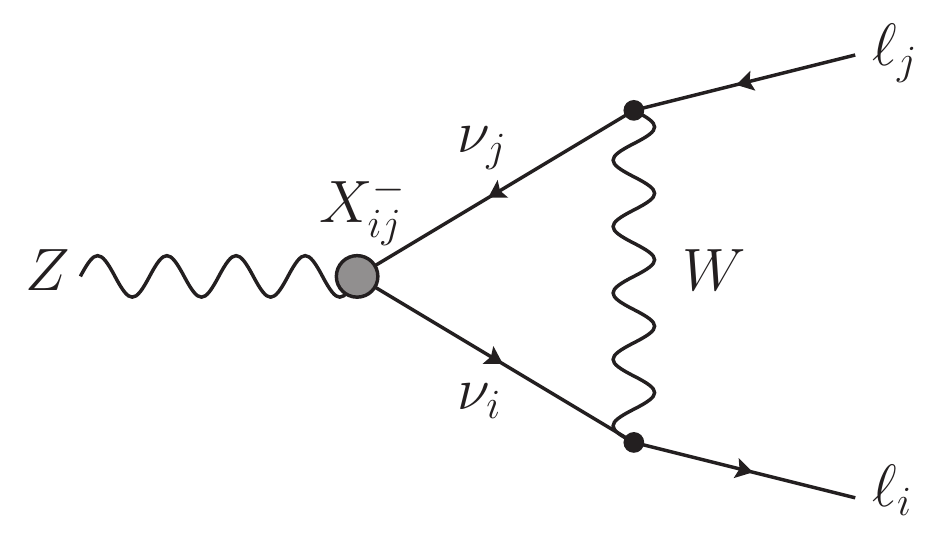}%
	}\hspace{5mm}
	\subfloat[]{%
		\includegraphics[width=0.4\textwidth]{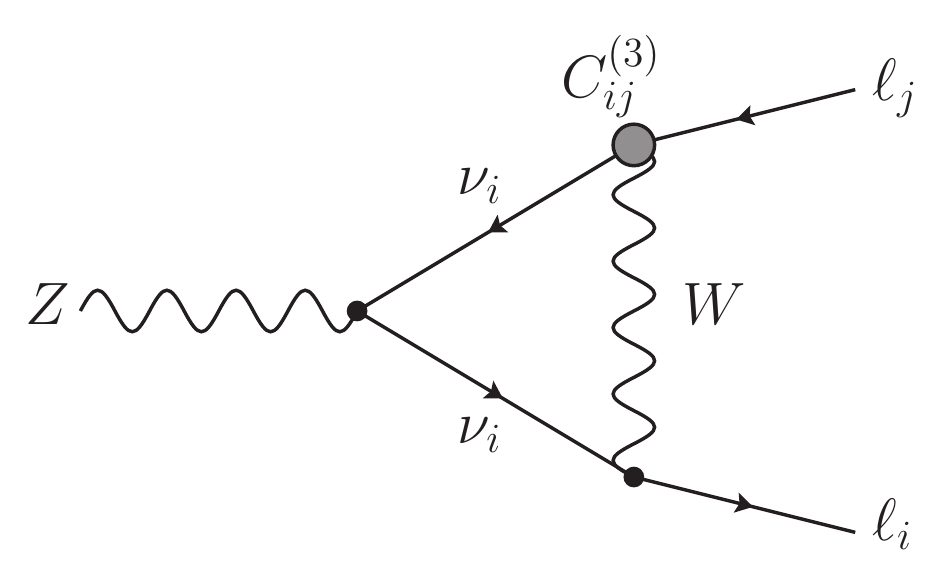}%
	}
	\hspace{5mm}
	\subfloat[]{%
		\includegraphics[width=0.4\textwidth]{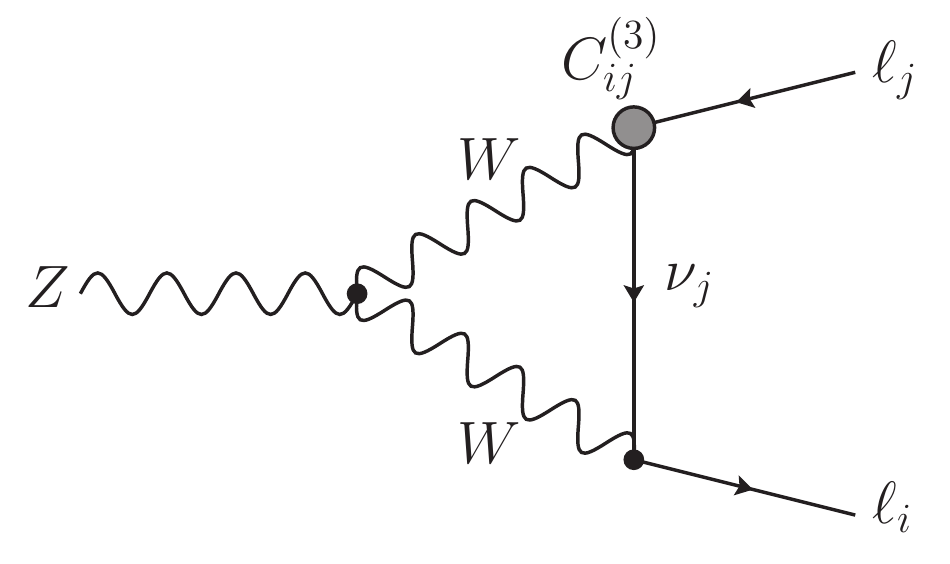}%
	}
	\hspace{5mm}
	\\
	\subfloat[]{%
		\includegraphics[width=0.4\textwidth]{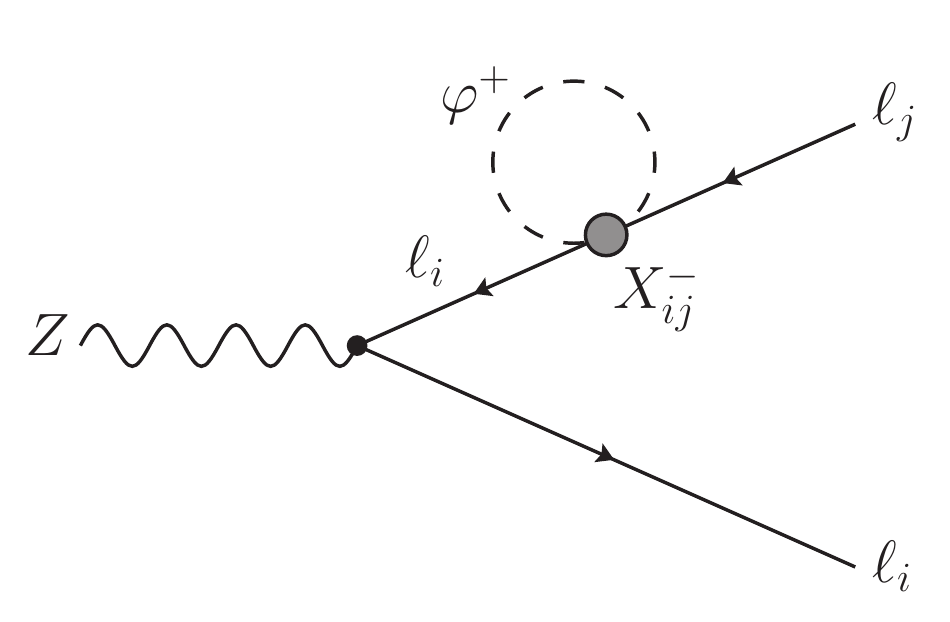}%
		}
	\hspace{5mm}
	\subfloat[]{%
		\includegraphics[width=0.4\textwidth]{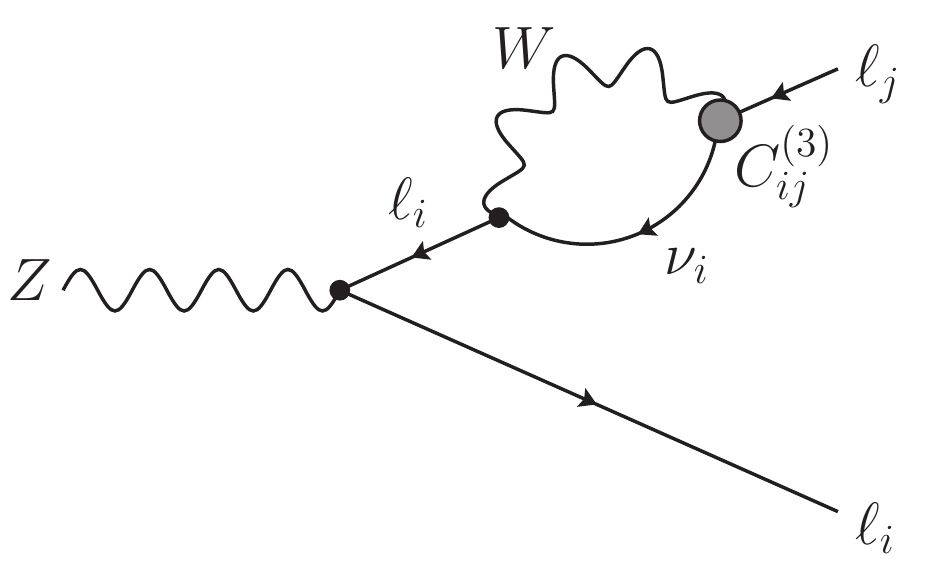}%
		}
   \caption{Insertions of $\Xpij{ij}$, $\Xmij{ij}$ and $C_{\varphi\ell,ij}^{(3)}$in loop diagrams contributing to the process $Z\to \ell\ell'$.}\label{fig:ZllDiagsSoft}	
\end{figure}

Below the electroweak scale, the diagrams with an off-shell $Z$ boson exchange are the ones shown in Figure~\ref{fig:ZllDiagsSoft}, with a fermion line attached to the $Z$ boson. They give
\begin{align}
\begin{split}
\tilde z^{\mathrm{V}LL}_{ij,kl}=& \,g^{\ell L}_{\rm SM}\delta_{kl}\tilde z_{ij}^L \\
\tilde z^{\mathrm{V}LR}_{ij,kl}=& \,g^{\ell R}_{\rm SM} \delta_{kl}\tilde z_{ij}^L \,,
\end{split}
\label{eq:ztV}
\end{align}
with
\begin{align*}
\tilde z_{ij}^L=-\frac{e^3\cw }{128 \pi^2 \sw^3 \MW^2}
\Bigg\lbrace 
&\sum _{a=1}^n M_{D,ia}\MRi{a}^{-2}M^*_{D,ja}
\left(8-\xi+\frac{6\xi\log\xi}{1-\xi}+6\log\left(\frac{\mu^2}{\MW^2}\right) \right)\Bigg\rbrace
\end{align*}
in $R_\xi$ gauge, and 
\begin{align*}
\tilde z_{ij}^L=-\frac{e^3\cw }{128 \pi^2 \sw^3 \MW^2}
\Bigg\lbrace 
&\sum _{a=1}^n M_{D,ia}\MRi{a}^{-2}M^*_{D,ja}
\left(1+6\log\left(\frac{\mu^2}{\MW^2}\right) \right)\Bigg\rbrace\,.
\end{align*}
in Feynman gauge.\smallskip

\begin{figure}[t]
	\center

		\includegraphics[scale=.8]{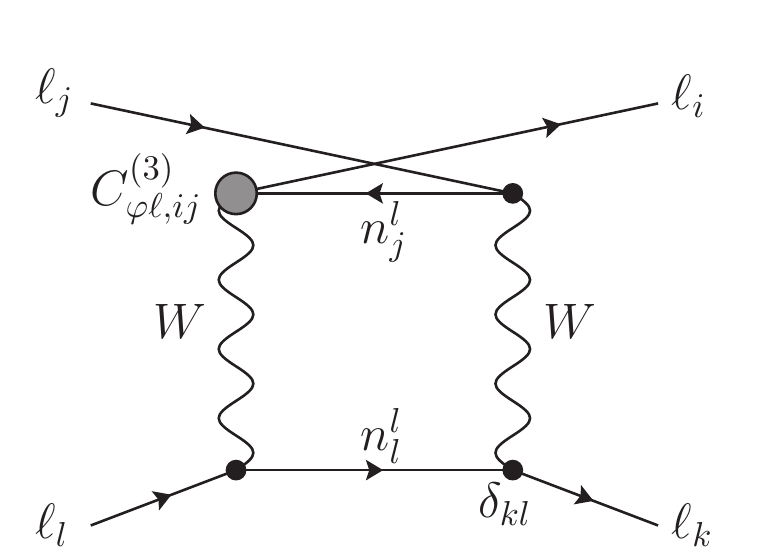}
   \caption{Box diagram contributing to four lepton processes with an (tree-level) operator insertion.}\label{fig:BoxDiags}	
\end{figure}

The box diagram with two W bosons in the loop (see Figure~\ref{fig:BoxDiags}) contributes as follows in the $R_\xi$-gauge:
\begin{align}
d_{ij,kl}^{\mathrm{V}LL}=-\frac{e^4}{64\pi^2\sw^4\MW^2}\sum_{a=1}^n\MDij{ia}\MRi{a}^{-2}\MDij{ja}^*\delta_{kl}\frac{1}{4}\left(-3+\xi-\frac{6\xi\log\xi}{1-\xi}\right)\,,
\label{eq:dVL}
\end{align}
In Feynman-gauge we have
\begin{align}
d_{ij,kl}^{\mathrm{V}LL}=-\frac{e^4}{64\pi^2\sw^4\MW^2}\sum_{a=1}^n\MDij{ia}\MRi{a}^{-2}\MDij{ja}^*\delta_{kl}\,.
\end{align}

\subsection{Two-Lepton-Two-Quark Interactions}
Next we consider $\mu \to e$ conversion in nuclei, whose amplitude can be written as
\begin{align}
\mathcal{M}_{\rm eff}= &
\left(a_{ij,qq}^{\mathrm{V}LL} +z_{ij,qq}^{\mathrm{V}LL} + d_{ij,qq}^{\mathrm{V}LL}\right)\left(\overline{\ell}_i \gamma_\mu P_L \ell_j\right)\left(\overline{q}\gamma_\mu P_L q\right)\notag\\
&+\left(a_{ij,qq}^{\mathrm{V}LR} +z_{ij,qq}^{\mathrm{V}LR} \right)\left(\overline{\ell}_i \gamma_\mu P_L \ell_j\right)\left(\overline{q}\gamma_\mu P_R q\right)\,.
\end{align}
with $q=u,d$.

The photon penguin contributions are of the same form of those given in \eqref{eq:aV}.
\begin{align}
\begin{split}
a_{ij,uu}^{\mathrm{V}LL}=&a_{ij,uu}^{\mathrm{V}LR}=-e Q_u a_{ij}^\mathrm{V}=-\frac{2}{3}e\,a_{ij}^\mathrm{V} \,,\\
a_{ij,dd}^{\mathrm{V}LL}=&a_{ij,dd}^{\mathrm{V}LR}=-e Q_d a_{ij}^\mathrm{V}=\frac{1}{3}e\,a_{ij}^\mathrm{V}\,.
\end{split}
\label{eq:aVLqq}
\end{align}
Similarly, the $Z$ boson penguins lead to 
\begin{align}
\begin{split}
\tilde z^{\mathrm{V}LL}_{ij,qq}=& \tilde z_{ij}^{L} \,g^{q L}_{\rm SM}\,,\\
\tilde z^{\mathrm{V}LR}_{ij,qq}=& \tilde z_{ij}^{L} \,g^{q R}_{\rm SM}\,,
\end{split}
\label{eq:zVLqq}
\end{align}
where $g^{q L}_{\rm SM}$, and $g^{q R}_{\rm SM}$, $q=u,d$, are the SM $Z$ boson couplings to left- and right-handed up- and down-type quarks,
\begin{align}
\begin{split}
g^{u L}_{\rm SM}=& -\frac{e}{2\sw \cw}\left(1-\frac{4}{3}\sw^2\right) \, ,\\
g^{u R}_{\rm SM}=&  \frac{2e \sw}{3\cw} \,,\\
g^{d L}_{\rm SM}=& \frac{e}{2\sw \cw}\left(1-\frac{2}{3}\sw^2\right) \, ,\\
g^{d R}_{\rm SM}=&  -\frac{e\sw}{3\cw} \,,
\end{split}
\label{eq:gZqq}
\end{align}
and $\tilde z_{ij}^{L}$ is as defined in \Eq{eq:ztV}.

In $R_\xi$-gauge we obtain the following box contributions:
\begin{align}
d_{ij,u_k u_l}^{\mathrm{V}LL}=&
-\frac{e^4}{256\pi^2\sw^4\MW^2}\Bigg\lbrace
\sum_{a=1}^n\sum_{g=1}^3\MDij{ia}\MRi{a}^{-2}\MDij{ja}^*V_{kg}^\mathrm{CKM}V_{lg}^\mathrm{CKM*}\notag\\
&\qquad\Bigg(\frac{-\left(-9+8\xi+\xi^2+6\xi\log \xi\right)\MW^2+\xi\left(-1+\xi+6\log\xi\right)m_{d_g}^2}{\left(-1+\xi\right)\left(\MW^2-m_{d_g}^2\right)}\notag\\
&\qquad\qquad+\frac{m_{d_g}^2}{\MW^2}\log\left(\frac{\MRi{a}^2}{\MW^2}\right)
+\frac{m_{d_g}^2\left(4\MW^2-m_{d_g}^2\right)^2}{\MW^2\left(\MW^2-m_{d_g}^2\right)^2}\log\left(\frac{\MW^2}{m_{d_g}^2}\right)
\Bigg)\Bigg\rbrace
\notag\\
d_{ij,d_k d_l}^{\mathrm{V}LL}=&
-\frac{e^4}{256\pi^2\sw^4\MW^2}\Bigg\lbrace
\sum_{a=1}^n\sum_{g=1}^3\MDij{ia}\MRi{a}^{-2}\MDij{ja}^*V_{gk}^\mathrm{CKM*}V_{gl}^\mathrm{CKM}\notag\\
&\qquad\Bigg(\frac{\left(3-4\xi+\xi^2+6\xi\log\xi\right)\MW^2-\xi\left(-1+\xi+6\log \xi \right)m_{u_g}^2}{\left(-1+\xi\right)\left(\MW^2-m_{u_g}^2\right)}\notag\\
&\qquad\qquad-\frac{m_{u_g}^2}{\MW^2}\log\left(\frac{\MRi{a}^2}{\MW^2}\right)
-\frac{m_{u_g}^2\left(4\MW^2-8 m_{u_g}^2\MW^2+m_{u_g}^4\right)}{\MW^2\left(\MW^2-m_{u_g}^2\right)^2}\log\left(\frac{\MW^2}{m_{u_g}^2}\right)\,.
\Bigg)\Bigg\rbrace
\label{eq:dVLqq}
\end{align}
The terms with two Dirac mass matrices are generated by double $W$ boson boxes, whereas the terms with four Dirac mass matrices are generated by double-Goldstone boxes.
In the following, we neglect the possibility of having a flavour transition in the quark line, since this effect is CKM-suppressed (when summing the contributions we set $\left(V^\mathrm{CKM}V^\mathrm{CKM\dagger}\right)_{kl}=\delta_{kl}$).

\section{Exact diagonalisation and/or $R_\xi$ Dependence}\label{sec:FullResults}
In the following, we give the full results, i.e. the sum of the soft and hard parts of the amplitudes, for the processes of interest in our work, with exact diagonalization of the neutrino mixing matrix (as described in Appendix~\ref{sec:FullNeutrinoMatrix}) and in the $R_{\xi}$ gauge. We sum over the flavour indices $a,b$ and $c$, which denote, respectively, internal neutrinos and charged leptons, while the indices $i,j,k,l$ denote external leptons, which are fixed. For the sake of simplicity, we express the result in terms of master integrals, reported in Appendix~\ref{sec:IntExp}.

\subsection{Anomalous Magnetic Moments and Radiative Leptonic Decays}
Defining
\begin{equation}
\mathcal{L}_{\rm eff}=
\mathcal{A}_{ij}^\mathrm{M}\lLRibar{i}\sigma_{\mu \nu} P_R \lLRi{j} F^{\mu \nu}+\hc\,,
\end{equation}
where $F^{\mu \nu}=\partial^\mu A^\nu-\partial^\nu A^\mu$ is the electromagnetic field strength tensor, we find
\begin{align}
\mathcal{A}^\mathrm{M}_{ij}=&\frac{e^3 m_{\ell_j}}{256 \pi ^2 \sw^2} \sum_{a=1}^n U_{ia} U_{ja}^*f^{\mathrm{M}}\left(\Mvi{a},\MW\right)\,,
\label{eq:AMfull}
\end{align}
where
\begin{align}
f^{\mathrm{M}}\left(\Mvi{a},\MW\right)=&\frac{
 \Mvi{a}^2}{\MW^2 \left(\Mvi{a}^2-\MW^2\right)^4}
 \Big(
 6 \Mvi{a}^2 \MW^2\left(A_0(\Mvi{a})-\MW^2 \right)\nonumber\\
&
 \qquad \qquad  \qquad\qquad-3 \Mvi{a}^4 \left(2 A_0(\MW)-\MW^2\right) +2 \Mvi{a}^6+\MW^6
 \Big)\,.
\end{align}

\subsection{$Z\to \ell \ell'$}
\label{sec:Zll}

The $Z\to \ell_i\overline{\ell}_j$ amplitude can be cast into the form
\begin{align*}
\mathcal{M}(Z\to \lLRi{i}\lLRibar{j})=-\frac{e^3}{16\pi^2\cw\sw^3}\;\overline{\mathcal{Z}}^{\mathrm{V}L}_{ij}\;
\lLRibar{j}\slashed{Z}P_L \lLRi{i}\,,
\end{align*}
with
\begin{align}
\overline{\mathcal{Z}}^{\mathrm{V}L}_{ij}=\;
& U_{ia}U_{ja}^*f_A^\mathrm{V}\left(\Mvi{a}^2\right)
+ U_{ia}\left(\sum_{k=1}^3 U_{ka}^*U_{kb}\right)U_{jb}^*f_B^\mathrm{V}\left(\Mvi{a}^2,\Mvi{b}^2\right)\notag\\
& +U_{ia}U_{ka}U^*_{kb}U^*_{jb} f_C^\mathrm{V}\left(\Mvi{a}^2,\Mvi{b}^2\right)\,,
\label{eq:Zllfull}
\end{align}
\begin{align}
f_A^\mathrm{V}\left(\Mvi{a}^2\right)=\Bigg\lbrace
&\frac{-2 \left(2 \MW^2-\MZ^2\right) }{\MZ^4 \left(\Mvi{a}^2-\MW^2\right)^2}\Big(4 \MW^4 \left(\MW^2+2 \MZ^2\right)-3 \MW^2\left(2 \MW^2+3\MZ^2\right) \Mvi{a}^2\notag\\
&\hspace{20mm}+4 \MZ^2 \Mvi{a}^4+2 \Mvi{a}^6\Big)A_0(\MW)\notag\\[2mm]
&+\frac{2 \left(2 \MW^2-\MZ^2\right)}{\MZ^4 \left(\Mvi{a}^2-\MW^2\right)^2} 
\Big(4 \MW^4 \left(\MW^2+2 \MZ^2\right)-6  \MW^2 \left(\MW^2+\MZ^2\right)\Mvi{a}^2\notag\\
&\hspace{20mm}+ \MZ^2\Mvi{a}^4+2 \Mvi{a}^6\Big) A_0(\Mvi{a})\notag\\[2mm]
&-\frac{2}{\MZ^4}\Big(\left(4 \MW^2 \MZ^2+4 \MW^4-\MZ^4\right)\Mvi{a}^2\notag\\
&\hspace{20mm}+2  \left(2 \MW^2-\MZ^2\right)\Mvi{a}^4\Big) B_0\left(\MZ^2;\MW,\MW\right)\notag\\[2mm]
&-\frac{4}{\MZ^4}\Big(
4 \MW^6\left(\MW^2+2 \MZ^2\right)
+ \MW^2\left(4 \MZ^4-5 \MW^2 \MZ^2-6 \MW^4\right)\Mvi{a}^2\notag\\
&\hspace{20mm}+\MZ^2\left(4 \MW^2 -\MZ^2\right)\Mvi{a}^4 
+\left(2 \MW^2-\MZ^2\right)\Mvi{a}^6 \Big)\notag\\
&\hspace{60mm}
\times C_0\left(0,0,\MZ^2;\MW,\Mvi{a},\MW\right)\notag\\[2mm]
&+\frac{2 \MW^2-\MZ^2}{\MZ^2 \left(\Mvi{a}^2-\MW^2\right)}\left(2 \MW^4+5 \Mvi{a}^2 \MW^2-\Mvi{a}^4\right)\Bigg\rbrace \frac{1}{16\MW^2}\,,
\label{eq:fA}
\end{align}
\begin{align}
f_B^\mathrm{V}\left(\Mvi{a}^2,\Mvi{b}^2\right)=\Bigg\lbrace& \MW^2 \left(-2 \MW^2-3 \MZ^2+\Mvi{a}^2+\Mvi{b}^2\right) B_0\left(\MZ^2;\Mvi{b},\Mvi{a}\right)\notag\\
&+\Big(-2 \MW^2 \left(\MW^2+\MZ^2\right)^2+ 2 \MW^2 \left(\MW^2+\MZ^2\right)\left(\Mvi{a}^2+\Mvi{b}^2\right)\notag\\
&\hspace{20mm}
- \left(2 \MW^2+\MZ^2\right)\Mvi{a}^2 \Mvi{b}^2
\Big)\notag\\
&\hspace{30mm}\times C_0\left(0,0,\MZ^2;\Mvi{a},\MW,\Mvi{b}\right)\Bigg\rbrace\frac{1}{4\MW^2\MZ^2}\,,
\label{eq:fB}
\end{align}
and
\begin{align}
f_C^\mathrm{V}\left(\Mvi{a}^2,\Mvi{b}^2\right)=\Bigg\lbrace &
\Big(-2\MW^2+\MZ^2+\Mvi{a}^2+\Mvi{b}^2\Big)B_0\left(\MZ^2;\Mvi{a},\Mvi{b}\right)\notag\\
&-2\Big(
\Mvi{a}^2 \Mvi{b}^2
- \MW^2 \left(\Mvi{a}^2+\Mvi{b}^2\right)
+\MW^2 \left(\MW^2+2 \MZ^2\right)
\Big)\notag\\
&\hspace{30mm}\times C_0\left(0,0,\MZ^2;\Mvi{a},\MW,\Mvi{b}\right)
   \Bigg\rbrace \frac{-\Mvi{a} \Mvi{b}}{8\MW^2 \MZ^2}\,.
\end{align}
Note that all three structures are both $\xi$-independent and UV-finite after taking into account the unitarity of the neutrino mixing matrix and \Eq{eq:UMUT}.

Performing the seesaw expansion (see \Eqs{eq:Vexp}-\eqref{eq:mhexp}), Eq.~(\ref{eq:Zllfull}) simplifies to Eq.~(\ref{eq:Zllseesaw}). The structure $U_{ia}U_{ca}U^*_{cb}U^*_{jb}$ vanishes by virtue of the inverse seesaw condition, Eq.~\eqref{eq:InverseSeesawCond}. Terms with this structure survive only in presence of sterile neutrino mass splitting and can be simplified as follows. 
Using the notation $Y^4|_\slashed{L} =\sum _{a,b=1}^N  \Yvij{ia}\MRi{a}^{-1}\left(\sum_{c=1}^3 \Yvij{ca}\Yvijconj{cb}\right)\MRi{b}^{-1}\Yvijconj{jb}$, we find
\begin{align}
    Y^4|_\slashed{L}
    \times &\left\lbrace -\frac{\MRi{a}^2}{\MRi{a}^2-\MRi{b}^2}\log\left(\frac{\mu^2}{\MRi{a}^2}\right)
    +\frac{\MRi{b}^2}{\MRi{a}^2-\MRi{b}^2}\log\left(\frac{\mu^2}{\MRi{b}^2}\right)
    \right\rbrace = \notag \\
    Y^4|_\slashed{L}
    \times &\left\lbrace +\frac{\MRi{a}^2}{\MRi{a}^2-\MRi{b}^2}\log\left(\MRi{a}^2\right)
    -\frac{\MRi{b}^2}{\MRi{a}^2-\MRi{b}^2}\log\left(\MRi{b}^2\right)
    \right\rbrace =\notag  \\
    Y^4|_\slashed{L}
    \times &\left\lbrace  
    \frac{1}{2}\frac{\MRi{a}^2-\MRi{b}^2}{\MRi{a}^2-\MRi{b}^2}\log\left(\MRi{a}^2\right)
    +\frac{\MRi{a}^2}{\MRi{a}^2-\MRi{b}^2}\log\left(\MRi{a}^2\right)
   \right.\notag \\
    &\left.\quad -\frac{1}{2}\frac{\MRi{a}^2-\MRi{b}^2}{\MRi{a}^2-\MRi{b}^2}\log\left(\MRi{b}^2\right)
    -\frac{\MRi{b}^2}{\MRi{a}^2-\MRi{b}^2}\log\left(\MRi{b}^2\right)
    \right\rbrace =\notag  \\
    Y^4|_\slashed{L}
    \times & 
    \left \lbrace \frac{1}{2}\frac{\MRi{a}^2+\MRi{b}^2}{\MRi{a}^2-\MRi{b}^2}\log\left(\frac{\MRi{a}^2}{\MRi{b}^2}\right)\right\rbrace = Y^4|_\slashed{L} \times  k(\MRi{a}^2,\MRi{b}^2) \,,
    \label{eq:MajoranaSimplification}
\end{align}
with the function $k$, as defined in Eq.~\eqref{eq:kFct}. 

\begin{align}
\overline{\mathcal{Z}}_{ij}^{\mathrm{V}L}(q_Z^2)=\;
&\sum _{a=1}^n \MDij{ia} \MRi{a}^{-2} \MDijconj{ja} f_2\left(\MRi{a}^2\right)\notag\\
&+\sum _{a,b=1}^n \MDij{ia}\left(\sum_{k=1}^3 \MDijconj{ka}\MDij{kb}\right)\MDij{jb}\; \; f_{4A}\left(\MRi{a}^2,\MRi{b}^2\right)\notag\\
&+\sum_{a,b=1}^n\MDij{ia}\MRi{a}^{-1}\left(\sum_{k=1}^3\MDij{ka}\MDijconj{kb}\right)\MRi{b}^{-1}\MDijconj{jb}\; \;f_{4B}\left(\MRi{a}^2,\MRi{b}^2\right)\,,
\label{eq:Zllseesaw}
\end{align}
The loop functions $f_2$ and $f_{4A}$ and $f_{4B}$, reduced to master integrals, are given by
\begin{align}
f_2\left(\MR^2\right)=
&
-\left(1-2 \cw^2\right)\frac{2 \MR^6+\MR^4 q_Z^2-6 \MR^2 \MW^2 \left(\MW^2+q_Z^2\right)+4
   \MW^4 \left(\MW^2+2 q_Z^2\right)}{8 \MW^2 q_Z^2 (\MR^2 -\MW^2 )^2}A_0(\MR)\notag\\[2mm]
& 
+\left(1-2 \cw^2\right)\frac{ \MR^2 \left(2 \MR^4-4 \MR^2 \left(\MW^2+q_Z^2\right)+7 \MW^2 q_Z^2+2
   \MW^4\right)}{8 \MW^2 q_Z^2 (\MR^2-\MW^2)^2}A_0(\MW)
\notag\\[2mm]
&
+\frac{1}{2 q_Z^2}\left(\MR^2-2 \MW^2-3 q_Z^2\right) B_0(q_Z^2;0,\MR)\notag
\\[2mm]
& +\frac{ \MR^2}{8 \MW^2 q_Z^2} \left(2\left(1-2 \cw^2\right) \MR^2-2 \left(1+2 \cw^2\right) \MW^2
  +\left(1-2 \cw^2\right)q_Z^2\right)B_0(q_Z^2;\MW,\MW)
\notag\\[2mm]
&
+\frac{1}{q_Z^2}\left(\MW^2+q_Z^2\right) \left(\MR^2-\MW^2-q_Z^2\right) C_0(0,0,q_Z^2,\MR,\MW,0)\notag\\[2mm]
&
+\frac{1}{4 \MW^2 q_Z^2}\Big(
\left(1-2 \cw^2\right)\MR^6
-\MR^4 \left(2 \MW^2-\left(1-2 \cw^2\right)q_Z^2\right)\notag\\
&
\hspace{25mm}+\MR^2 \MW^2\left(\left(1+6 \cw^2\right) \MW^2-4 \left(1-\cw^2\right) q_Z^2\right)\notag\\
&
\hspace{25mm}-4\cw^2\MW^4\left(\MW^2+2q_Z^2\right)
\Big)C_0(0,0,q_Z^2,\MW,\MR,\MW)\notag\\[2mm]
&+\frac{\left(1-2 \cw^2\right) \MR^2 \left(\MR^2-7 \MW^2\right)}{16 \MW^2 (\MR^2-\MW^2)}\,,
\label{eq:fMD2offshell}
\end{align}
and
\begin{align}
f_{4A}(\MRi{a},\MRi{b})=&-\frac{1}{4 \MW^2}C_0(0,0,q_Z^2,\MRi{a},\MW,\MRi{b})
\label{eq:fMD4Aoffshell}\\
f_{4B}\left(\MRi{a}^2,\MRi{b}^2\right)=&-\frac{1}{8\MW^2 q_Z^2\MRi{a} \MRi{b}}\notag\\
&\times \Bigg\lbrace 
\Big(\MRi{a}^2+\MRi{b}^2-2\MW^2+q_Z^2\Big)B_0\left(q_Z^2;\MRi{a},\MRi{b}\right)\notag\\
& \qquad-2\Big(
\MRi{a}^2 \MRi{b}^2- \MW^2 \left(\MRi{a}^2+\MRi{b}^2\right)
+\MW^2 \left(\MW^2+2 q_Z^2\right)\Big)\notag\\
&\hspace{30mm} \times C_0\left(0,0,q_Z^2;\MRi{a},\MW,\MRi{b}\right)
   \Bigg\rbrace 
\end{align}
$q_Z^2$ corresponds to the momentum squared of the $Z$ boson. 
For on-shell $Z$, $q_Z^2$ is to be substituted with $\MZ^2$.

\subsection{$\ell\to\ell'\overline{\ell'}\ell'''$}
The amplitudes of $\ell\to 3\ell^{\prime}$ processes can be cast into the form
\begin{align}
\mathcal{M}\left(\lLRi{j}\to\lLRi{i}\lLRi{k}\lLRibar{l}\right)=&
\left(\mathcal{A}^{\mathrm{V}LR}_{ij,kl}+\mathcal{Z}^{\mathrm{V}LR}_{ij,kl}\right)
\left(\lLRibar{i} \gamma_\mu P_L \lLRi{j}\right)\left(\lLRibar{k} \gamma_\mu P_R\lLRi{l}\right)\notag\\
&+\left[\left(\mathcal{A}^{\mathrm{V}LL}_{ij,kl}+\mathcal{Z}^{\mathrm{V}LL}_{ij,kl}+\mathcal{D}^{\mathrm{V}LL}_{ij,kl}\right)
\left(\lLRibar{i} \gamma_\mu P_L \lLRi{j}\right)\left(\lLRibar{k} \gamma_\mu P_L \lLRi{l}\right)\right.\notag\\
&\left.\qquad+\left(\left(ij,kl\right)\to\left(kl,ij\right)\right)\right.\notag\\
&\left.\qquad+\left(\left(ij,kl\right)\to\left(kj,il\right)\right)\right.\notag\\
&\left.\qquad+\left(\left(ij,kl\right)\to\left(il,kj\right)\right)\right]\notag\\
& + \mathcal{D}^{\mathrm{S}LR}_{jl,ik}\left(\lLRibar{i} P_L \lLRi{k}\right)\left(\lLRibar{j} P_R \lLRi{l}\right)\,.
\end{align}

with $\mathcal{A}_{ij,kl}^{\mathrm{V}L(L/R)}$,
 $\mathcal{Z}_{ij,kl}^{\mathrm{V}L(L/R)}$,  $\mathcal{D}_{ij,kl}^{\mathrm{V}LL}$ and $\mathcal{D}_{ij,kl}^{\mathrm{S}LR}$ as defined in the following subsections.

\subsubsection{Photon Penguin Contributions}

From the photon penguins, we obtain 
\begin{align}
&\mathcal{A}_{ij,kl}^{\mathrm{V}L}=\frac{e^3 }{384 \pi ^2 \MW^2 \sw^2} U_{ia}\Mvi{a}^2 U^*_{ja} \delta _{k,l}\Bigg\lbrace
\frac{3 \xi  \left(6 \Mvi{a}^2-(7-\xi ) \xi  \MW^2\right)}{(1-\xi )
   \left(\Mvi{a}^2-\xi  \MW^2\right)^2} \log \left(\frac{\Mvi{a}^2}{\xi  \MW^2}\right)
\\
&\qquad-\frac{(28-3 \xi) \Mvi{a}^6-(43+19 \xi ) \MW^2  \Mvi{a}^4+(34 \xi +9)  \MW^4\Mvi{a}^2-6 \xi \MW^6}{\left(\Mvi{a}^2-\MW^2\right)^3 \left(\Mvi{a}^2-\xi  \MW^2\right)}\notag\\
&\qquad+\frac{2 (8 \xi +1) \Mvi{a}^6-(41 +13 \xi) \MW^2
   \Mvi{a}^4+6 (11-2 \xi ) \MW^4 \Mvi{a}^2-3 (7-\xi ) \MW^6}{(1-\xi) \left(\Mvi{a}^2-\MW^2\right)^4}\log
   \left(\frac{\Mvi{a}^2}{\MW^2}\right)
   \Bigg\rbrace\,.\label{eq:AVfull}
\end{align}
\subsubsection{$Z$ Penguin Contributions}
The $Z$ penguin contributions give
\begin{align}
\mathcal{Z}^{\mathrm{V}LR}_{ij,kl}=& g_\mathrm{SM}^{\ell L}\mathcal{Z}_{ij}^{\mathrm{V}L}\delta_{kl}\,,\notag\\
\mathcal{Z}^{\mathrm{V}LL}_{ij,kl}=& g_\mathrm{SM}^{\ell R}\mathcal{Z}_{ij}^{\mathrm{V}L}\delta_{kl}\,,
\label{eq:VVfull}
\end{align}
with the couplings $g_\mathrm{SM}^{\ell L}$ and $g_\mathrm{SM}^{\ell R}$, as defined in Eq.~\eqref{eq:glSM}, and 
\begin{align}
\left.\mathcal{Z}^{\mathrm{V}L}_{ij,kl}\right\vert_{R_\xi}=\left.\mathcal{Z}^{\mathrm{V}L}_{ij,kl}\right\vert_\mathrm{FG} + \left.\mathcal{Z}^{\mathrm{V}L}_{ij,kl}\right\vert_\mathrm{\xi}\,,
\end{align}
where $\left.\mathcal{Z}^{\mathrm{V}L}_{ij,kl}\right\vert_\mathrm{R_\xi} $ denotes the result in $R_\xi$ gauge, $\left.\mathcal{Z}^{\mathrm{V}L}_{ij,kl}\right\vert_\mathrm{FG} $ is the result in Feynman gauge and $\left.\mathcal{Z}^{\mathrm{V}L}_{ij,kl}\right\vert_\mathrm{\xi}$ collects the $\xi$-dependent terms:
\begin{align}
\left.\mathcal{Z}_{ij}^{\mathrm{V}L}\right\vert_\mathrm{FG}=&
\frac{e^3\cw}{128 \pi ^2 \sw^3 \MW^2}
\frac{U_{ia}U^*_{ja}}{\left(\Mvi{a}^2-\MW^2\right)^2}
\Bigg\lbrace
\Mvi{a}^4 \left(9-10 \log \left(\frac{\Mvi{a}^2}{\MW^2}\right)\right)
 -8 \MW^2 \Mvi{a}^2
 -\MW^4\Bigg\rbrace
\notag\\
&- \frac{e^3\cw}{64 \pi ^2 \sw^3 \MW^2} U_{ia}U^*_{ma}U_{mb}U^*_{jb} \notag\\
&\qquad\times
\frac{\Mvi{a}^4 \big(\Mvi{b}^2-\MW^2\big)^2 \log \bigg(\frac{\Mvi{a}^2}{\MW^2}\bigg)
-\Mvi{b}^4 \big(\Mvi{a}^2-\MW^2\big)^2 \log \bigg(\frac{\Mvi{b}^2}{\MW^2}\bigg)
}{\MW^2
   \big(\Mvi{a}^2-\Mvi{b}^2\big) \big(\Mvi{a}^2-\MW^2\big) \big(\Mvi{b}^2-\MW^2\big)}\notag\\
& - \frac{e^3\cw}{128 \pi ^2 \sw^3 \MW^2} U_{ia} U_{ma}U^*_{mb} U^*_{jb}\notag\\
&\qquad\times\left\lbrace\rule{0cm}{10mm}\right.
\frac{\Mvi{a}^3 \Mvi{b} \big(\Mvi{a}^2-4 \MW^2\big) \log \bigg(\frac{\Mvi{a}^2}{\MW^2}\bigg)}{\MW^2 \big(\Mvi{a}^2-\Mvi{b}^2\big)\big(\Mvi{a}^2-\MW^2\big)}
-\frac{\Mvi{a} \Mvi{b}^3 \left(\Mvi{b}^2-4 \MW^2\right) \log
   \bigg(\frac{\Mvi{b}^2}{\MW^2}\bigg)}{\MW^2 \big(\Mvi{a}^2-\Mvi{b}^2\big) \big(\Mvi{b}^2-\MW^2\big)}
   \left.\rule{0cm}{10mm}\right\rbrace
\, ,
   \label{eq:ZVLfullFG}
\end{align}
and
\begin{align}
\left.\mathcal{Z}_{ij}^{\mathrm{V}L}\right\vert_\xi=\frac{e^3 \cw }{128 \pi ^2 \sw^3} & U_{ia}U^*_{ja}\notag \\
 \times  
 \left\lbrace\rule{0cm}{8mm}\right.&
\frac{7}{\Mvi{a}^2-\MW^2}-\frac{\xi ^2}{\Mvi{a}^2-\xi  \MW^2}\notag \\
&+\left(\frac{\xi ^2 \Mvi{a}^2}{\left(\Mvi{a}^2-\xi  \MW^2\right)^2}-\frac{2 \xi ^2 (1+2 \xi )}{(1-\xi ) \left(\Mvi{a}^2-\xi  \MW^2\right)}\right) \log
   \left(\frac{\Mvi{a}^2}{\xi  \MW^2}\right)\notag \\
   &-\left(\frac{3 (5-9 \xi )}{2 (1-\xi) \left(\Mvi{a}^2-\MW^2\right)}+\frac{7 \left(\Mvi{a}^2+\MW^2\right)}{2 \left(\Mvi{a}^2-\MW^2\right)^2}+\frac{4 (1-\xi)}{\MW^2}\right) \log \left(\frac{\Mvi{a}^2}{\MW^2}\right)
      \left.\rule{0cm}{8mm}\right\rbrace\notag \\
  +\frac{ e^3\cw}{64 \pi ^2 \sw^3} & U_{ia}U^*_{ka}U_{kb}U^*_{jb}\notag \\
    \times  \left\lbrace\rule{0cm}{10mm}\right.&
    \frac{\Mvi{a}^2}{\MW^2} \left(\frac{\xi  \log \bigg(\frac{\Mvi{a}^2}{\xi  \MW^2}\bigg)}{\Mvi{a}^2-\xi  \MW^2}-\frac{\log \bigg(\frac{\Mvi{a}^2}{\MW^2}\bigg)}{\Mvi{a}^2-\MW^2}\right)
      +\frac{\Mvi{b}^2}{\MW^2}
   \left(\frac{\xi  \log \bigg(\frac{\Mvi{b}^2}{\xi  \MW^2}\bigg)}{ \Mvi{b}^2-\xi  \MW^2 }-\frac{\log \bigg(\frac{\Mvi{b}^2}{\MW^2}\bigg)}{\Mvi{b}^2-\MW^2}\right)
      \left.\rule{0cm}{10mm}\right\rbrace
   \label{eq:ZVLfullRxi}
\end{align}

\subsubsection{Box Contributions}
For general neutrino mixing matrices, we find two different classes of box contributions: vectorial, lepton number conserving boxes that we will denote $\mathcal{D}^{\mathrm{V}LL}_{ij,kl}$, as well as scalar, lepton number violating boxes, which we denote $\mathcal{D}^{\mathrm{S}LR}_{ij,kl}$ and vanish in the inverse seesaw limit in presence of degenerate sterile neutrinos. 
The full box contributions in the $R_\xi$ gauge can be split into the result in Feynman gauge and the $\xi$-dependent terms:
\begin{align}
\left.\mathcal{D}^{\mathrm{V}LL}_{ij,kl}\right\vert_{R_\xi}=&\left.\mathcal{D}^{\mathrm{V}LL}_{ij,kl}\right\vert_\mathrm{FG} + \left.\mathcal{D}^{\mathrm{V}LL}_{ij,kl}\right\vert_\mathrm{\xi}\,,\notag\\
\left.\mathcal{D}^{\mathrm{S}LR}_{ij,kl}\right\vert_{R_\xi}=&\left.\mathcal{D}^{\mathrm{S}LR}_{ij,kl}\right\vert_\mathrm{FG} + \left.\mathcal{D}^{\mathrm{S}LR}_{ij,kl}\right\vert_\mathrm{\xi}\,.\\
\end{align}
We find
\begin{align}
\left.\mathcal{D}^{\mathrm{V}LL}_{ij,kl}\right\vert_\mathrm{FG}=&\frac{e^4}{256\pi^2\sw^4\MW^4}\sum_{a,b=1}^n U_{ia}U^*_{ja}U_{kb}U^*_{lb}
\frac{1}{(\Mvi{a}^2-\Mvi{b}^2)(\Mvi{a}^2-\MW^2)^2(\Mvi{b}^2-\MW^2)^2}\notag\\
& \times \Bigg\lbrace 
 \Mvi{a}^6 \Mvi{b}^6 \log \left(\frac{\Mvi{a}^2}{\Mvi{b}^2}\right)\notag\\
& \qquad - \MW^2\Mvi{a}^6 \Mvi{b}^4 \left(7+2 \log \left(\frac{\Mvi{a}^2}{\MW^2}\right)-8 \log \left(\frac{\Mvi{b}^2}{\MW^2}\right)\right)\notag\\
& \qquad + \MW^2\Mvi{a}^4 \Mvi{b}^6 \left(7+2 \log \left(\frac{\Mvi{b}^2}{\MW^2}\right)-8 \log \left(\frac{\Mvi{a}^2}{\MW^2}\right)\right)\notag\\  
& \qquad + 20\MW^4 \Mvi{a}^4 \Mvi{b}^4 \log \left(\frac{\Mvi{a}^2}{\Mvi{b}^2}\right)\notag\\
&  \qquad  + \MW^4\Mvi{a}^2 \Mvi{b}^2 
      \left(\Mvi{a}^4 \left(7+\log \left(\frac{\Mvi{a}^2}{\MW^2}\right)\right)
      -\Mvi{b}^4 \left(7+\log \left(\frac{\Mvi{b}^2}{\MW^2}\right)\right)\right)\notag\\
& \qquad   -\MW^6\Mvi{a}^2 \Mvi{b}^2 
      \left(\Mvi{a}^2 \left(3+16 \log \left(\frac{\Mvi{a}^2}{\MW^2}\right)\right)
      -\Mvi{b}^2 \left(3+16 \log \left(\frac{\Mvi{b}^2}{\MW^2}\right)\right)\right)\notag\\
& \qquad   -4 \MW^8 \left(\Mvi{a}^4 \left(1-\log \left(\frac{\Mvi{a}^2}{\MW^2}\right)\right)
               -\Mvi{b}^4 \left(1-\log \left(\frac{\Mvi{b}^2}{\MW^2}\right)\right)\right)\notag\\
& \qquad   +4 \MW^{10}
   \left(\Mvi{a}^2-\Mvi{b}^2\right)
\Bigg\rbrace\,,
   \label{eq:DVfull}
\end{align}
\begin{align}
\left.\mathcal{D}^{\mathrm{V}LL}_{ij,kl}\right\vert_\xi= &\frac{e^4}{1024 \pi ^2 \sw^4}\sum_{a,b=1}^n U_{ia}U^*_{ja}U_{kb}U^*_{lb}\notag\\
\times\left\lbrace\rule{0cm}{8mm}\right.& 
-\frac{4 \xi ^2 \left( 1-\xi + 6 \log \bigg(\frac{\Mvi{a}^2}{\xi  \MW^2}\bigg)\right)}{(1-\xi ) \big(\Mvi{a}^2-\xi  \MW^2\big)}
-\frac{4 \xi ^2 \left(1-\xi +6 \log \bigg(\frac{\Mvi{b}^2}{\xi\MW^2}\bigg)\right)}{(1- \xi) \big(\Mvi{b}^2-\xi  \MW^2\big)}
   \notag\\
&+\frac{4 \xi ^2 \Mvi{a}^2 }{\big(\Mvi{a}^2-\xi  \MW^2\big)^2}\log \left(\frac{\Mvi{a}^2}{\xi 
   \MW^2}\right)
   +\frac{4 \xi ^2
   \Mvi{b}^2 }{\big(\Mvi{b}^2-\xi  \MW^2\big)^2}\log \left(\frac{\Mvi{b}^2}{\xi  \MW^2}\right)
   \notag\\
   &-\frac{(7-19 \xi) \log \bigg(\frac{\Mvi{a}^2}{\MW^2}\bigg)-14 (1-\xi )}{(1-\xi) \Mvi{a} (\Mvi{a}-\MW)}
   -\frac{(7-19 \xi ) \log \bigg(\frac{\Mvi{b}^2}{\MW^2}\bigg)-14 (1-\xi)}{(1-\xi) \Mvi{b}  (\Mvi{b}-\MW)}
   \notag\\   
   &-\frac{(7-19 \xi ) \log
   \bigg(\frac{\Mvi{a}^2}{\MW^2}\bigg)-14 (1-\xi)}{(1-\xi) \Mvi{a} (\Mvi{a}+\MW)}
   -\frac{(7-19 \xi ) \log \bigg(\frac{\Mvi{b}^2}{\MW^2}\bigg)-14 (1-\xi)}{(1-\xi) \Mvi{b} (\Mvi{b}+\MW)}
   \notag\\
   &-\frac{7 \log \bigg(\frac{\Mvi{a}^2}{\MW^2}\bigg)}{(\Mvi{a}-\MW)^2}
   -\frac{7 \log \bigg(\frac{\Mvi{b}^2}{\MW^2}\bigg)}{(\Mvi{b}-\MW)^2}
   -\frac{7 \log
   \bigg(\frac{\Mvi{a}^2}{\MW^2}\bigg)}{(\Mvi{a}+\MW)^2}
   -\frac{7 \log
   \bigg(\frac{\Mvi{b}^2}{\MW^2}\bigg)}{(\Mvi{b}+\MW)^2}
   \left.\rule{0cm}{8mm}\right\rbrace\,,
\end{align}
\begin{align}
\left.\mathcal{D}^{\mathrm{S}LR}_{jl,ik}\right\vert_\mathrm{FG}=& -\frac{e^4}{32\pi^2\sw^4\MW^4} \sum_{a,b=1}^n U^*_{ja}U^*_{la}U_{ib}U_{kb}
\frac{\Mvi{a}\Mvi{b}}{(\Mvi{a}^2-\Mvi{b}^2)(\Mvi{a}^2-\MW^2)^2(\Mvi{b}^2-\MW^2)^2}\notag\\
& \times\Bigg\lbrace
\Mvi{a}^6 \Mvi{b}^4 \left(1-\log\left(\frac{\Mvi{b}^2}{\MW^2}\right)\right)-
\Mvi{a}^4 \Mvi{b}^6 \left(1-\log \left(\frac{\Mvi{a}^2}{\MW^2}\right)\right)\notag\\
&\qquad- \MW^2 \Mvi{a}^2 \Mvi{b}^2\left(\Mvi{a}^4+4 \Mvi{a}^2 \Mvi{b}^2 \log \left(\frac{\Mvi{a}^2}{\Mvi{b}^2}\right)-\Mvi{b}^4\right)\notag\\
&\qquad+ \MW^4\Mvi{a}^4 \Mvi{b}^2 \left(3+5 \log \left(\frac{\Mvi{a}^2}{\MW^2}\right)-4 \log \left(\frac{\Mvi{b}^2}{\MW^2}\right)\right)\notag\\
&\qquad- \MW^4 \Mvi{a}^2 \Mvi{b}^4\left(3+5 \log \left(\frac{\Mvi{b}^2}{\MW^2}\right)-4 \log \left(\frac{\Mvi{a}^2}{\MW^2}\right)\right)\notag\\
&\qquad-8 \MW^6 \Mvi{a}^2 \Mvi{b}^2 \log \left(\frac{\Mvi{a}^2}{\Mvi{b}^2}\right)\notag\\
&\qquad-2 \MW^6\left(\Mvi{a}^4 \left(1+\log\left(\frac{\Mvi{a}^2}{\MW^2}\right)\right)
     -\Mvi{b}^4\left(1+\log \left(\frac{\Mvi{b}^2}{\MW^2}\right)\right)\right)\notag\\
&\qquad+2 \MW^8 \left(\Mvi{a}^2 \left(1+2 \log \left(\frac{\Mvi{a}^2}{\MW^2}\right)\right)-\Mvi{b}^2 \left(1+2\log \left(\frac{\Mvi{b}^2}{\MW^2}\right)\right)\right)
   \Bigg\rbrace\,,
   \label{eq:DSfull}
\end{align}
and
\begin{align}
\left.\mathcal{D}^{\mathrm{S}LR}_{jl,ik}\right\vert_\xi=&\frac{e^4  }{128 \pi ^2 \sw^4}\sum_{a,b=1}^n U^*_{ja}\Mvi{a}U^*_{la}U_{ib}\Mvi{b}U_{kb}\\
\times\left\lbrace\rule{0cm}{8mm}\right.& 
-\frac{4 \xi ^2 \left(1-2 \log \bigg(\frac{\Mvi{a}^2}{\xi  \MW^2}\bigg)\right)}{\Mvi{a}^4-\xi  \Mvi{a}^2 \MW^2}
-\frac{4 \xi ^2 \left(1-2 \log \bigg(\frac{\Mvi{b}^2}{\xi  \MW^2}\bigg)\right)}{\Mvi{b}^4-\xi  \Mvi{b}^2
   \MW^2}
   \notag\\
&+\frac{4 \xi ^2 }{\big(\Mvi{a}^2-\xi 
   \MW^2\big)^2}\log \left(\frac{\Mvi{a}^2}{\xi  \MW^2}\right)
   +\frac{4 \xi ^2 }{\big(\Mvi{b}^2-\xi  \MW^2\big)^2}\log \left(\frac{\Mvi{b}^2}{\xi  \MW^2}\right)
   \notag\\
   &+\frac{4 }{\Mvi{a}^2 \MW^2}\left(1-\xi+2 \xi  \log \left(\frac{\Mvi{a}^2}{\xi  \MW^2}\right)-2 \log \left(\frac{\Mvi{a}^2}{\MW^2}\right) \right)
      \notag\\
   &+\frac{4 }{\Mvi{b}^2 \MW^2}\left(1-\xi+2 \xi  \log \left(\frac{\Mvi{b}^2}{\xi  \MW^2}\right)-2 \log
   \left(\frac{\Mvi{b}^2}{\MW^2}\right)\right)
   \notag\\
   &+\frac{2-5 \log
   \bigg(\frac{\Mvi{a}^2}{\MW^2}\bigg)}{\Mvi{a}^3 (\Mvi{a}+\MW)}
+\frac{2-5 \log \bigg(\frac{\Mvi{b}^2}{\MW^2}\bigg)}{\Mvi{b}^3 (\Mvi{b}+\MW)}
   \notag\\
   &+\frac{2-5 \log \bigg(\frac{\Mvi{a}^2}{\MW^2}\bigg)}{\Mvi{a}^3 (\Mvi{a}-\MW)}
+\frac{2-5 \log
   \bigg(\frac{\Mvi{b}^2}{\MW^2}\bigg)}{\Mvi{b}^3 (\Mvi{b}-\MW)}
   \notag\\
   &-\frac{1}{\Mvi{a}^2 (\Mvi{a}-\MW)^2}\log \left(\frac{\Mvi{a}^2}{\MW^2}\right)
   -\frac{1}{\Mvi{b}^2 (\Mvi{b}+\MW)^2}\log \left(\frac{\Mvi{b}^2}{\MW^2}\right)
   \notag\\
   &-\frac{1}{\Mvi{b}^2 (\Mvi{b}-\MW)^2}\log
   \left(\frac{\Mvi{b}^2}{\MW^2}\right)
   -\frac{1}{\Mvi{a}^2
   (\Mvi{a}+\MW)^2}\log \left(\frac{\Mvi{a}^2}{\MW^2}\right)
   \left.\rule{0cm}{8mm}\right\rbrace\,.
\end{align}
In the sums $\left(\mathcal{A}^{\mathrm{V}LR}_{ij,kl}+\mathcal{Z}^{\mathrm{V}LR}_{ij,kl}\right)$ and $\left(\mathcal{A}^{\mathrm{V}LL}_{ij,kl}+\mathcal{Z}^{\mathrm{V}LL}_{ij,kl}+\mathcal{D}^{\mathrm{V}LL}_{ij,kl}\right)$, which are the relevant contributions to $\ell\to 3\ell^{\prime}$ processes in the case of lepton number conservation, the $\xi$ dependence drops out by unitarity of the neutrino mixing matrix ($UU^\dagger=\id{}$). 
In $\mathcal{D}^{\mathrm{S}LR}$, the $\xi$ dependence drops out by virtue of \Eq{eq:UMUT}, which is a consequence of the $SU(2)_L$ invariance of the Lagrangian (see \Eq{eq:LN}). Note that the structure $U_{ia} U_{ma}U^*_{mb} U^*_{jb}$ in the $Z$ penguin contribution, \Eq{eq:ZVLfullFG}, and the structure $U^*_{ja}U^*_{la}U_{ib}U_{kb}$ in the scalar boxes, \Eq{eq:DSfull}, arise from lepton number violating contributions. These structures vanish in the inverse seesaw limit (see \Eqs{eq:WeinberglessLimit2} and \eqref{eq:WeinberglessLimit}) if the sterile neutrino mass splitting is small.

\subsection{$\mu \to e$ Conversion in Nuclei}
Next we consider $\mu \to e$ conversion in nuclei. We define
\begin{align}
\mathcal{M}_\mathrm{eff}= \sum_{q=u,d}&
\left(\mathcal{A}^{\mathrm{V}LR}_{ij,qq}+\mathcal{Z}^{\mathrm{V}LR}_{ij,qq}\right)
\left(\lLRibar{i} \gamma_\mu P_L \lLRi{j}\right)\left(\bar{q} \gamma_\mu P_L q\right)\notag\\
&+\left(\mathcal{A}^{\mathrm{V}LL}_{ij,qq}+\mathcal{Z}^{\mathrm{V}LL}_{ij,qq}+\mathcal{D}^{\mathrm{V}LL}_{ij,qq}\right)
\left(\lLRibar{i} \gamma_\mu P_L \lLRi{j}\right)\left(\bar{q} \gamma_\mu P_R q\right)\,.
\label{eq:mueconvfull}
\end{align}

with $\mathcal{A}_{ij,qq}^{\mathrm{V}LL/LR}$, $\mathcal{Z}_{ij,qq}^{\mathrm{V}LL/LR}$ and $\mathcal{D}_{ij,qq}^{\mathrm{V}LL}$ as defined in the following subsections.\\

Taking the sum in Eq.~\eqref{eq:mueconvfull} and the quark flavour-diagonal limit of the box contributions, $\mathcal{D}_{ij,qq}^{\mathrm{V}AB}\equiv \mathcal{D}_{ij,q_k q_l}^{\mathrm{V}AB}\delta_{kl}$, where $q=u,d$, we obtain the leading contribution to $\mu\to e$ conversion, which is given in Eq.~\eqref{eq:mueconvFormFactors}.

\subsubsection{Photon Penguin Contributions}
The photon penguin contributions to $\mu\to e$ conversion in nuclei are of the same form as those contributing four lepton processes.
\begin{align}
\begin{split}
\mathcal{A}_{ij,uu}^{\mathrm{V}LL}=&\mathcal{A}_{ij,uu}^{\mathrm{V}LR}=Q_u\mathcal{A}_{ij}^{\mathrm{V}L}=\frac{2}{3}\mathcal{A}_{ij}^{\mathrm{V}L}\\
\mathcal{A}_{ij,dd}^{\mathrm{V}LL}=&\mathcal{A}_{ij,dd}^{\mathrm{V}LR}=Q_d\mathcal{A}_{ij}^{\mathrm{V}L}=-\frac{1}{3}\mathcal{A}_{ij}^{\mathrm{V}L}
\end{split}
\end{align}
with the form factor $\mathcal{A}_{ij}^{\mathrm{V}L}$ given in \Eq{eq:AVfull}.

\subsubsection{$Z$ Penguin Contributions}
Similarly, the $Z$ boson penguins lead to the contributions
\begin{align}
\begin{split}
\mathcal{Z}^{\mathrm{V}LL}_{ij,qq}=& \mathcal{Z}_{ij}^L \,g^{q L}_{\rm SM}\\
\mathcal{Z}^{\mathrm{V}LR}_{ij,qq}=& \mathcal{Z}_{ij}^L \,g^{q R}_{\rm SM}
\end{split}
\label{eq:vVqqLR}
\end{align}
where $g^{q L}_{\rm SM}$, and $g^{q R}_{\rm SM}$, $q=u,d$, are the SM $Z$ boson couplings to left- and right-handed up- and down-type quarks, which are given in \Eq{eq:gZqq},
and $\mathcal{Z}_{ij}^L$ is the form factor given in \Eq{eq:VVfull}.

\subsubsection{Box Contributions}
Since to order $v^2/\MR^2$ there are no box diagrams contributing to the matching onto $\mathcal{O}_{\ell q}^{(1)}$ and $\mathcal{O}_{\ell q}^{(3)}$, the only contributions are those given in \Eq{eq:dVLqq}. Neglecting the possible flavour effects on the quark line, we define
\begin{align}
    \mathcal{D}_{ij,qq}^{\mathrm{V}AB}\equiv \mathcal{D}_{ij,q_k q_l}^{\mathrm{V}AB}\delta_{kl} = d_{ij,q_k q_l}^{\mathrm{V}AB}\delta_{kl}\,,\qquad q=u,d\,.
\end{align}

\section{Integrals}\label{sec:IntExp}
Assuming the hierarchy $\MW^2, \MZ^2, q_Z^2\sim v^2\ll \MR^2$, $|\MRi{a}-\MRi{b}|\ll|\MRi{a}+\MRi{b}|$, we can use the following expansions of the master integrals:
\begin{align}
B_0(q_Z^2;0,\MR)=&1+\frac{q_Z^2}{2 \MR^2}+\frac{1}{\varepsilon}+\log \left(\frac{\mu ^2}{\MR^2}\right)
\notag\\
B_0(q_Z^2;\MW,\MW)=&2+\frac{1}{\varepsilon}+\log \left(\frac{\mu^2}{\MW^2}\right)
- \arctan \left(\frac{\sqrt{q_Z{}^2\left(4 \MW^2 -q_Z{}^2\right)}}{2 \MW^2-q_Z{}^2}\right)\sqrt{4 \frac{\MW^2}{q_Z^2} -1}
\notag\\
B_0\left(q_Z^2;\MRi{a},\MRi{b}\right)=& 1+\frac{1}{\varepsilon } 
      +\frac{1}{2} \log\left(\frac{\mu ^2}{\MRi{a}^2}\right)+\frac{1}{2} \log \left(\frac{\mu ^2}{\MRi{b}^2}\right)
      -\frac{\MRi{a}^2+\Mvi{b}^2 }{2 \left(\MRi{a}^2-\Mvi{b}^2\right)}\log \left(\frac{\MRi{a}^2}{\MRi{b}^2}\right)
\notag\\
C_0(0,0,q_Z^2,\MR,\MW,0)=&-\frac{1}{2 \MR^4}\left(\MW^2-q_Z^2+\left(2 \left(\MR^2+\MW^2\right)+q_Z^2\right) \log
   \left(\frac{\MR^2}{\MW^2}\right)\right)
\notag
\end{align}
\begin{align}
C_0(0,0,&q_Z^2,\MW,\MR,\MW)=\notag\\
=&-\frac{1}{36 \MR^6}\Bigg\lbrace
36 \MR^4
-9 \MR^2 \left(q_Z^2-4 \MW^2\right)
+4 \left(q_Z^2-3 \MW^2\right)^2\notag\\
&-6 \left(6 \MR^4+3 \MR^2 \left(4 \MW^2-q_Z^2\right)+2\left(q_Z^2-3 \MW^2\right)^2\right)
 \log \left(\frac{\MR^2}{\MW^2}\right)\notag\\
& +\bigg(6 \left(\MR^4+\MR^2 \MW^2+\MW^4\right)-q_Z^2 \left(3 \MR^2+8 \MW^2\right)+2 q_Z^4\bigg)\notag\\
&\qquad\times 
6  \arctan\left(\frac{\sqrt{q_Z{}^2\left(4 \MW^2 -q_Z{}^2\right)}}{2 \MW^2-q_Z{}^2}\right)\sqrt{\frac{4 \MW^2}{q_Z^2}-1}
\Bigg\rbrace \notag\\[2mm]
C_0(0,0,&q_Z^2,\MRi{a},\MW,\MRi{b})=
-\frac{1}{\MRi{a}^2-\MRi{b}^2}\log \left(\frac{\MRi{a}^2}{\MRi{b}^2}\right)
\label{eq:PVIexp}
\end{align}

\clearpage

\bibliographystyle{JHEP}
\bibliography{main}
\end{document}

%% file: Conventions.tex
\newcommand{\nc}{\newcommand}

\nc{\cmm}[1]{\ensuremath{#1}\ifmmode\else{}\fi}
\nc{\nmc}[2]{\newcommand{#1}{\cmm{#2}}}

\def\hc{\mathrm{h.c.}}

\def\hc{\text{h.c.}\ }

\nc\Eq[1]{Eq.~\eqref{#1}}
\nc\Eqs[1]{Eqs.~\eqref{#1}}

\nc{\id}[1]{\mathbbm{1}_{{#1}}}
\nc{\niente}[1]{\mathbb{O}_{{#1}}}

\nc{\sw}{s_\mathrm{W}}
\nc{\cw}{c_\mathrm{W}}
\nc{\MZ}{M_Z}
\nc{\MW}{M_W}

\nc{\LN}{\mathcal{L}_N}
\nc{\Db}{\mbox{${\raisebox{2mm}{\boldmath ${}^\leftarrow$}\hspace{-4mm} D}$}}
\nc{\Dfb}{\mbox{$\raisebox{2mm}{\boldmath ${}^\leftrightarrow$}\hspace{-4mm} D$}}
\nc{\DfbImu}{\mbox{$\raisebox{2mm}{\boldmath ${}^\leftrightarrow$}\hspace{-4mm} D_\mu^{\,I}$}}

\nc{\LL}{L} 
\nc{\eR}{e} 
\nc{\LLbar}{\bar L}
\nc{\eRbar}{\bar e}

\nc{\LLi}[1]{L_{{#1}}} 
\nc{\eRi}[1]{e_{{#1}}} 
\nc{\LLibar}[1]{\bar L_{{#1}}}
\nc{\eRibar}[1]{\bar e_{{#1}}}

\nc{\lLR}{\ell}
\nc{\lLRbar}{\bar\ell}
\nc{\lL}{\ell_L} 
\nc{\lR}{\ell_R} 
\nc{\lLbar}{\bar\ell_L} 
\nc{\lRbar}{\bar\ell_R} 

\nc{\lLRi}[1]{\ell_{{#1}}}
\nc{\lLRibar}[1]{\bar\ell_{{#1}}}
\nc{\lLi}[1]{\ell_{L{#1}}} 
\nc{\lRi}[1]{\ell_{R{#1}}} 
\nc{\lLibar}[1]{\bar\ell_{L{#1}}} 
\nc{\lRibar}[1]{\bar\ell_{R{#1}}} 

\nc{\QL}{Q} 
\nc{\qR}{q} 
\nc{\QLbar}{\bar Q}
\nc{\qRbar}{\bar q}
\nc{\uR}{u} 
\nc{\uRbar}{\bar u} 
\nc{\dR}{d} 
\nc{\dRbar}{\bar d} 

\nc{\QLi}[1]{Q_{{#1}}} 
\nc{\qRi}[1]{q_{{#1}}} 
\nc{\uRi}[1]{u_{{#1}}} 
\nc{\dRi}[1]{d_{{#1}}} 

\nc{\QLibar}[1]{\bar Q_{{#1}}}
\nc{\qRibar}[1]{\bar q_{{#1}}}
\nc{\uRibar}[1]{\bar u_{{#1}}}
\nc{\dRibar}[1]{\bar d_{{#1}}}
\nc{\qLR}{q} 
\nc{\qLRbar}{\bar q} 
\nc{\qLRi}[1]{q_{{#1}}} 
\nc{\qLRibar}[1]{\bar q_{{#1}}}

\nc{\uLR}{u}
\nc{\uLRbar}{\bar u}
\nc{\uLRi}[1]{u_{{#1}}}
\nc{\uLRibar}[1]{\bar u_{{#1}}}
\nc{\dLR}{d} 
\nc{\dLRbar}{\bar d} 
\nc{\dLRi}[1]{d_{{#1}}} 
\nc{\dLRibar}[1]{\bar d_{{#1}}} 

\nc{\vL}{\nu _L} 
\nc{\vLc}{\nu _L^c} 
\nc{\vR}{N_R} 
\nc{\vRc}{N_R^c} 
\nc{\vLbar}{\bar \nu _L} 
\nc{\vLcbar}{\bar \nu _L^c} 
\nc{\vRbar}{\bar N_R} 
\nc{\vRcbar}{\bar N_R^c} 

\nc{\vLi}[1]{\nu _{{#1}}}
\nc{\vLic}[1]{\nu _{{#1}}^c}
\nc{\vLibar}[1]{\bar \nu _{{#1}}}
\nc{\vLicbar}[1]{\bar\nu _{{#1}}^c}

\nc{\nL}{n_R^c} 
\nc{\nR}{n_R} 
\nc{\nLi}[1]{n_{R,{#1}}^c} 
\nc{\nRi}[1]{n_{R,{#1}}} 

\nc{\nLbar}{\bar n_R^c} 
\nc{\nRbar}{\bar n_R} 
\nc{\nLibar}[1]{\bar n_{R,{#1}}^c} 
\nc{\nRibar}[1]{\bar n_{R,{#1}}} 

\nc{\nAL}{n_R^{{\rm l},c}} 
\nc{\nAR}{n_R^{\rm l}} 
\nc{\nSL}{n_R^{{\rm h},c}} 
\nc{\nSR}{n_R^{\rm h}} 
\nc{\nALi}[1]{n_{R,{#1}}^{{\rm l},c}} 
\nc{\nARi}[1]{n_{R,{#1}}^{\rm l}} 
\nc{\nSLi}[1]{n_{R,{#1}}^{{\rm h},c}} 
\nc{\nSRi}[1]{n_{R,{#1}}^{\rm h}} 

\nc{\nALbar}{\bar n_R^{{\rm l},c}} 
\nc{\nARbar}{\bar n_R^{\rm l}} 
\nc{\nSLbar}{\bar n_R^{{\rm h},c}} 
\nc{\nSRbar}{\bar n_R^{\rm h}} 
\nc{\nALibar}[1]{\bar n_{R,{#1}}^{{\rm l},c}} 
\nc{\nARibar}[1]{\bar n_{R,{#1}}^{\rm l}} 
\nc{\nSLibar}[1]{\bar n_{R,{#1}}^{{\rm h},c}} 
\nc{\nSRibar}[1]{\bar n_{R,{#1}}^{\rm h}} 

\nc{\n}{n} 
\nc{\nA}{n^{\rm l}} 
\nc{\nS}{n^{\rm h}} 

\nc{\OlL}{V^{\ell L}}
\nc{\OlLconj}{V^{\ell L,*}}
\nc{\OlLtransp}{V^{\ell L,T}}
\nc{\OlLdag}{V^{\ell L,\dagger}}

\nc{\OlR}{V^{\ell R}}
\nc{\OlRconj}{V^{\ell R,*}}
\nc{\OlRtransp}{V^{\ell R,T}}
\nc{\OlRdag}{V^{\ell R,\dagger}}

\nc{\Ov}{V}
\nc{\Ovconj}{V^{*}}
\nc{\Ovtransp}{V^{T}}
\nc{\Ovdag}{V^{ \dagger}}
\nc{\Ovscript}{V}
\nc{\OvLO}{V^{LO}}

\nc{\OvAA}{V_{AA}}
\nc{\OvAAconj}{V_{AA}^{*}}
\nc{\OvAAtransp}{V_{AA}^{T}}
\nc{\OvAAdag}{V_{AA}^{\dagger}}

\nc{\OvAS}{V_{AS}}
\nc{\OvASconj}{V_{AS}^{*}}
\nc{\OvAStransp}{V_{AS}^{T}}
\nc{\OvASdag}{V_{AS}^{\dagger}}

\nc{\OvSA}{V_{SA}}
\nc{\OvSAconj}{V_{SA}^{*}}
\nc{\OvSAtransp}{V_{SA}^{T}}
\nc{\OvSAdag}{V_{SA}^{\dagger}}

\nc{\OvSS}{V_{SS}}
\nc{\OvSSconj}{V_{SS}^{*}}
\nc{\OvSStransp}{V_{SS}^{T}}
\nc{\OvSSdag}{V_{SS}^{\dagger}}

\nc{\Uv}{U_{\text{PMNS}}}
\nc{\Uvconj}{U_{\text{PMNS}}^*}
\nc{\Uvtransp}{U_{\text{PMNS}}^T}
\nc{\Uvdag}{U_{\text{PMNS}}^\dagger}
\nc{\Uvscript}{U_{\text{PMNS}}}

\nc{\Vv}{U_{SS}}
\nc{\Vvconj}{U_{SS}^*}
\nc{\Vvtransp}{U_{SS}^T}
\nc{\Vvdag}{U_{SS}^\dagger}
\nc{\Vvscript}{U_{SS}}

\nc{\U}{U}
\nc{\Uconj}{U^*}
\nc{\Utransp}{U^T}
\nc{\Udag}{U^\dagger}
\nc{\Uscript}{U}

\nc{\MR}{M_R}
\nc{\MRi}[1]{M_{R,{#1}}}

\nc{\MD}{M_D}
\nc{\MDconj}{M_D^*}
\nc{\MDtransp}{M_D^T}
\nc{\MDdag}{M_D^\dagger}
\nc{\MDij}[1]{M_{D,{#1}}}
\nc{\MDijconj}[1]{M_{D,{#1}}^*}

\nc{\Mv}{M_\nu}
\nc{\Mvi}[1]{M_{\nu,{{#1}}}}
\nc{\Mdiag}{M_\nu^{\rm diag}}
\nc{\Mvblock}{\tilde{M}_\nu}
\nc{\MAblock}{\tilde{m}^{\rm l}}
\nc{\MSblock}{\tilde{m}^{\rm h}}
\nc{\MA}{m^{\rm l}}
\nc{\MS}{m^{\rm h}}
\nc{\MAi}[1]{m_{A,{#1}}}
\nc{\MSi}[1]{M_{S,{#1}}}

\nc{\Yl}{Y^{\ell}}
\nc{\Yv}{Y^{\nu}}
\nc{\Yvconj}{Y^{\nu *}}
\nc{\Yvtransp}{Y^{\nu T}}
\nc{\Yvdag}{Y^{\nu \dagger}}
\nc{\Yli}[1]{Y^{\ell}_{{{#1}}}}
\nc{\Yvij}[1]{Y^{\nu}_{{#1}}}
\nc{\Yvijconj}[1]{Y^{\nu *}_{{#1}}}

\nc{\Om}{\mathcal{O}^-}
\nc{\Op}{\mathcal{O}^+}
\nc{\Xm}{X^-}
\nc{\Xp}{X^+}
\nc{\Ouno}{\mathcal{O}_{\varphi \ell}^{(1)}}
\nc{\Cuno}{C_{\varphi \ell}^{(1)}}
\nc{\Otre}{\mathcal{O}_{\varphi \ell}^{(3)}}
\nc{\Ctre}{C_{\varphi \ell}^{(3)}}

\nc{\Omij}[1]{\mathcal{O}^-_{{#1}}}
\nc{\Opij}[1]{\mathcal{O}^+_{{#1}}}
\nc{\Xmij}[1]{X^-_{{#1}}}
\nc{\Xpij}[1]{X^+_{{#1}}}
\nc{\Ounoij}[1]{\mathcal{O}^{(1)}_{\varphi \ell ,{{#1}}}}
\nc{\Cunoij}[1]{C^{(1)}_{\varphi \ell ,{{#1}}}}
\nc{\Otreij}[1]{\mathcal{O}^{(3)}_{\varphi \ell ,{{#1}}}}
\nc{\Ctreij}[1]{C^{(3)}_{\varphi \ell ,{{#1}}}}

\nc{\Sij}[1]{S_{{#1}}}
\nc{\Tij}[1]{T_{{#1}}}